\documentclass[twocolumn,pra,aps]{revtex4}\usepackage{psfrag}
\usepackage{epsfig}
\usepackage{amsmath}
\usepackage{amssymb}
\usepackage{bbm}
\usepackage{pifont}
\usepackage{pstricks,pst-node}
\usepackage{psfrag}
\newcommand{\bra}[1]{\ensuremath{\langle#1|}}
\newcommand{\ket}[1]{\ensuremath{|#1\rangle}}
\newcommand{\braket}[2]{\ensuremath{\langle #1|#2\rangle}}
\newcommand{\ketbra}[2]{\ensuremath{| #1 \rangle \langle #2 |}}

\newcommand{\be}{\begin{equation}}
\newcommand{\ee}{\end{equation}}

\newcommand{\avg}[1]{\ensuremath{\langle #1 \rangle}}

\newcommand{\id}{\openone}

\newcommand{\im}{\text{i}}
\newcommand{\adop}{a^{\dagger}}
\newcommand{\aop}{a}
\newcommand{\aodop}{a_0^{\dagger}}
\newcommand{\aoop}{a_0}

\newcommand{\ain}{a_{\text{in}}}
\newcommand{\adin}{a^\dagger_{\text{in}}}

\newcommand{\bdop}{b^{\dagger}}
\newcommand{\bop}{b}

\newcommand{\hc}{\text{H.c.}}

\newcommand{\ketcat}{\ensuremath{|\Psi \rangle}}

\newcommand{\ie}{{\it i.e.}}
\newcommand{\eg}{{\it e.g. }}

\newcommand{\etal}{{\it et al.}}

\newcommand{\Htot}{H_{\text{tot}}}
\newcommand{\Hfm}{H_{\text{m}}^{\text{f}}}
\newcommand{\Hfc}{H_{\text{cav}}^{\text{f}}}
\newcommand{\Hfout}{H_{\text{out}}^{\text{f}}}
\newcommand{\Hffree}{H_{\text{free}}^{\text{f}}}
\newcommand{\Hicavout}{H_{\text{cav-out}}^{\text{i}}}
\newcommand{\Hidiel}{H_{\text{diel}}^{\text{i}}}

\newcommand{\Etot}{E(\xx)}
\newcommand{\Ec}{E_\text{cav}(\xx)}
\newcommand{\Ecc}{\mathcal{E}_\text{cav}(\xx)}
\newcommand{\Et}{\mathcal{E}_\text{tw}(\xx)}

\newcommand{\Ef}{E_\text{free}(\xx)}

\newcommand{\HLM}{H_{\rm LM}}

\newcommand{\HLC}{H_{\rm LC}}

\newcommand{\HSC}{H_{\rm sc}}
\newcommand{\HSG}{H^\Gamma_{\rm sc}}
\newcommand{\HSK}{H^\kappa_{\rm sc}}
\newcommand{\HST}{H^\Gamma_{\rm tw}}
\newcommand{\HS}{H_{\rm sh}}
\newcommand{\HOM}{H_{\rm OM}}
\newcommand{\Htotr}{H^\text{r}_{\rm tot}}
\newcommand{\Htotb}{H^\text{b}_{\rm tot}}
\newcommand{\HRN}{H_{\rm rn}}

\newcommand{\Ls}{\mathcal{L}_{\text{sc}}[\rho]}
\newcommand{\Lm}{\mathcal{D}_{\text{m}}[\rho]}
\newcommand{\Lc}{\mathcal{L}_{\text{cav}}[\rho]}

\newcommand{\Htotp}{H'_{\text{tot}}}

\newcommand{\xx}{\bold{x}}
\newcommand{\kk}{\bold{k}}
\newcommand{\rr}{\bold{r}}

\newcommand{\uu}{\bold{u}}
\newcommand{\vv}{\bold{v}}
\newcommand{\pp}{\bold{p}}

\begin{document}

\title{Optically Levitating Dielectrics in the Quantum Regime: Theory and Protocols}

\author{O. Romero-Isart$^{1,*}$ \footnote[0]{\scriptsize$^{\rm *}$ These authors have contributed equally to this work.}}
\author{A. C. Pflanzer$^{1,*}$}
\author{M. L. Juan$^2$}
\author{R. Quidant$^{2,3}$}
\author{N. Kiesel$^{4}$}
\author{M. Aspelmeyer$^4$}
\author{J. I. Cirac$^1$}
\affiliation{$^1$Max-Planck-Institut f\"ur Quantenoptik,
Hans-Kopfermann-Strasse 1,
D-85748, Garching, Germany.}
\affiliation{$^2$ICFO--Institut de
Ci\`encies Fot\`oniques, E-08860, Castelldefels, Spain}
\affiliation{$^3$ICREA-- Instituci\'o Catalana  de Recerca i
Estudis Avan\c cats, E-08010, Barcelona, Spain}
\affiliation{$^4$ Faculty of Physics, University of Vienna, Strudlhofgasse 4, A-1090 Vienna, Austria  }

\begin{abstract}
We provide a general quantum theory to describe the coupling of light with the motion of a dielectric object inside a high finesse optical cavity. In particular, we derive the total Hamiltonian of the system as well as a master equation describing the state of the center of mass mode of the dielectric and the cavity field mode. In addition, a quantum theory of elasticity is used in order to study the coupling of the center of mass motion with internal vibrational excitations of the dielectric. This general theory is applied to the recent proposal of using an optically levitating nanodielectric as a cavity optomechanical system~\cite{romeroisart10,chang10}. On this basis, we also design a light-mechanics interface to prepare non-Gaussian states of the mechanical motion, such as quantum superpositions of Fock states. Finally, we introduce a direct mechanical tomography scheme to probe these genuine quantum states by time of flight experiments.
\end{abstract}

\maketitle

\section{Introduction}

The field of optical trapping and manipulation of small neutral particles using the radiation pressure force of lasers was originated in 1970 by the seminal experiments of Ashkin~\cite{Ashkin70}. Over the course of the next 40 years, the techniques of optical trapping and manipulation have stimulated revolutionary developments in the fields of atomic physics, biological sciences, and chemistry~\cite{ashkinbook}. In physics, the progress in optical cooling and manipulation of single atoms opened up a plethora of novel perspectives. The precise control over the atomic degrees of freedom has created applications ranging from atom interferometry~\cite{Cronin09}, quantum simulations of condensed matter systems with ultracold gases~\cite{bloch08}, and the implementation of quantum gates for quantum computation purposes~\cite{Zoller05}. 

More recently, the possibility to apply the techniques of optical cooling and manipulation to the mechanical degree of freedom of larger objects, such as micromirrors or cantilevers, has established a very active research field -- cavity quantum optomechanics~\cite{Asp08, Kippenberg08, Marquardt09, Karrai09, genes09, Aspelmeyer10, Thourhout10}. 
Future applications range from ultra-high sensitivity detectors of mass- or force~\cite{Giscard09,Geraci10} and quantum transducers for quantum computation purposes~\cite{Rabl10, Stannigel2010, Hammerer09, Cleland04}, to their potential of being an ideal testbed for the investigation of fundamental aspects of quantum mechanics, such as the quantum-to-classical transition~\cite{Bouwmeester03, kleckner08}. In most optomechanical systems, the mechanical oscillator is unavoidably attached to its suspension providing a thermal contact that limits the isolation of the mechanical motion -- thus preventing longer coherence times.
\begin{figure}
 \includegraphics[width=\linewidth]{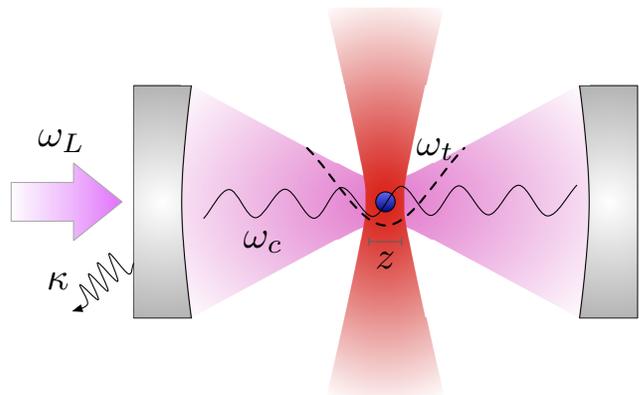}
\caption{Schematic representation of the setup. A nanodielectric is confined by optical tweezers which provides a trapping frequency of $\omega_t$. The nanodielectric is placed inside an optical cavity with resonance frequency $\omega_c$, decay rate $\kappa$, and is driven by a laser at a frequency $\omega_L$.}
\label{Fig:Cavitytweezers}
\end{figure}
A potential improvement to better isolate the system is to use optically levitating nanodielectrics as a cavity quantum optomechanical system~\cite{romeroisart10, chang10} (see also~\cite{Barker10, Singh10}). This consists in optically trapping a nanodielectric by means of optical tweezers inside a high finesse optical cavity, see Fig.~\ref{Fig:Cavitytweezers} for an illustration. Using standard optomechanical techniques~\cite{Mancini98, Marquardt07, WilsonRae07, WilsonRae08, Genes08}, the center of mass (CM) motion could be cooled to its quantum ground state in the harmonic potential created by the optical tweezers. Owing to the fact that it is levitating, the dielectric is not attached to any mechanical object, allowing for a very good thermal isolation, even at room temperature.  More recently, both theoretical and experimental research along this direction has been reported. In \cite{Geraci10}, levitating nanospheres placed close to a surface have been proposed to test forces at very small scales in order to explore corrections to the Newtonian force. Remarkably, an experiment measuring the instantaneous velocity of the Brownian motion of a particle, a glass bead levitated in air, has been reported in~\cite{tongcang10}. Besides, other aspects have been investigated, such as the possibility of Doppler cooling a microsphere \cite{Barker10B}, a scheme to measure the impact of air molecules into the nanodielectric \cite{Yin10}, and the possibility to use a ring cavity to cool and trap polarizable particles \cite{Schulze10}. It can thus be foreseen that a new generation of exciting experiments, aiming at bringing levitating dielectrics into the quantum regime, will eventually take place in the near future. Indeed, from a broad perspective, this project aims at extending the techniques developed during the last decades of optical cooling and manipulation of atoms (\eg like in cavity QED with single atoms and molecules~\cite{Gan99, Vul00, Vul01, miller05}), back to the nanodielectrics that were first used in the times of birth of optical trapping~\cite{Ashkin71, Ashkin74, Ashkin76}. This  experimental challenge, if successful, would allow to test quantum mechanics at unprecedented scales. On this basis, a general quantum theory to describe and predict the phenomena to be encountered in these potential experiments is timely. 

In this article, we aim at contributing to this goal by developing a general quantum theory to describe the coupling of light with the mechanical motion of dielectrics in high finesse optical cavities. Starting from the total Hamiltonian of the system, we derive a master equation which takes into account the effects caused by the scattering of light: decoherence of the mechanical motion, decrease of the cavity finesse, and a non-negligible renormalization of the scattering force. Additionally, we utilize a quantum elasticity theory to describe the effect of elastic deformations of the dielectric object. This provides us with a quantitative expression of  the coupling between the center of mass motion and the internal vibrational modes. This theory is applied to the particular proposal of cavity optomechanics with levitating nanodielectrics~\cite{romeroisart10, chang10}.
We then focus on ``post ground state"~\footnote{By post ground state optomechanics we mean the eventual experimental situation when the ground state of the mechanical oscillator has been achieved by laser cooling (see the recent experiment reporting the preparation of the ground state in a high-frequency mechanical oscillator~\cite{O'Connell2010a}). In this situation, one can think about applications and protocols starting from the ground state of the mechanical oscillator.} optomechanics, and design a light-mechanics interface to prepare non-Gaussian states of the mechanical system, such as a quantum superpostion of Fock states. In particular, we develop three different protocols with different features, together with a formalism which is required to describe these input-ouptut problems in the Schr\"odinger picture. The non-Gaussian light-mechanics interface can be interpreted as an effective way to have non-linear effects in optomechanical systems~\cite{sankey10}. Finally, we introduce a scheme to perform direct full tomography of the mechanical state by imaging the nanodielectric after some time of flight. 

The article is organized as follows: in Section~\ref{sec:results}, we provide a detailed summary of the results. The derivation of the total Hamiltonian as well as the master equation of the theory is addressed in Sec.~\ref{sec:interaction} (some details regarding light scattering are left to App.~\ref{app:scattering}). The derivation of the trapping using optical tweezers and the optomechanical coupling can be found in App.~\ref{app:trappingandcoupling}. The theory part is completed by the introduction of the quantum theory of elasticity in Sec.~\ref{sec:elasticity}. We briefly discuss ground state cooling in App.~\ref{app:cooling}, and the typical experimental parameters in App.~\ref{app:experimental}. 
The second part of the article, the description of the protocols, can be found in Sec.~\ref{sec:protocols} (and in App.~\ref{sec:outputmodes}), where three different ways to interface light with the center of mass motion of the nanodielectric are introduced. The article is rounded off by a proposal to perform full tomography of the mechanical state in Sec.~\ref{sec:tomography}, and we finish by stating the conclusions and discussing further directions in Sec.~\ref{sec:conclusions}.

\section{Summary of results}~\label{sec:results}

This section aims at providing a general roadmap, summarizing the results presented in this manuscript without providing the proofs or mathematical derivations. 

\subsection{Theory}

Here we develop a quantum theory to describe the coupling of light with the mechanical motion of dielectric objects in high finesse optical cavities, see Fig.~\ref{Fig:Cavitytweezers}. The assumptions that are made in the theory are the following:
\begin{enumerate}
\item The dielectric object has a constant relative dielectric constant $\epsilon_r$, as well as, a homogeneous density $\rho$.
\item The electric fields are assumed to be scalar, that is, we do not consider polarizations. We assume a three dimensional field for the free electric field outside the cavity, and a one-dimensional field along the cavity axis for the output field of the cavity. 
\item The object is assumed to be absorption-free and therefore only elastic scattering processes are considered. The effects of light absorption are thus neglected~\cite{chang10,romeroisart10}.
\end{enumerate}
These assumptions are made in order to ease the derivation of the theory and do not imply any fundamental simplification. Indeed, non-homogeneity and polarizations can be incorporated easily. Moreover, as it is shown in the quantum theory of elasticity, the center of mass mode is decoupled from the internal vibrations for sufficiently small objects, and therefore can be treated independently.

Thus, the total Hamiltonian of the system can be written as a sum of free (interacting) terms, labeled with the super-index f (i),%
\be
\Htot = \Hfm + \Hfc + \Hfout + \Hffree + \Hicavout + \Hidiel.
\ee
The first term $\Hfm=p^2/2M$ is the kinetic energy of the center of mass position along the cavity axis. The motion along the transverse direction of the cavity is not relevant in the theory~\footnote{For experimental purposes, the motion along the transverse action should be cooled by external means (\eg feedback cooling) in order to keep the trap stable.}.  The energy of the cavity mode $\aop$ is given by $\Hfc=\omega_c \adop \aop$ (we assume $\hbar=1$), where $\omega_c$ is its resonance frequency. The energy of the free modes is given by $ \Hffree = \int d\bold{k}|\kk| \adop(\bold{k})  \aop(\bold{k})$, and the energy of the ouptut modes of the cavity by $\Hfout = \int_0^\infty d\omega \omega \aodop(\omega) \aoop(\omega)$~\footnote{As usually done in cavity QED, we are indeed double counting the output modes by writing them separately from the free modes, which is correct since they have zero measure.}.
The interaction between the cavity mode and the output modes is described by the usual term  
$\Hicavout =\im \int_0^\infty d\omega \gamma(\omega) \left(\adop \aoop(\omega)-\hc \right)$,
where  the coupling strength is approximated by $\gamma(\omega)\approx \kappa/\pi$, around the resonance frequency, where $\kappa$, is the decay rate of the cavity \cite{gardinerbook}. 

The last term $\Hidiel $ is the crucial one describing the interaction between the electric field and the dielectric object. In the most general form, it can be written as
\be \label{eq:idielI}
\Hidiel =-\frac{1}{2}  \int_{V(\bold{r})} d\bold{x} P(\bold{x}) E(\bold{x}),
\ee 
where $P(\bold{x})$ is the polarization of the object and the integration is performed over the volume of the dielectric object $V$ with center of mass coordinate $\bold{r}$. This Hamiltonian is the starting point for the theoretical discussion. Assuming $P(\xx)= \alpha_p E(\xx)$, one obtains 
\be \label{eq:interactionI}
\Hidiel = -\frac{\epsilon_{\rm c} \epsilon_0}{2}\int_{V(\bold{r})} d^3x [E(\bold{x})]^2,
\ee
where $\epsilon_c=3(\epsilon_r-1)/(\epsilon_r+2)$, $\epsilon_r$ being the relative dielectric constant of the nanodielectric. This can be obtained by connecting the quantum expression of the polarization field in the object with the classical relation. This part of the Hamiltonian is the key ingredient of the theory, and applies for any size and shape of the object. The total electric field inside the object can be now written as a sum of three parts $\Etot=\Ec+\Ef+\Et$, where $\Ec$ contains the cavity modes, $\Ef$ the free modes, and $\Et$ is the classical part of the electric field describing the optical tweezers. By plugging $\Etot$ into Eq.~\eqref{eq:interactionI} the following terms are obtained:
\begin{enumerate}
\item  $[\Et]^2$ creates a harmonic trap with a frequency 
\be
\omega_t^2=\frac{4  \epsilon_{\rm c}}{\rho c} \frac{I}{W_t^2},
\ee
where $I$ is the laser intensity, $\rho$ the density of the dielectric object, $c$ the speed of light, and $W_t$ the laser waist. For the typical experimental parameters discussed in App.~\ref{app:experimental} it is of the order of MHz. This field provides the harmonic trap of the mechanical oscillator with the Hamiltonian $ \omega_t \bdop \bop$, where $\bop$ ($\bdop$) is the annihilation (creation) operator of the center of mass phonon mode along the cavity axis.

\item The cavity field $[\Ec]^2$ gives rise to the optomechanical coupling $g_0 \adop \aop (\bdop + \bop)$, where the coupling strength is given by
\be
g_0=-z_{0} \frac{\epsilon_{\rm c}\omega_c^2 }{ c} \frac{V}{V_c}.
\ee
Here $z_0=(2 M \omega_t)^{-1/2}$ is the zero point motion of the ground state, $V_c=L \pi W_c^2/4$ the cavity volume, $L$ the cavity length, and $W_c$ the laser waist of the cavity. For the experimental parameters discussed in App.~\ref{app:experimental},  $g_0$ is of the order of tens of Hz.

\item The contribution $2\Ef (\Ec+\Et)$ is responsible for scattering processes. This term describes the process of elastic scattering of cavity photons, as well as photons of the tweezers, into free modes. The term $2\Et \Ec $ leads to scattering events already taken into account in $2\Ef \Et$ as well as a shift in both the trapping frequency and the equilibrium position of the object.

\item The term $[\Ef]^2$ yields a negligible coupling between the center of mass mode and the vacuum fluctuations of the electromagnetic field which is negligible. 
\end{enumerate}

Starting from the total Hamiltonian $\Htot$ and the terms given by the total electric field $\Etot$, one can derive a master equation describing the state of the cavity mode $\aop$ and the mechanical mode $\bop$, given by the density matrix $\rho$, by tracing out the free modes $\aop(\kk)$ and the ouptut modes $\aoop(\omega)$. The master equation is given by:
\be
\dot \rho (t)=\im [\rho,\HOM' + \HRN]  + \Lc+\Ls + \Lm,
\ee
with the following contributions:
\begin{enumerate}
\item The optomechanical Hamiltonian in the non-displaced frame~\footnote{As shown later in the article, the derivation of the master equation is done by extracting the classical part of the cavity field and by displacing the cavity operators $\aop \rightarrow \aop + \alpha$. Note however that the cavity operators appearing in $\HOM$ are not displaced.} is given by
\be
\HOM=\omega_t \bdop \bop + \omega_c \adop \aop   +g_0 \adop \aop (\bdop+\bop),
\ee
describing the coherent coupling between the cavity mode and the mechanical mode.
\item The dissipation term
\be
\Lc = \kappa(2 \aop \rho \adop - \adop \aop \rho- \rho \adop \aop),
\ee
describing the photon losses, at a rate $\kappa$, due to the imperfection of the cavity mirrors.
\item The new cavity field dissipation term
\be
\Ls = \kappa_\text{sc}(2 \aop \rho \adop - \adop \aop \rho- \rho \adop \aop),
\ee
due to losses, at a rate $\kappa_\text{sc}$ caused by the scattering of cavity photons out of the cavity. Although the theory is valid for any size of the object, we provide in the article the expression of $\kappa_\text{sc}$ for objects smaller than the optical wavelength. Indeed, in order to keep the high finesse of the cavity, that is, $\kappa_\text{sc}/ \kappa < 1$, the objects have to be of the order of $100$ nm in case of spherical objects~\cite{chang10}.
\item The mechanical  diffusion term
\be
\Lm= \Gamma_\text{sc} \left[ \bop + \bdop, \left[ \bop + \bdop, \rho \right] \right],
\ee
which, although it does not yield mechanical damping, does generate decoherence of the motional mechanical state due to light scattering. We also provide in this article the expression of $\Gamma_\text{sc}$ for sub-wavelength objects, which contains the contribution of both the optical tweezers and the cavity field. For spherical objects of the order of $100$ nm, $1/\Gamma_\text{sc} \sim 0.1$ ms.

\item Finally, an additional coherent term $\HRN$ is obtained, which renormalizes the optomechanical Hamiltonian due to QED effects. The effects of this term are discussed in more detail in~\cite{pflanzer11}. 
\end{enumerate}

This theory is complemented by a quantum theory of elasticity. Starting from the classical expression of the Lagrangian density of an elastic object, the deformable field is expressed in terms of normal modes for the case of a vanishing external potential. Then, by plugging in the external potential given by the light matter interaction, one obtains an expression describing coupling between the normal modes. This can be quantized canonically and provides a quantum description of the coupling of the center of mass mode with the internal vibrational modes, as well as the coupling of the light with the internal modes. This theory can be applied to objects at the micron scale, where the internal modes have frequencies of the order of $10^{11}$ Hz, and thus are decoupled from the $10^6$ Hz center of mass mode. This allows us to adiabatically eliminate the internal modes, merely leading to a renormalization of the trapping frequency. This  correction is many orders of magnitude smaller than $\omega_{\rm t}$ and consequently represents a negligible effect. This justifies the separate treatment of the center of mass degree of freedom in the developed theory which is applicable to objects at the the micron scale.

\subsection{Protocols}

In the second part of the article we focus on how to bring these systems into the quantum regime. In particular we design a light-mechanics interface which consists in injecting non-Gaussian states  of light, such as superposition of Fock states, to the mechanical oscillator.

First of all, as usually done in optomechanics, we describe the effect of driving the cavity with a strong driving field at frequency $\omega_L$~\cite{Marquardt07, dobrindt08, WilsonRae08}. We transform the total Hamiltonian of the system into a displaced frame which describes the states on top of the steady state of the cavity field, the mechanical state, as well as the output modes of the cavity. We particularly emphasize the displacement that one needs to do in the output modes in order to be able to describe input-output problems in the Schr\"odinger picture. The main change in the Hamiltonian is in the optomechanical coupling. In particular, when the driving is red-detuned $\Delta=\omega_c -\omega_L=\omega_t$, the coupling term, in the resolved sideband regime and after the rotating wave approximation, has the beam-splitter interaction form $g(\adop \bop + \aop \bdop)$. Here, $g=\sqrt{n_\text{ph}}g_0$ is an effective optomechanical coupling, which is enhanced by the square root of steady state cavity photons. This allows one to reach the strong coupling regime $g > \kappa$. When driving the cavity with the blue-detuned field $\Delta=\omega_c -\omega_L=-\omega_t$, one induces the two mode squeezing interaction term $g(\adop \bdop + \aop \bop)$. With these tools, we design and derive different protocols to perpare non-Gaussian states.

The first protocol, called the reflected one-photon, consists in sending one resonant photon on top of the driving field and measuring the reflected part. More specifically, the cavity is driven with a red-detuned field in order to induce the beam-splitter interaction. The mechanical object is assumed to be in its ground state. Now, on top of the driving field, a one-photon pulse centered at the resonance frequency is sent into the cavity. Impinging the cavity, part of it enters and part is reflected. At the time $t_h$, where the part of the beam that has entered the cavity is transferred to the mechanical oscillator through the beam-splitter interaction, the light field is switched off. Consequently, the light mode corresponding to the reflected photon is entangled with the mechanical system inside the cavity. We can obtain the exact form of the state by solving the input-output problem in the Schr\"odinger picture. The state in the displaced frame is given by
\be
\ket{\psi(t_h)}= c_b(t_h)\ket{10\Omega}+\int_{-\omega_L}^\infty   c(\omega,t_h) a_0^\dagger(\omega)d \omega \ket{00\Omega},
\ee
where $\ket{n_b n_a \Omega}$ describes a state with $n_b$ phonons, $n_a$ photons, and all the output modes in the vacuum state. Here, the coefficients $c_b(t)$ and $c(\omega,t)$ are obtained analytically. This makes clear that, by measuring the quadrature of the output mode of the photon, one prepares a superposition state of zero and one phonon with coefficients given by the outcome of the measurement. Some technical issues are addressed in detail for this protocol in the manuscript, such as the fact that in the original frame, the state $\ket{\psi(t_h)}$ is displaced by a considerable amount. This makes it challenging to obtain a significant signal-to-noise ratio in the measurement of the output mode.

An extension of the reflected one-photon protocol is the perfect mapping protocol. In this protocol, the possibility to time-modulate the laser intensity, and consequently the optomechanical coupling, is exploited. Then, by imposing the condition that the output field, with the transformed Hamiltonian, is zero, we can obtain the equation of motion for the optomechanical coupling $g(t)$. 
%
%
Its solution yields a modulation of $g(t)$ such that the light pulse sent on top of the driving field is perfectly absorbed and therefore the non-Gaussian state of the light is transferred to the mechanical oscillator. In this section we also discuss some technical details regarding the transformation of the Hamiltonian that has to be performed carefully since there are time-dependent displacements.

The two protocols require a moderately strong coupling $g \sim \kappa$. As an alternative, we also derive a protocol, called teleportation in the bad-cavity limit, which does not require the strong coupling regime. Once the mechanical oscillator is in the ground state, it consists in driving the cavity with a blue-detuned field, such that the two mode squeezing interaction is induced inside the cavity. This Hamiltonian creates a two mode squeezed state between the mechanical oscillator and the light field leaking out of the cavity. The squeezing parameter of this state is a measure of the degree of entanglement. 
This entangled state can then be used to teleport a non-Gaussian light state from outside of the cavity to the mechanical oscillator. In this section, we also discuss in detail the effect of the driving field and the possibility to choose the initial state to be teleported in order to prepare a particular state in the mechanical oscillator.

This part of the article is concluded by providing a direct method to perform full tomography of the state of the mechanical oscillator. In general optomechanical systems, tomography can in principle be done by coupling the mechanical resonator to a well-controlled quantum system (\eg a qubit), and then measuring the quantum system. The method we provide here performs direct tomography of the mechanical oscillator. It is well known that by measuring the rotated quadrature phase operator
\be
\mathcal{X} (\theta) = e^{\im \theta} \bdop + e^{-\im \theta} \bop,
\ee
for all $\theta$, one can reconstruct the Wigner function and therefore obtain all the information about the state of the harmonic oscillator~\cite{Lvovsky01}. Our protocol achieves that by measuring the position of the nano-dielectric after some time of flight. In the Heisenberg picture, the momentum operator in the harmonic potential evolves like
\be
p(t)= \im p_m (\bdop e^{\im \omega_t t} - \bop e^{-\im \omega_t t}),
\ee
where $p_m=(M \omega_t /2)^{1/2}$. Thus the momentum operator $p(t)/p_m= \mathcal{X} (\omega_t t_e + \pi/2)$ is directly related to the rotated quadrature phase operator. By switching off the optical tweezers at $t_e$, letting the nanodielectric fall, and measuring the position at some later time $t_f$, one obtains $z(t_e+t_f)\approx(t_f-t_e)p(t_e)/M$, which is a measurement of the momentum operator. By repeating the experiment at different times $t_e$ full tomography of the mechanical state can be performed. In this section, we discuss the experimental parameters and an extension of the protocol to amplify the oscillation prior to the time of flight.

\section{Total Hamiltonian and master equation of the theory} \label{sec:interaction}
In this section we develop the main part of the theory to describe the coupling of light with the mechanical motion of dielectric  objects in high finesse optical cavities. In particular, in Sec.~\ref{sec:setup} we derive the total Hamiltonian. We focus on the light-matter interaction term in Sec.~\ref{Sec:LMint}, and make the connection between the microscopic theory with the macroscopic parameters such as the dielectric constant of the object. In Sec.~\ref{sec:hamiltonian} we show how to obtain the optomechanical Hamiltonian for the case of levitating objects inside a cavity. 
Finally in Sec.~\ref{sec:scattering} we derive the effects of scattering of light embedded in the total Hamiltonian of the theory. Indeed, we show how to derive a master equation in order to describe the time evolution of the state of both the cavity and mechanical mode. We provide the quantitative expression of the light scattering parameters for subwavelength spheres.
 
\subsection{Setup and general Hamiltonian} \label{sec:setup}

We consider a dielectric object with center of mass position $\rr$ and a dielectric constant 
\be
\epsilon_r(\xx) = \left\{ \begin{array}{ll}
         \epsilon_r & \mbox{if $\xx \in V(\rr)$};\\
        1 & \mbox{if $\xx \not \in V(\rr)$},\end{array} \right.
\ee
where $V(\rr)$ is the spatial region of the object, centered at $\rr$, with volume $V$, density $\rho$ and mass $M=\rho V$. The homogeneity of the dielectric constant inside the object is chosen for simplicity, the non-homogeneous case can be incorporated easily. As shown in Sec.~\ref{sec:elasticity}, the center of mass degree of freedom of dielectrics at the micron-scale is decoupled from its relative modes. Hence, we will only consider the center of mass degree of freedom in the following analysis. We suppose that the dielectric object is inside an optical cavity. We define the cavity mode, characterized by an annihilation (creation) operator $\aop$ ($\adop$), the modes coupled to the cavity mirror $\aoop(\omega)$ ($\aodop(\omega)$), we call them output modes, and the other free modes of the electromagnetic field $\aop(\kk)$ ($\adop(\kk)$). We use a 1D theory to describe the modes coupled to the cavity mode, therefore denoting them by $\omega=k$ (we use $c=1$ in the protocols part of the article) and a scalar 3D theory for the rest of the modes. The effects of polarization can also be easily incorporated but will be neglected for simplicity. 

The total Hamiltonian of the dielectric object inside the optical cavity can be written as
 \begin{equation}
 \begin{split}
 \label{eq:Hamtot}
\Htot =&\Hfm + \Hfc + \Hfout + \Hffree  + \Hicavout + \Hidiel.
 \end{split}
 \end{equation}
The superscript $f$ ($i$) labels free (interacting) terms.
The first term, $\Hfm=p^2/2M$ describes the kinetic energy of the center of mass
mode along the cavity axis. The motion along other directions is not considered since it decouples from the motion along the cavity axis, as it will become clear in the following. Note however that in analogy to trapping and cooling of ions, these other modes are assumed to be cooled by external means (\eg by feedback cooling) in order to make the trap stable (see a recent article for a 3D ground state cooling scheme based on using different cavity modes~\cite{Yin10}). The next three terms describe the free radiation parts
 \begin{eqnarray}\Hfc&=& \omega_c a^\dagger a,\\
\Hfout &=& \int_0^\infty d\omega \omega \aodop(\omega) \aoop(\omega),\\
 \Hffree &=& \int d\bold{k}|\kk| \adop(\bold{k})  \aop(\bold{k}), 
  \label{eq:Hfreeterms}
 \end{eqnarray}
 of the cavity mode $\aop$ with the mode frequency $\omega_{\rm c}$, the output modes $\aoop(\omega)$ and the free modes $\aop(\kk)$. As it is usually done in the context of cavity QED~\cite{dutrabook}, we double-count some of the modes by considering the ones coupled to the cavity mode separately. However, since they have zero measure, this does not affect the correctness of the description. The interaction between the cavity mode and the free modes is described by~\cite{gardinerbook}
 \begin{eqnarray}
\Hicavout &=&\im \int_0^\infty d\omega \gamma(\omega) \left(\adop \aoop(\omega)-\hc \right),
  \end{eqnarray}
where the coupling strength is described by $\gamma(\omega)$ and can be assumed to be constant over a large
frequency interval centered around $\omega_c$ with a value $\gamma(\omega)=\sqrt{\kappa/\pi}$, where $\kappa$ is the decay rate of the cavity~\cite{gardinerbook}.

Finally, $\Hidiel$ describes the interaction between the light field and the dielectric object which can be written as
\be \label{eq:idiel}
\Hidiel =-\frac{1}{2}  \int_{V(\bold{r})} d\bold{x} P(\bold{x}) E(\bold{x}).
\ee 
Here, $P(\bold{x})$ is the polarization field and the volume integral is performed over the volume of the object $V$ around the center of mass position $\rr$. This term is the central equation in the rest of subsections: in Sec.~\ref{Sec:LMint}, we develop this interaction term by relating the polarization field with the electric field; in Sec.~\ref{sec:hamiltonian}, we consider the proposal of using an optically levitating nanodielectric as a quantum optomomechanical system, and derive the optomechanical Hamiltonian;  and in Sec.~\ref{sec:scattering}, we show how this term can be used to derive the effects induced by light scattering.

As a final remark and for later convenience, let us define the light mechanics (LM) and light-cavity (LC) part of the Hamiltonian as
\be 
\begin{split} \label{eq:LMLC}
\HLM&= \Hfc+\Hffree +\Hidiel,\\
\HLC &= \Hfout + \Hicavout,
\end{split}
\ee
such that $\Htot=\Hfm+\HLM+\HLC$.

\subsection{Light-matter interaction Hamiltonian}\label{Sec:LMint}

Let us here focus on the key part of the total Hamiltonian, the light-matter interaction term $\Hidiel$, Eq.~\eqref{eq:idiel}. First of all, note that for the typical light intensities, the polarization field responses linearly to the electric field, such that $P(\xx)=\alpha_p E(\xx)$. The parameter $\alpha_p$ can in principle be computed by performing a quantum theory of the object by considering its atomic properties. However, this involved task is not necessary since one can relate $\alpha_p$ to macroscopic properties of the object, such as the dielectric constant $\epsilon_r$.  
Comparing the resulting relation between the polarization and the electric field for the macroscopic~\cite{born_wolf,jackson} and microscopic case:
\be
\begin{split}
&\text{macroscopic:} \hspace{1em}  P(\xx)=3\epsilon_0\frac{ \epsilon_r-1}{\epsilon_r+2} E(\xx) \equiv \epsilon_c \epsilon_0 E(\xx) \\
&\text{microscopic:} \hspace{1em}  P(\xx)=\alpha_p E(\xx) \\
\end{split}
\ee
one can identify the microscopic constant to the macroscopic one, 
\be
\alpha_p=\epsilon_c \epsilon_0.
\ee
Then, plugging this back into the Hamiltonian Eq.~\eqref{eq:idiel}, we obtain the final form of the light-matter interaction Hamiltonian 
\be \label{eq:Hamint}
\Hidiel=-\frac{\epsilon_c \epsilon_0}{2}\int_{V({\bold{r}})}d\xx[E(\xx)]^2.
\ee
In App.~\ref{app:scattering}, we show how from this Hamiltonian one can derive the scattering equation that can be used to compute the electric field inside the object.


\subsection{Optomechanical Hamiltonian} \label{sec:hamiltonian}

The expression for $\Hidiel$ obtained in the above section, see Eq.~\eqref{eq:Hamint}, is the key ingredient to develop our theory. Let us now apply it to the particular proposal of using levitated objects in a cavity as an optomechanical system~\cite{romeroisart10,chang10}. This experimental setup consists of an external optical tweezers as well as a laser driving the cavity at frequency $\omega_L$, see Fig.~\ref{Fig:Cavitytweezers}. The total electric field inside the object can be decomposed into
\be
\Etot=\Ec+\Ef,
\ee
where the $\Ec$ is the cavity electric field and $\Ef$ the free electric field. The free electric field contains a classical part due to the optical tweezers generated by the laser, which can be incorporated as $\EfÊ\to \Ef+\Et$, where $\Et$ describes the optical tweezers~\cite{Ashkin86}, see App.~\ref{app:trappingandcoupling} for its expression. The implementation of the driving laser is carefully done in Sec.~\ref{sec:transformed} where, in order to keep the structure of the total Hamiltonian, we will have to displace all the output modes $\aoop(\omega$) as well
as the cavity mode and the mechanical mode. Let us here discuss the terms that will be obtained  when plugging the total electric field
\be
\Etot=\Ec+\Ef+\Et
\ee
in $\Hidiel$, Eq.~\eqref{eq:Hamint}. By doing so, one obtains six different terms, which have been discussed in the summary of results section, and therefore are only summarized here. The term $[\Et]^2$ which accounts for the harmonic trapping, see App.~\ref{app:trappingandcoupling},  with a trapping frequency
\be
\omega_t^2=\frac{4  \epsilon_{\rm c}}{\rho c} \frac{I}{W_t^2},
\ee
where $I$ is the field intensity, $W_t\approx \lambda/(\pi \mathcal{N})$ the laser waist, $\mathcal N$ the numerical aperture, and $k$ the wave vector number.
This allows us to quantize the CM motion along the $z$ axis as $z=z_0 (\bdop+\bop)$, where $z_0=(2 M\omega_t)^{-1/2}$. The term $[\Ec]^2$ describes the coupling of the
cavity mode and the motional state, see  App.~\ref{app:trappingandcoupling}. By considering the center of mass position of the object to be placed at the maximum slope of the standing wave one obtains the standard optomechanical coupling $g_0 \adop \aop (\bdop+\bop)$. The coupling strength is given by
\be
g_0=-z_{0} \frac{\epsilon_{\rm c}\omega_c^2 }{ c} \frac{V}{V_c},
\ee
where $V$ is the volume of the object, $V_c=L \pi W_c^2/4$ the cavity volume, $L$ the cavity length, and $W_c=[\lambda L/(2 \pi)]^{1/2}$ the waist of the cavity field. This term also yields a shift of the resonance frequency of the cavity, see App.~\ref{app:trappingandcoupling}. The two scattering terms $2 \Ef (\Et+\Ec)$, which describe the scattering of cavity photons and the laser light from the optical tweezers, are addressed in Sec.~\ref{sec:scattering}. The term $2\Ec \Et$, which yields a shift of both the trapping frequency and the equilibrium position as well as some scattering processes already taken into account in the term $2 \Ef \Et$, is discussed in the next section. Finally, the term $[\Ef]^2$ accounts for a negligible coupling of the center of mass motion with vacuum fluctuations of the electromagnetic field.

Hence, the total Hamiltonian can be now written as
\be \label{eq:Htot}
\begin{split}
\Htot
&=\HOM +\HSC + \HLC + \HS,
\end{split}
\ee
where $\HOM$ is the standard optomechanical Hamiltonian
\be
\HOM=\omega_t \bdop \bop + \omega_c \adop \aop  +g_0 \adop \aop (\bdop + \bop).
\ee
The term $\HLC$ was already introduced in Eq.~\eqref{eq:LMLC} and we have defined the scattering Hamiltonian
\be \label{eq:scattering}
\HSC= \Hffree - \epsilon_c \epsilon_0 \int_{V(\rr)} d\xx\Ef(\Ec + \Et), 
\ee
which is studied in Sec.~\ref{sec:scattering}. The shift term is given by $\HS=- \epsilon_c \epsilon_0 \int\Ec \Et d\xx$.
Finally, note that by tracing out the output modes of the cavity,  $\aoop(\omega)$ in $\HLC$, one obtains the usual master equation
\be \label{eq:master}
\dot \rho (t)=\im [\rho,\HOM+\HLC+\HS] + \Lc 
\ee
where
\be
\Lc = \kappa(2 \aop \rho \adop - \adop \aop \rho- \rho \adop \aop).
\ee
This term describes the photon losses (at the decay rate of the cavity $\kappa$) through the end mirrors of the cavity,  and is treated in detail in, e.g., \cite{gardinerbook}. 


%
\subsection{Light Scattering} \label{sec:scattering}

In the present setup both trapping and cooling is achieved by light, yielding an optomechanical system without thermal contact to other mechanical objects. However, the effect of scattering of light has to be considered~\cite{chang10} . In this section we provide a framework to study elastic light scattering within a quantum theory.  While the derivation of this framework for arbitrary dielectric objects will be provided elsewhere~\cite{pflanzer11}, here we will discuss the general theory and present the results obtained for objects smaller than the optical wavelength. 

The key term of the total Hamiltonian describing the effects of light scattering is $\HSC$,  defined in Eq.~\eqref{eq:scattering}. First, we are interested in the case in which the cavity is strongly driven. Then a coherent steady state is present in the cavity field, which we explicitly consider by decomposing the cavity field into a quantum part plus a coherent (classical) part, $\Ec \to \Ec + \Ecc $, where $\Ecc$ is the classical one. A detailed discussion of this transformation is given in Sec.~\ref{sec:transformed}. In this framework, the scattering term of the Hamiltonian can be written as
\be
\begin{split}
\HSC=\Hffree + \HSG+\HSK+\HST,
\end{split}
\ee
where we have defined:
\be
\begin{split} \label{eq:HS}
\HSG&=- \epsilon_c \epsilon_0 \int_{V(\rr)} d\xx\Ef \Ecc\\
\HSK&=- \epsilon_c \epsilon_0 \int_{V(\rr)} d\xx\Ef \Ec \\
\HST&=- \epsilon_c \epsilon_0 \int_{V(\rr)} d\xx\Ef \Et.
\end{split}
\ee
Additionally, the displacement of the cavity field also modifies the shift term of the Hamiltonian, which now reads $\HS=- \epsilon_c \epsilon_0 \int(\Ec+\Ecc) \Et d\xx$. Whereas the first term is already included in $\HST$ (since in $\Ef$ we integrate over all the electromagnetic modes without excluding the cavity mode), the second term yields a shift of the trapping frequency as well as of the equilibrium position, as discussed in App.~\ref{app:shift}.
  
The time evolution of the density matrix describing the center of mass motion $\rho$ is determined by tracing out the free modes using a Markovian master equation approach. Its utilization is justified for the following reasons: first, due to the fact that the reservoir of free modes of the electromagnetic field is very large, the bath density operators are not significantly changed by the interaction, such that one can always assume that its density matrices remain constant in time $\rho_{E}\approx\rho_{E}(0)$. Second, the Markov assumption, stating that the decay of correlations is much faster than any other time scale in the system, $\tau_{\mathrm{corr}}\ll\tau_{\mathrm{S}}$, is fulfilled: the Hamiltonian operator contains terms $\propto \int \exp(-\im\omega_kt)d^3k$, with a distribution of $\omega_k$ peaked around $\omega_{L}$, which is the fastest time scale in the system. Any correlations in the bath scale with the mode frequencies $\omega_k$ and thus decay very quickly. The details of derivation will be provided in~\cite{pflanzer11}. In here we just report the final result,
\be
\dot \rho (t)=\im [\rho,\HOM' + \HRN]  + \Lc+\Ls + \Lm.
\ee
Comparing to the case without scattering, Eq.~\eqref{eq:master}, the new terms are the following: first, two dissipation terms
\be
\begin{split}
\Ls &= \kappa_\text{sc}(2 \aop \rho \adop - \adop \aop \rho- \rho \adop \aop), \\
\Lm &= \Gamma_\text{sc} \left[ \bop +\bdop,\left[ \bop + \bdop, \rho \right]\right].
\end{split}
\ee
The first one $\Ls$ describes cavity photon losses due to events in which cavity photons are scattered out of the cavity. This term, which contributes to the decay rate of the cavity, is obtained from $\HSK$. For spherical objects smaller than the wavelength, $\kappa_\text{sc}$ is given by
 \be
\begin{split} \label{eq:kscs}
\kappa_{\rm sc}&=\frac{\epsilon_c^2V^2k_c^4c}{16 \pi V_c},
\end{split}
\ee
where we have assumed the sphere to be trapped at the maximum slope of the standing wave, $k_cz\approx \pi/4$, and the cavity volume is defined as $V_c=\pi/(4dW_c^2)$. As a check of the theory, one can compare this expression with the decay rate one would estimate using the Rayleigh cross section $\sigma_R$. With this, the optical finesse is estimated as $\mathcal{F}_{\rm R}=\pi W_c^2/\sigma_R$, and consequently the decay rate $\kappa_R=c\sigma_R/(4 V_c)$. The Rayleigh cross section neglecting the different polarization of the incoming and scattered light (to be consistent with the rest of the article) is
\be
\sigma_R=\frac{4 \pi}{9}  k_c^4 R^6 \epsilon_c^2.
\ee
By plugging $\sigma_R$ into $\kappa_R=c\sigma_R/(4 V_c)$ one recovers the same expression as derived in the theory, Eq.~\eqref{eq:kscs}.

The term $\Lm$ describes recoil heating due to elastic scattering out of the cavity. This is obtained from $\HSG$ and $\HSG$. The heating rate $\Gamma_\text{sc}=\Gamma^{\text{cav}}_\text{sc}+\Gamma^{\text{tw}}_\text{sc}$ has two contributions from the cavity photons and from the tweezers light. For sub-wavelength dielectric spheres, it reads
\be
\Gamma_\text{sc} =\frac{\epsilon_c^2 k_c^6 V }{6 \pi \rho \omega_t} \left( \frac{P_t}{\omega_L \pi W_t^2} + \frac{ n_\text{ph}c }{2 V_c} \right),
\ee
where the first(second) term is $\Gamma^{\text{tw}}_\text{sc}$($\Gamma^{\text{cav}}_\text{sc}$) and $P_t$ denotes the power of the trapping laser. Here, $n_\text{ph}$ is the number of steady state photons in the cavity due to the driving field. We remark that surprisingly these results are in agreement with the ones obtained in the standard theory of decoherence~\cite{joos85, joos_decoherence, schlosshauer_decoherence}. Another important remark is that the dissipative term $\Lm$ does not create any mechanical damping of the oscillator, but only diffusion. Hence, in this case it might be misleading to discuss the mechanical quality factor of the harmonic oscillator. We think that it is more appropriate to consider  coherent times $1/\Gamma_\text{sc}$ as usually done in the case of ions. The harmonic oscillator can oscillate  without mechanical damping~\footnote{A non-zero mechanical damping will of course be induced by other sources of decoherence, such as the background gas pressure, see \cite{romeroisart10}, laser shot noise, or from other sources, see supplementary information in~\cite{chang10} for an explicit analysis of all of them. These mechanical damping sources yield very high mechanical quality factors $> 10^{10}$, as predicted in~\cite{romeroisart10, chang10}.} for very long times, however, a quantum state prepared in the harmonic oscillator will lose coherence in a time scale of  $1/\Gamma_\text{sc}$. Using typical numbers, see App.~\ref{app:experimental}, this corresponds to time scales of the order of $0.1$ ms for nanospheres.

Finally, $\HSG$ and $\HST$ also yield 
 an additional force $\HRN$, which for subwavelength spheres modifies the trapping frequency by $\omega_t \rightarrow \omega_t (1-\epsilon_c)$, and the optomechanical coupling by $g_0 \rightarrow g_0 (1-\epsilon_c)$. This contribution has to be added to the optomechanical Hamiltonian $\HOM$, and represents a non-negligible QED renormalization of the Hamiltonian due to virtual photon exchange. This QED effects will be addressed in Pflanzer \etal, {\em in preparation}, where we will show that higher perturbative terms, e.g., corresponding to emission and reabsorption of two photons are suppressed by several orders of magnitude for small spheres.

\section{Quantum elasticity}\label{sec:elasticity}

Let us now address the coupling of the center of mass (CM) motion mode to other internal vibrational modes. One can model the dielectric as an object containing $N$ constituents, in this case atoms, that are coupled to each other by mutual interactions, here modeled by springs. The entire nanodielectric inherits $N$ different modes, one of them is the center of mass mode; a collective movement of all the system's constituents into the same direction. The other modes can be described as movements of the different constituents relative to each other, mediated by the springs. All of these different modes are also coupled to each other, which, in turn, influences their form and lifetime. In principle, one can couple any of these modes to light, especially if the object is sufficiently large. We are particularly interested in the CM mode in this article. We will focus on investigating the influence of the relative modes, also denoted as vibrational modes, on the center of mass mode treating them as a source of decoherence: the vibrational modes can in principle take the role of a thermal bath and prevent ground state cooling of the CM degree of freedom. In order to investigate this source of noise, we use an elasticity theory for quantum systems in this section. After introducing a field characterizing the object's deformation, we determine the vibrational eigenmodes in Sec.~\ref{sec:eigenmode}. Thereafter, we analyze the effect of an additional external potential and the induced interactions between CM and vibrational modes in Sec.~\ref{sec:addpot}. Finally, in Sec.~\ref{sec:smallsphere} we discuss the effect for small objects and obtain an effective Hamiltonian by adiabatically eliminating the internal modes.  
\subsection{Vibrational eigenmodes}\label{sec:eigenmode}
\begin{figure} [t]
  \includegraphics[width=.8\linewidth]{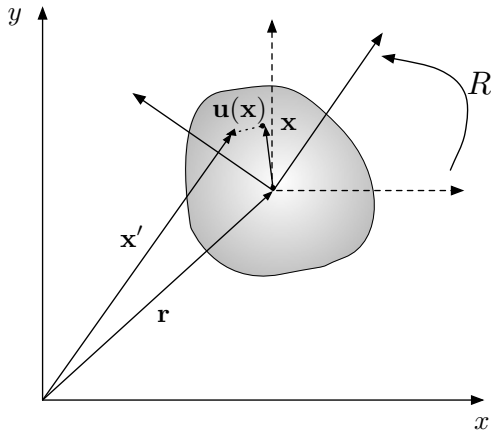}
\caption{Coordinates used to describe a position $\bold{x}'$ within an arbitrary dielectric object given by $\bold{x}'=\bold{r}+\hat{R}(\phi_1, \phi_2, \phi_3)\left(\uu(\bold{x})+\bold{x}\right)$, where $\bold{r}$ denotes the center of mass, $\uu(\bold{x})$ a small displacement from the equilibrium position $\bold{x}$ and $R(\phi_1,\phi_2, \phi_3)$ the Euler rotation matrix acting on the entire object.}
\label{Fig:coordinates}
\end{figure} 

Let us start by defining the coordinate $\xx'$, which describes a point in the dielectric object. As illustrated in Fig.~\ref{Fig:coordinates}, this can be written in the most general form as
\be
\bold{x}'=\bold{r}+\hat{R}(\phi_1, \phi_2, \phi_3)\left(\uu(\bold{x})+\bold{x}\right),
\ee
where $\bold{r}$ denotes the center of mass position. In the coordinate system centered at the center of mass position,  $\xx$ is the coordinate describing an equilibrium point and $\uu(\bold{x})$ its deformation field. The term $\hat{R}(\phi_1, \phi_2, \phi_3)$ is the Euler rotation matrix with the Euler angles $\phi_1, \phi_2, \phi_3$ that is used to rotate the coordinates $\bold{x}$ and $\uu(\bold{x})$. Note that the center of mass position can be defined as $\bold{r}=\int d^3x \rho(\bold{x})\bold{x}'/(\int d^3x \rho(\bold{x}))$, with $\rho(\bold{x})$ denoting the system's density distribution. Therefore, $\int d^3x \rho(\bold{x})[\xx+\uu(\bold{x})]=0$. In order to guarantee that $\rr$ remains the CM coordinate in case of a vanishing deformation field, \ie~$\uu(\xx)=0$,  one requires $\int d^3x \rho(\bold{x})\xx=0$, and consequently, the deformation field always has to fulfill $\int d^3x \rho(\bold{x}) \uu(\bold{x})=0$.
 
The system's Lagrangian in the presence of a general three-dimensional potential $V(\bold{x}')$ reads \cite{landaubook7, goldsteinbook}%
\be \label{eq:lagrange0}
\mathcal{L}=\int_V d^3x \left[\frac{1}{2}\rho(\bold{x})\dot{\bold{x}'}^2-V(\bold{x}')-V_\text{E}(\xx)\right].
\ee
The elasticity potential is given by
\be
V_\text{E}(\xx)=\frac{1}{2}\sum_{i,j}\sigma_{ij}(\xx)\epsilon_{ij}(\xx),
\ee
 where
 \be
 \begin{split}
 \epsilon_{ij} (\xx)&=\frac{1}{2}\left(\frac{\partial u_i(\xx)}{\partial x_j}+\frac{\partial u_j(\xx)}{\partial x_i}\right) \\
 \sigma_{ij}(\xx)&=2\mu\epsilon_{ij}(\xx)+\lambda\delta_{ij}\sum_k\epsilon_{kk}(\xx)
 \end{split}
 \ee
 are the elasticity and the stress tensor. The Lam\'e constants are defined as $\lambda=\sigma Y[(1+\sigma)(1-2\sigma)]^{-1}$ and $\mu=Y [2(1+\sigma)]^{-1}$, with $\sigma$ being the Poisson ratio and $Y$ the Young modulus characterizing the elastic properties of the material. One can now replace the expression of $\xx'$ in the kinetic part of Lagrangian and obtains
 \be
\begin{split} \label{eq:lagrange}
\mathcal{L}&=\frac{1}{2}M\dot{\bm{r}}^2+\frac{1}{2}\sum_iI_i\dot{\phi}_i^2+\frac{1}{2}\int_V d^3x\rho(\bold{x})\dot{\uu}(\bold{x})^2\\
&-\int_V d^3x\left[V(\bold{x}')+V_\text{E}(\xx)\right],
\end{split}
\ee
where the dots denote time derivatives and $I_i$ is the object's moment of inertia. We have used that in the kinetic part of the Lagrangian, the rotational, vibrational, and center of mass degrees of freedom decouple~\cite{goldsteinbook}. 

Let us now determine the unperturbed vibrational eigenmodes of the system, that is, the modes obtained without considering the potential density $V(\bold{x}')$. In the following, we will assume for simplicity the homogenous case $\rho(\xx)=\rho$, the non-homogeneous case can be incorporated easily. Also, we will omit the rotational modes since they decouple without the presence of the external potential. Let us first derive the Hamiltonian by defining the CM momentum as $p_{i}=\partial \mathcal{L}/\partial\dot{r_i}$ and the momentum density as $v_i(\xx)=\partial \mathcal{L}/\partial \dot u_i(\xx) $, leading to
\be\label{eq:hamtot0}
H_0=\frac{\pp^2}{2M}+ \int_Vd^3x \left( \frac{[\vv(\xx)]^2}{2\rho}+V_\text{E}(\xx) \right).
\ee 
One can determine the vibrational eigenmodes by separating variables in the corresponding equation of motion for $\uu(\bold{x},t)$, which reads~\cite{landaubook7}
\be
\rho \ddot{\uu}(\bold{x},t)=\mu \nabla^2 \uu(\bold{x},t)+\frac{1}{2}\lambda \nabla[\nabla \cdot \uu(\bold{x},t)].
\label{wave_general}
\ee
Here,  $\uu(\bold{x},t)$ can be separated into transversal and longitudinal oscillation modes, $\uu(\bold{x},t)=\uu_\perp(\bold{x},t)+\uu_{||}(\bold{x},t)$, where $\nabla \cdot \uu_\perp(\bold{x},t)=0$ and $\nabla \times \uu_{||}(\bold{x},t)=0 $, and either open or periodic boundary conditions can be used. The longitudinal modes describe compression waves propagating at velocity $c_{||}=[(\lambda+2\mu)/\rho]^{1/2}$ and the transversal modes torsion wave propagating at $c_\perp=[\mu/\rho]^{1/2}$. In the following we will only consider the longitudinal modes along the cavity axis. By expanding the elasticity field along the cavity axis for these eigenmodes $u^0_n(z)$ (which are normalized as $\int_V d^3 x u^0_n(z) u^0_m(z)=\delta_{nm} V $), one has 
\be
\begin{split}
u(z,t)&=\sum_n u^0_n(z)Q_n(t) \\
v(z,t)&=\sum_n u^0_n(z)P_n(t),
\end{split}
\ee
where $P_n(t)=\rho \,\dot Q_n(t)$. By plugging this decomposition into the Hamiltonian Eq.~\eqref{eq:hamtot0}, one obtains after some algebra
 \be
 H_{0}=\frac{p^2}{2M}+\sum_n \left[ \frac{P_n^2}{2 M}+\frac{1}{2} M \omega_n^2 Q_n^2\right],
 \label{eq:hamode}
 \ee
where the frequency of the internal modes is given by
\be \label{eq:vibfrequency}
\omega^2_n =\frac{\lambda (1-\sigma)}{M\sigma}\int_V d^3x \left[\frac{d }{dz} u_n(z) \right]^2. 
\ee
The eigenmodes $u^0_n(z)$ have to be chosen accordingly to the geometry of the object. We will discuss the specific form of the mode and the value of the parameters in Sec.~\ref{sec:smallsphere}.

At this position, it is straightforward to perform a canonical quantization of the eigenmodes $Q_n$, by considering them as operators fulfilling the canonical commutation rules $\left[ Q_n, P_m \right]=\im$. As already done in the previous sections, the momentum operator of the CM will also be quantized with the external harmonic trap.
 
\subsection{Effect of the external potential}\label{sec:addpot}

The external potential $V(\xx')$ can in principle effect a coupling between the  rotational, the center of mass, and the vibrational degrees of freedom. In case of a purely isotropic harmonic potential, it can be easily verified that the coupling vanishes. On the other hand, for arbitrarily shaped objects, the external anharmonic part of the potential effects some coupling  between all degrees of freedom. In the following we assume spherical objects, for which the direct coupling between the CM and the rotational degrees of freedom vanishes. Even in the case of a prolate spheroid, the coupling is negligible~\cite{chang10}. For spherical objects, there is only an indirect coupling between the CM and the rotations, mediated by the vibrational modes, which is negligible and will be omitted hereafter. Therefore, with these assumptions one can consider the center of mass mode to be decoupled from the rotational motion, and we consequently omit the rotational modes in the rest of the section. One can then focus on the one-dimensional case derived in the previous section by only considering the longitudinal modes. 

The total Hamiltonian, including the external potential, is hence given by
\be
H=H_0 + \int_{V} d^3 x V(z') 
\ee
Assuming that the deformations $u(z)$ are small and that the object is trapped at $r\approx0$, one can expand $V(z'=z+u(z)+r)$ to second order in $r$ and $u(z)$, which leads to 
\begin{equation} \label{eq:Hdiel}
\begin{split}
H =&H_0+ r \int_{V}d^3x V'(z) +\frac{r^2}{2} \int_{V} d^3xV''(z)\\
&+\frac{1}{2}\sum_{n,m} Q_nQ_m\int_{V}d^3x u^0_n(z) u^0_m(z)V''(z) \\
&+\sum_n Q_n  \int_{V} d^3x u^0_n(z) V'(z)\\
&+r \sum_n Q_n  \int_{V}d^3x u^0_n(z) V''(z),
\end{split}
\end{equation}
where the primes denote spatial derivatives. By recalling that the external potential is, in our case, given by the light-matter interaction term Eq.~\eqref{eq:Hamint}, that is  $V(\bold{x}')=-\epsilon_c \epsilon_0 [E(\bold{x}')]^2/2$, one can understand the terms appearing in Eq.~\eqref{eq:Hdiel} as follows:
\begin{enumerate}
\item The term $r \int_{V}d^3x V'(z) $ yields the optomechanical coupling of the center of mass mode as described in App.~\ref{Sec:optocoupl}.
\item  The second term $r^2 \int_{V} d^3xV''(z)/2$ describes the harmonic trap of CM mass given by the optical tweezers, as described in App.~\ref{exttrap}.
\item The term $Q_nQ_m\int_{V}d^3x u^0_n(z) u^0_m(z) V''(z)/2$ gives a correction to the harmonic trap for the internal modes as well as a coupling between internal modes.
\item The first new interesting term is $Q_n  \int_{V} d^3x u^0_n(z) V'(z)$, which describes an optomechanical coupling between the internal modes and the cavity field.
\item Finally, the most relevant term for our purposes is $r Q_n  \int_{V}d^3x u^0_n(z) V''(z)$, which describes the coupling between the vibrational degrees of freedom $Q_n$  and the center of mass mode $r$.
\end{enumerate}

Taking into consideration these terms, one can now write the center of mass mode as $r=z_0 (\bdop+\bop)$, where $z_0$ is the ground state size, as used in Sec.~\ref{sec:interaction}, and the internal modes as 
$Q_n=q_{0,n} \left(c_n+ c^{\dagger}_n \right)$,
with $q_{0,n}=(2 M \omega'_n)^{-1/2}$. Note that, due to the additional external traping with frequency $\omega_{\rm t}$, the effective vibrational frequencies are changed to $\omega'_n=(\omega_{\rm t}^2+\omega_n^2)^{1/2}$ (we will omit the prime hereafter). The new part that has to be added to the total Hamiltonian $\Htot$, see Eq.~\eqref{eq:Htot}, which takes into account the presence of internal modes, is given by  
\be \label{eq:elasticity}
\begin{split}
H_\text{E}&=\sum_n \omega_n c_n^{\dagger} c _n+\sum_n g_n(\aop,\adop) ({c}_n+{c}_n^{\dagger}) \\
&+\sum_{n,m}^{\infty}\xi_{nm}({c}_n+{c}_n^{\dagger})({c}_m+{c}_m^{\dagger})\\&+\sum_n^{\infty}\gamma_{n}({c}_n+{c}_n^{\dagger})({b}+{b}^{\dagger}).
\end{split}
\ee
The coupling between the cavity field (which depends on the cavity mode $\aop$) and the vibrational modes  is given by $g_n(\aop, \adop)=q_{0,n}\int_V d^3xV'(z)u^0_n(z)$. The coupling between the internal modes is $\xi_{nm}=q_{0,n}q_{0,m}\int_{V}d^3x u^0_n(z) u^0_m(z)V''(z)/2$. Finally, the coupling between the CM mode and the vibrational modes is given by
\be \label{eq:gammaelast}
\gamma_n=z_0q_{0,n}\int_V d^3x V''(z)u^0_n(z).
\ee
Summing up this subsection, we have derived the quantized Hamiltonian describing the coupling between the CM and the vibrational modes in the presence of an external potential density. It can be shown that for a harmonic external potential, the CM mode is decoupled from the internal ones since $V''(z)$ is constant and by recalling that $\int_V d^3x u^0_n(z)=0$, one obtains $\gamma_n=0$. In the next section, we estimate the order of magnitude of the parameters for objects smaller than the optical wavelength in the presence of the anharmonic potential given by the standing wave.

\subsection{Sub-wavelength spheres}\label{sec:smallsphere}
 
First of all, let us estimate the order of magnitude of the internal vibrational frequencies, see Eq.~\eqref{eq:vibfrequency}, for the case of a sphere of radius $R$. To get an estimation of the order of magnitude, for simplicity one can just use the eigenmode $u^0_n(z)=\cos(k_n z)$ with $k_n=n\pi/(2R)$, obtained for a cube of length $2R$ and with open free periodic boundary conditions. Then, using typical values of the Young's elasticity module $Y$ and the Poisson constant $\sigma$ (see App.~\ref{app:experimental}), the vibrational frequences are of the order $\omega_n \approx 10^{11}\rm{ Hz}$ ($\omega_n \sim n c_{||}/R$). Note that comparing this to the typical values of the CM frequency $\omega_t \sim 10^6$ Hz, the internal frequencies are five orders of magnitude larger for objects of the order of $100$ nm.

This large difference in frequencies between the CM modes and the internal modes enable us to adiabatically eliminate the vibrational energy levels. It can be shown that this approximation is justified by solving the equation of motion for the CM and vibrational operators by applying Laplace transformations. The solution obtained in this way contains parts oscillating at frequencies $\omega_{\rm t}$ and $\omega_{n}$, where all terms oscillating at $\omega_{n}$ are suppressed by a factor $\omega_{\rm t}/\omega_n \ll 1$. Thus, it is well-justified to neglect these terms and to perform an adiabatic elimination. One can perform this by eliminating the vibrational levels on top of the steady state, yielding the result that the only effect is a shift of the trapping frequency of the CM mode given by
\be
\left(\frac{\omega'_t}{\omega_t}\right)^2=1-\sum_n\frac{4 \gamma_n^2}{\omega_t(\omega_n-\omega_t)}(2 \avg{c_n^\dagger c_n}+1),
\ee
where $\avg{c_n^\dagger c_n}$ is the occupation number of phonons in the vibrational mode $n$. By plugging in typical numbers, one gets a correction to the trapping frequency of $(\omega_t'-\omega_t)/\omega_t\approx 10^{-12}\rm$, which shows that the CM mode is decoupled from the internal modes for objects smaller than the optical wavelength. 

\section{Light-mechanics interface} \label{sec:protocols}

One of the most fascinating perspectives of quantum optomechanics is the possibility to prepare superposition states of objects containing billions of atoms, and therefore,  to  test quantum mechanics at larger scales. Already in the early days of this research area, several groups proposed to create non-classical states of a movable mirror~\cite{bose97, mancini97, bose99}. The idea behind these proposals is to use the optomechanical interaction to entangle a small quantum system with the macroscopic object. By observing the state of the small quantum system, the creation and loss of the non-classical state in the macroscopic system can be monitored. This idea was also used in~\cite{armour02}, where the coupling between a micromechanical resonator and a Copper box is proposed in order to prepare entanglement between the quantum system (Copper box) and the cantilever.  We remark that an experiment has been recently reported in~\cite{O'Connell2010a}, where coherent control of single phonon has been achieved in a high-frequency micromechanical oscillator. In Marshall \etal ~\cite{Bouwmeester03} (see also \cite{kleckner08}) a scheme to prepare a superposition state of two distinct locations of a mirror through the optomechanical interaction with a single photon has been proposed. All these ideas pose a major challenge to an experimental realization mainly due to the following reasons: (i) the coupling between the small quantum system and the macroscopic mechanical system is not strong enough and (ii) the mechanical system suffers from its fast decoherence due to the thermal contact.

In this section, we show a possible way to circumvent these two restrictions. We propose two protocols to strongly couple a non-Gaussian light state to a mechanical object. This is achieved by using a driving field which enhances the interaction into the strong-coupling regime (the interaction time has to be faster than the decoherence times). This enhancement of the optomechanical coupling by the driving field was suggested in~\cite{Marquardt07, dobrindt08} and experimentally observed in~\cite{Groeblacher09b}. Then, on top of the driving field, which is red-detuned, a quantum light state is sent into the cavity which is transferred to the mechanical system by the strong coupling. This idea has been introduced in \cite{romeroisart10} (see also \cite{akram10, Khalili10}). Additionally, we propose an alternative protocol that uses the weak coupling regime to prepare non-Gaussian states. These protocols, which can be applied to general optomechanical systems, are ideally suitable for optically levitating nanodielectrics, since they do not have a thermal contact~\cite{romeroisart10, chang10}, and thus possess longer coherent times.

These general light-mechanics interface protocols allow us to prepare non-Gaussian states by using a Gaussian Hamiltonian. Their key ingredient is that one uses non-Gaussian input states (similar ideas have been used in the context of quantum computation ~\cite{lloyd99, knill01}). Hence, these protocols represent an effective and simple way to produce non-linearities in optomechanical systems, a goal that is intensively pursued (see for instance \cite{sankey10}).

Finally, we remark that in case of a levitating object light scattering yields decoherence of the mechanical state with a rate given by $\Gamma_\text{sc}$. For sufficiently small objects, this can be made much smaller than $\kappa$. In the following, where we are interested in designing the protocols, we will neglect the effects of light scattering (more precisely, the term $\HSC$) by assuming that the protocols can be realized on a time scale much shorter than $1/\Gamma_\text{sc}$. For other optomechanical setups, decoherence in the mechanical system could be incorporated easily into the protocols.

This section is organized as follows: first, in Sec.~\ref{sec:transformed} we transform the total Hamiltonian of the system in order to account for the driving field of the laser. We then present three different protocols in Secs.~\ref{subsec:reflected}, \ref{subsec:perfectmapping}, and \ref{subsec:teleportation}.

\subsection{Driving field: displaced frame \& initial state} \label{sec:transformed}

In this section we transform the Hamiltonian $\Htot$, see Eq.~\eqref{eq:Htot}, into $\Htotp$ in order to incorporate the driving field of the cavity and keep a close structure of the Hamiltonian. This allows us to describe the quantum states on top of the steady state which will be used in the protocols. Throughout the article we will use either the original frame, in which states are described according to $\Htot$, or the transformed or displaced frame, in which states are related to $\Htotp$. Then, we describe in both frames the form of the total initial state that one obtains after cooling  the mechanical oscillator to the ground state.

\subsubsection{Displaced frame}

In this section we will perform the standard transformation~\cite{Marquardt07, dobrindt08, WilsonRae08} done in quantum optomechanics in order to shift the coherent part of the states obtained when driving the cavity with a laser. However, in contrary to what is usually done, here we also need to displace the output modes since we use them in the light-mechanics interface. 

First, one moves the cavity and the output field to the frame rotating with the laser frequency $\omega_L$. This is described by the unitary operator
\be
U_{\text r}(t)= \exp \left[ -\im \omega_L \left(\adop \aop+\int_0^\infty \aodop(\omega) \aoop(\omega) d\omega \right)t \right].
\ee
To ease the notation, after this transformation we redefine the $\aoop(\omega)$ and $\gamma(\omega)$ such that $\aoop(\omega)\equiv \aoop(\omega+\omega_L)$, and $\gamma(\omega)Ê\equiv \gamma(\omega+\omega_L)$. The total Hamiltonian (ignoring the scattering part $\HSC$ and the shift $\HS$ which is discussed later) reads
\be
\begin{split} \label{eq:basicH}
\Htot =& \Delta_0 \adop \aop+ \omega_t \bdop \bop + g_0 \adop \aop (\bdop + \bop)+ \int_{-\omega_L}^\infty \omega \aodop(\omega)\aoop(\omega)  \\
&+ \im \int_{-\omega_L}^\infty \gamma(\omega)(\adop \aoop(\omega)-\hc),
\end{split}
\ee
where $\Delta_0=\omega_c-\omega_L$.
Then, one displaces the cavity field with the displacement operator $D_a(\alpha)$, the mechanical field with $D_b(\beta)$, and the output modes with $D_\text{out}(\alpha_\omega)$, that is, 
\be
\begin{split}
& D^\dagger_a(\alpha) \aop D_a(\alpha)= \aop + \alpha, \\
& D^\dagger_b(\beta)\bop D_b(\beta) = \bop + \beta, \\
& D^\dagger_\text{out}(\alpha_\omega) \aoop(\omega) D_\text{out}(\alpha_\omega)=\aoop(\omega) + \alpha_\omega.   \\
\end{split}
\ee
 After applying this transformation to the Hamiltonian, one fixes $\alpha$, $\beta$, and $\alpha_\omega$, such that 
the terms in the Hamiltonian that have only one creation or annihilation operator vanish.  This corresponds to solving the following set of equations:
\be
\begin{split}
& \Delta_0 \alpha + 2g_0 \alpha\beta + \im \int_{-\omega_L}^\infty \gamma(\omega) \alpha_\omega d\omega=0, \\
& \omega_t \beta + g_0 |\alpha|^2 =0, \\
&  \int_{-\omega_L}^\infty \omega \aodop(\omega) \alpha_\omega - \im \int_{-\omega_L}^\infty \gamma(\omega) \aodop(\omega) \alpha=0,
\end{split}
\ee
which have the solutions
\be
\begin{split} \label{eq:displacements}
\alpha&= \frac{\Omega_L}{\im \Delta+\kappa}, \\
\beta&= -\frac{g_0|\alpha|^2}{\omega_t}, \\
\alpha_\omega&=  \left(\frac{\Omega_L}{\gamma(0)}- \pi\alpha \gamma(0) \right) \delta(\omega) +\im \alpha \gamma(\omega)\mathcal{P}(\omega^{-1}).
\end{split}
\ee
Here, $\Delta=\Delta_0+ 2 g_0 \beta $, and  $\alpha_0=\Omega_L /\gamma(0)
$, where $\Omega_L= \sqrt{2 P_c \kappa / \omega_L}$, $P_c$ being the laser power. The symbol $\mathcal{P}$ denotes the principal part, and we have used that $\gamma^2(\omega)\approx\kappa/\pi$ in a finite region around $\omega=0$~\cite{gardinerbook} in order to perform the integral 
$\mathcal{P}\int_{-\infty}^\infty\omega^{-1} d\omega = 0$. In the next subsection Sec.~\ref{sec:initialstate},  we show how to obtain the expression of $\alpha_\omega$ from a more physical perspective.%

To sum up, the transformation applied to the Hamiltonian can be defined as $\mathcal{D}\equiv  D_\text{out}(\alpha_\omega) D_b (\beta) D_a (\alpha)$, and the transformed Hamiltonian is given by
\be\label{eq:htottrans}
\begin{split}
\Htot'=& \mathcal{D}^\dagger \Htot \mathcal{D}= \HOM' + \HLC',
\end{split}
\ee
where
\be
\HOM'=\omega_t \bdop \bop + \Delta \adop \aop   +g (\adop +  \aop)(\bdop + \bop)
\ee
is the enhanced optomechanical Hamiltonian, and $\HLC$ is transformed into
\be
\begin{split}
\HLC' =&  \int_{-\omega_L}^\infty \omega \aodop(\omega) \aoop(\omega)d\omega\\ &+ \im \int_{-\omega_L}^\infty\gamma(\omega) (\adop  \aoop(\omega) -\hc)d\omega.
\end{split}
\ee
 Note that Eq.~\eqref{eq:htottrans} has the same structure as Eq.~\eqref{eq:basicH} with the only replacement $\Delta_0 \rightarrow \Delta$, and $g_0 \adop \aop (\bdop +\bop)\rightarrow g (\adop+\aop)(\bdop+\bop)$.
We have defined $g=g_0|\alpha|$, and $\xi =\arg(\alpha)$, and we have redefined  the $a$ ($\aoop(\omega)$) operators as $a'= a e^{-\im \xi}$ ($\aoop'(\omega)= \aoop(\omega) e^{-\im \xi}$) (we omit the tilde hereafter). A crucial remark is that the optomechanical coupling $g$ is enhanced by $\alpha$, which is the square root of the mean number of photons inside the cavity in the steady state. This will allow us to reach the strong coupling $g \sim \kappa$ (where $\kappa$ is the decay rate of the cavity) in the light-mechanics interface.

We remark that in case of using levitating objects, the shift to the trapping frequency as well as the shift in the equilibrium position, given by the Hamiltonian $\HS$, should be taken into account in the $\HOM'$ Hamiltonian. As discussed in App.~\ref{app:shift}, this would imply to change the trapping frequency to $\omega_t \rightarrow \omega_t + \omega_\text{sh}$, and the displacement of the cavity mode to $\beta \rightarrow \beta+ \xi_\text{sh}/\omega$, where $ \omega_\text{sh}$ and $\xi_\text{sh}$ are given in App.~\ref{app:shift}. However, to keep the section in a general form, so that it can also be applied to other optomechanical systems, we will omit this effect hereafter.

The transformed Hamiltonian can now be written in the interaction picture (assuming that the free part is $H_0=\omega_t \bdop \bop + \Delta \adop \aop +  \int_{-\omega_L}^\infty \omega \aodop(\omega) \aoop(\omega)d\omega$) as
\be
\begin{split}
 \Htot^{I}=& g (\adop e^{\im \Delta t} +  \aop e^{-\im \Delta t} )(\bdop e^{\im \omega_t t}  + \bop e^{-\im \omega_t t} ) \\&+  \im \int_{-\omega_L}^\infty \! \! \! \! \! \! \gamma(\omega) (\adop \aoop(\omega) e^{\im ( \Delta t - \omega t)}  -\hc)d\omega.
 \end{split}
\ee
Now, by choosing a red-detuned driving $\Delta=\omega_t$, one can perform  the rotating wave approximation (valid at $ \omega_t \gg g$), and obtain the beam-splitter interaction form of the total transformed Hamiltonian in the Schr\"odinger picture
\be \label{eq:bs}
\begin{split}
\Htotr= \omega_t (\adop\aop+\bdop \bop) +g (\adop  \bop + \aop \bdop )+ \HLC'.
 \end{split}
\ee
Analogously, one can consider a blue-detuned driving $\Delta=-\omega_t$ in order to get the two mode squeezing interaction Hamiltonian:
\be \label{eq:tms}
\begin{split}
 \Htotb=&-\omega_t(\adop \aop-\bdop \bop)+ g (\adop \bdop  +  \aop \bop  ) +\HLC'.
 \end{split}
\ee
These two types of interaction will be used in Sec.~\ref{sec:protocols} to design different protocols  in the light-mechanics interface.

\subsubsection{Initial state} \label{sec:initialstate}

All the protocols that we shall discuss in the next section assume that the initial state is the ground state cooled by the red-detuned field ($\Delta=\omega_t$). As discussed in the previous section and in App.~\ref{sec:outputmodes}, this state is given by
\be
\ket{\text{in}}=\ket{\beta}_\text{b} \otimes\ket{\alpha}_\text{a} \otimes \int_{-\omega_L}^\infty D_\text{out}(\alpha_\omega)d\omega \ket{\Omega}_\text{out}= \mathcal{D} \ket{00\Omega},
\ee
where ``b (a)" labels the subspace of the mechanical mode (cavity mode), ``out" the subspace of the output modes, and $\Omega$ the vacuum state for the output modes. The displacements $\alpha$, $\beta$, and $\alpha_\omega$ are defined in Eqs.~\eqref{eq:displacements}. 

Note that $\ket{\text{in}}$ is an eigenstate of the total Hamiltonian $\Htot$, see Eq.~\eqref{eq:basicH}. This can be trivially demonstrated by using that $\mathcal{D}^\dagger\Htot \mathcal{D}=\Htotr$ (for the red-detuned case Eq.~\eqref{eq:bs}), and that $\Htotr \ket{00 \Omega}=0$, since then one has
\be
\Htot \ket{\text{in}} =\mathcal{D}\mathcal{D}^\dagger \Htot \mathcal{D} \ket{00 \Omega}=\mathcal{D}\Htotr \ket{00 \Omega}=0.
\ee
The state $\ket{\text{in}}$ (reading $\ket{00\Omega}$ in the displaced frame) will be considered as the initial state upon which the protocols are designed using either the beam splitter interaction Eq.~\eqref{eq:bs} or the two mode sequeezing interaction Eq.~\eqref{eq:tms}.

\subsection{Reflected One-photon} \label{subsec:reflected}

In this section, we will present a protocol which strongly couples a one photon state to the mechanical motion of the oscillator. This protocol is general and can be applied to various optomechanical systems. Let us remark that it has already been introduced by some of the authors in  \cite{romeroisart10} and that  related ideas have been reported in \cite{akram10, Khalili10}. In this section we will provide a thorough analysis. In particular, we develop a formalism to solve the input-output formalism in the Schr\"odinger picture in order to be able to describe the final state of the protocol. 

Let us start by sketching the different steps of the protocol:
\begin{enumerate}

\item Cool the mechanical motion to the ground state by the red-detuned driving field.

\item Keep the strong driving field switched on such that the beam-splitter interaction is induced inside the cavity.

\item Impinge the cavity with a resonant single-photon
state, sent on top of the driving field as a result of parametric down conversion followed by a detection of 
a single photon~\cite{Lvovsky01}.

\item When impinging the cavity, part
of the field is reflected and part transmitted~\cite{Duan04}.

\item  The beam-splitter interaction Eq.~\eqref{eq:bs}  caused by the red--detuned laser, swaps the state of light inside the cavity 
to the state of the mechanical motion.

\item  By tuning the width of the light pulse appropriately, one finds that at time $t_h$, one has a maximum mean number of phonons of $1/2$ in the mechanical system. At that time, the driving field is switched off. Then, the entangled state
$\ket{E}_{\text{out},\text{b}} \sim \ket{\tilde 0}_\text{out} \ket{1}_\text{b} +e^{\im \phi} \ket{\tilde 1}_\text{out} \ket{0}_\text{b} $ is prepared. Here $\text{out}$($\text{b}$) stands for the 
reflected cavity field (mechanical motion) of the system, and $ \ket{\tilde 0(\tilde 1)}_\text{out}$
is a displaced vacuum (one photon) light state in the output mode of the cavity $A_\text{out}$.  The phase $\phi$, given by the light-mechanics interaction, is always fixed.

\item At a later time, once the reflected photon is far away from the cavity, a
balanced homodyne measurement of the output mode is performed. The motional state collapses 
into the superposition state $\ketcat_\text{b}=c_0 \ket{0}_\text{b} + c_1 e^{\im \phi}\ket{1}_\text{b}$, where the coefficients 
$c_{0(1)}$ depend on the measurement results.
\end{enumerate}
 
In the following we will analyze carefully the important steps of the protocol. In the shifted frame, the initial state (according to Sec.~\ref{sec:initialstate}), consisting of a photon on top of the ground state of the mechanical oscillator, is given by
\be
\ket{\Psi(0)} =\int_{-\omega_L}^{\infty} \phi^\star_\text{in}(\omega)\aodop(\omega)\ket{00\Omega},
\ee
where $\phi^\star_\text{in}(\omega)$ is the shape of the photon pulse which is assumed to be Gaussian  
\be \label{eq:gaussian}
\phi_\text{in}(\omega)=\left(\frac{2}{\pi \sigma^2}\right)^{1/4}e^{-(\omega-\Delta)^2/ \sigma^2} e^{-\im \omega x_\text{in}}.
\ee
Here, $x_\text{in}$ is the position from which the pulse has been sent (it is considered to be large, $x_\text{in}\gg 0$). $\Delta=\omega_c-\omega_L=\omega_t$ is the detuning, which shows that in the non-rotating frame the pulse is centered at the resonance frequency of the cavity. Note also that one can express the mode function in position space by the Fourier transform $\tilde \phi_\text{in} (x) = \int d\omega \phi_{\text{in}}(\omega) e^{\im \omega x} /\sqrt{2\pi}$.

The time evolved state with the beam-splitter interaction Eq.~\eqref{eq:bs}, $\ket{\psi(t)}=\exp[-\im \Htotr t] \ket{\psi(0)}$ can be expanded in the following basis,
\be
\begin{split} \label{eq:evolvedstate}
\ket{\psi(t)}= &c_b(t) \ket{10\Omega}+c_a(t) \ket{01\Omega}\\&+\int_{-\omega_L}^\infty c(\omega,t) a_0^\dagger(\omega)d\omega  \ket{00\Omega}.
\end{split}
\ee
The time-dependence of the coefficients can be obtained using the Wigner-Weisskopf formalism. By using the Schr\"odinger equation, one obtains
\be
\begin{split}
\dot c_b(t)&= -\im \omega_t c_b(t) -\im g c_a(t),\\
\dot c_a(t)&= -\im \omega_t c_a(t)-\im g c_b(t) +\int_{-\omega_L}^\infty  \gamma(\omega)c(\omega,t) d\omega, \\
\dot c(\omega,t) &=-\im \omega c(\omega,t)-Ê\gamma(\omega)c_a(t).  \\
\end{split}
\ee
This system can be further simplified by formally solving the differential equation corresponding to $c(\omega,t)$, plugging it into the equation for $\dot c_a(t)$, and by using the approximation $\gamma(\omega) \approx \gamma(0) = \sqrt{\kappa/\pi}$. One gets (analagous manipulations have been explicitly done in App.~\ref{sec:outputmodes}):
\be
\begin{split} \label{eq:equationsc}
\dot c_b(t)&= -\im \omega_t c_b(t) -\im g c_a(t), \\
\dot c_a(t)&= -(\im \omega_t+ \kappa) c_a(t)-\im g c_b(t)\\&+  \int_{-\omega_L}^\infty \gamma(\omega)e^{-\im \omega t}  c(\omega,0) d\omega,  \\
\dot c(\omega,t) &=-\im \omega c(\omega,t) -Ê\gamma(\omega)c_a(t). 
\end{split}
\ee
This system of differential equations can be solved by using that $c_a(0)=c_b(0)=0$ and $c(\omega,0)= \phi_\text{in}^\star(\omega)$. After some effort, one obtains
\be
\begin{split}
c_a(t)&= \sqrt{2\kappa} \int_0^t p_-(t-\tau) \tilde \phi^\star_\text{in}(\tau) e^{-\im \omega_t (t-\tau)} d\tau,\\
c_b(t)&= \sqrt{2\kappa} \int_0^t q(t-\tau) \tilde \phi^\star_\text{in}(\tau) e^{-\im \omega_t(t- \tau)} d\tau. \\
\end{split}
\ee
The functions $q(t)$ and $p_{-}(t)$ are defined by
\be \label{eq:pandq}
\begin{split}
p_{\pm}(t) &= e^{-\kappa t/2} \left[ \cosh(\chi t) \pm \frac{\kappa}{2 \chi} \sinh(\chi t)\right],\\
q(t) &= - \im \frac{g}{\chi} e^{-\kappa t/2} \sinh(\chi t),
\end{split}
\ee
where $\chi=\sqrt{\kappa^2/4-g^2}$.
\begin{figure}
\centering
\includegraphics[scale=0.7]{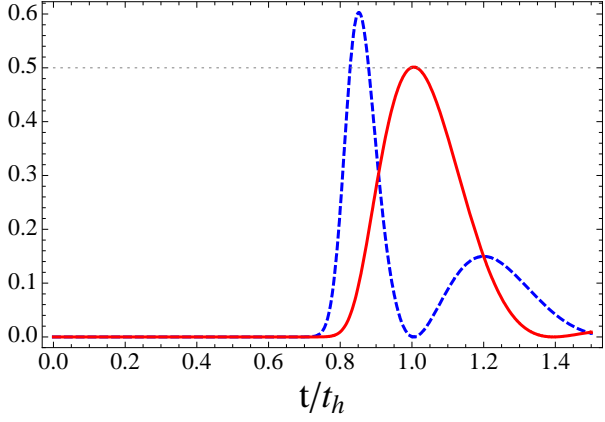}
\renewcommand{\figurename}{Fig.}
\caption{Input-output dynamics after sending a one-photon pulse centered at $x_\text{in}= 5 /\kappa$ from the cavity at $t=0$.  A Gaussian pulse of width $\sigma =5.6 \kappa$ is used. We plot the mean number of phonons in the mechanical system $\bar n_b(t) =|c_b(t)|^2$ (red solid line) and the mean number of cavity photons $\bar n_a(t) =|c_a(t)|^2$ (blue dashed line). We consider the strong coupling regime $g=\kappa$, and tune the width of the pulse so that the maximum mean number of phonons is $\sim 1/2$ (dotted grey line) at $t=t_h$.}\label{Figure21A}
\end{figure}
One can now plot the mean number of phonons $\bar n_b(t) =|c_b(t)|^2$ and photons $\bar n_a(t) =|c_a(t)|^2$, see Fig.~\ref{Figure21A} for some parameters given in its caption.
Note that at $t=t_h$, where
\be
t_h=x_\text{in} +\frac{ \arccos(\kappa/2 g)}{\sqrt{g^2-\kappa^2/4}},
\ee
the mean number of phonons $\bar n_b$ is maximal. By tuning the width of the initial pulse, one obtains that $c_a(t_h) \approx 0$ and $|c_b(t_h)| \approx1/\sqrt{2}$. In this case,  the total state at $t_h$ is given by
\be
\begin{split}
\ket{\psi(t_h)}=& c_b(t_h)\ket{10\Omega}+\int_{-\omega_L}^\infty   c(\omega,t_h) a_0^\dagger(\omega)d \omega \ket{00\Omega}.
\end{split}
\ee
This is an entangled state between the ouptut photon mode, described by the pulse shape $c(\omega, t_h)$, and the mechanical phonon mode. In the non-displaced frame, the state at $t_h$ is described by $\ket{\psi'(t_h)}=\mathcal{D} \ket{\psi(t_h)}$.

At $t=t_h$ the driving field is switched off. However, at this time, there is still a large number of photons $|\alpha|^2$ present inside the cavity. They will leak out of the cavity reducing the classical force that they were exerting on the mechanical system, which is described by the displacement of the mechanical system, $\beta$. In order to compensate this effect, one could move the center of the trap $m \omega_t^2 (x-x_t(t))^2/2$ accordingly, which yields a force term $-m \omega_t^2 x_t(t) x_0 (\bop+\bdop)$, in order to keep the ground state of the harmonic oscillator. Another effect of this leaking out of photons is that the coefficient $c_b(t_h)$ will be decreased at some later time. Note however, that one could send a pulse that generates  $|c_b(t_h)|>1/\sqrt{2}$ such that, after the decrease due to the leaking out of the coherent photons, one obtains 
$|c_b(t>t_h)|=1/\sqrt{2}$. The discussion on how to compute and estimate this effect is done in App.~\ref{app:off}.

Here, we just approximate the state at $t \gg t_h$ as
\be
\ket{\psi(t)}= c_b(t_h) e^{-\im \omega_t (t-t_h)} D_\text{out} \ket{10 \Omega}+ D_\text{out} A^\dagger_\text{out,t} \ket{00\Omega},
\ee
where $D_\text{out}$ only displaces the output modes with $\alpha_\omega$. Also, we have defined the output mode of the cavity as
 \be
 A^\dagger_\text{out,t}=\int_{-\omega_L}^\infty  \phi_\text{out}(\omega,t) a_0^\dagger(\omega)d \omega,
 \ee 
where $  \phi_\text{out}(\omega,t)=c(\omega,t_h)  e^{-\im \omega (t-t_h)}$. Note that the displacement is only in the output modes since the photons inside the cavity, and the consequent radiation force into the mechanical object, are not present at times $t \gg t_h$ since the driving field is switched off.

\subsubsection{Measurement of the output mode}

The final step of the protocol is the measurement of the quadrature of the output mode $A_\text{out,t}$,  that is
\be
X_\text{out,t} =A^\dagger_\text{out,t} + A_\text{out,t}.
\ee
This measurement consists in integrating the signal of a continuous measurement with the mode shape given by $\phi_\text{out}(\omega,t)$. 

More generally, the output operator $A_\text{out}=\int_{-\omega_L}^\infty \phi(\omega)\aoop(\omega)$ can be written as a combination of mode operators at position $x$ by using $\aoop(\omega)=\int_0^\infty dx e^{-\im \omega x} \aoop(x) dx /\sqrt{2\pi}$, leading to
\be
A_\text{out}=\int_0^\infty \tilde \phi(x) \aoop(x) dx.
\ee
Note that now the mode $\aoop(x)$ can be measured at the position $x=x_d$ of the detector at time $t$, by the relation $\aoop(x_d,t)=\aoop(x=x_d-t,0)$. Then, by a continuous measurement of $\aoop(x_d,t)$, one has access to the measurement of all $\aoop(x)$ and consequently, also to $A_\text{out}$ by integrating the signal over $\tilde \phi(x)$ (note that $A_\text{out}$ is a linear combination of the independent modes $\aoop(x)$).

After the continuous measurement, let us assume one obtains the value $x_\text{out}$. Then, the superposition state in the mechanical object,  given by
\be
\ketcat=\frac{1}{\sqrt{2}}\left( c_0 \ket{0}_b + c_1 \ket{1}_b \right),
\ee
is prepared, where $c_{0(1)}=\bra{x_\text{out}}  1(0)\rangle$.  The measurement of the quadrature poses an experimental challenge. If we define the two orthogonal states $\ket{\pm}_0 =\ket{\Omega} \pm  A^\dagger_\text{out,t} \ket{\Omega}$ and their displaced states $ \ket{\pm}=D_\text{out} \ket{\pm}_0$, one obtains that the mean value and fluctuations of $X_\text{out}$ are given by
\be
\begin{split}
\avg{X_\text{out,t}}_\pm &= \alpha_x +\avg{X_\text{out,t}}_{\pm,0}, \\
\avg{X^2_\text{out,t}}_\pm &= \alpha_x^2 +2 \alpha_x \avg{X_\text{out,t}}_{\pm,0} + \avg{X^2_\text{out,t}}_{\pm,0},
\end{split}
\ee
where we have defined $\alpha_x=D^\dagger_\text{out,t} X_\text{out,t} D_\text{out} - X_\text{out,t}$. Thus, $\avg{\Delta X_\text{out,t}}_\pm=\avg{\Delta X_\text{out,t}}_{\pm,0}$. This shows, that from the theoretical point of view, the two displaced states $\ket{\pm}$ are as distinguishable as the non displaced ones $\ket{\pm}_0$. From the experimental point of view, the problem is that the signal to noise ratio in a balanced homodyne measurement is too low. Although the displacement $\alpha_x$ can be computed by using $\phi_\text{out}(\omega,t)$ and $\alpha_\omega$ (see Eq.~\eqref{eq:displacements}), the final expression is not very illustrative. Instead, in App.~\ref{sec:photon}, we will analyze the problem of the measurement of the output field when a photon on top of the coherent field was prepared inside the cavity. This general problem is more enlightening and the conclusions apply directly to the reflected-photon protocol, that is, that the displacement of the output field is of the order of $\alpha \sim 10^4$.

In order to circumvent this experimental challenge we envisage the following ways out: i) subtract the coherent part by destructively interfering a coherent beam with the same phase; ii) use an optomechanical system where the detuning between the resonant photon and the red-detuned driving is much larger (since  $\Delta=\omega_t$, this would correspond to a large-frequency mechanical oscillator). This must be done without compromising the strong coupling requirement which is based on the enhanced coupling $g=|\Omega_L| g_0 /\sqrt{\Delta^2+\kappa^2}$; iii) use a scheme similar to the one proposed in \cite{Khalili10}, where the photon is sent in the dark port of an interferometer; iv) design a scheme where the light pulse is perfectly absorbed in the cavity and therefore no measurement is needed. In the following  section we present a protocol which follows solution iv). 

\subsection{Perfect absorption} \label{subsec:perfectmapping}

Here we present a protocol which circumvents the challenging step of measuring the displaced output mode in the reflected one-photon protocol, as discussed in the previous section (this protocol was announced in~\cite{romeroisart10}). The goal is to perfectly absorb the light pulse, which is in a non-Gaussian state, into the cavity, and therefore transfer it to the mechanical system. This is achieved by using a time modulation of the optomechanical coupling $g(t)$, which can be implemented by varying the intensity of the driving field. Similar ideas have been proposed in the context of quantum communication \cite{Cirac97}, and recently in quantum optomechanical transducers~\cite{Stannigel2010}. 

\subsubsection{Time-dependent displacement}

In this section, one cannot use the beam-splitter interaction~\eqref{eq:bs}, since the laser intensity is time-dependent. Therefore, a time-dependent displacement has to be performed carefully.
Let us start with the basic Hamiltonian in the non-displaced frame Eq.~\eqref{eq:basicH}. From there, one can derive the evolution equations for $\aop$, $\bop$, and $\aoop(\omega)$
\be
\begin{split} \label{eq:perfect1}
&\dot \aop=-\im \Delta_0 \aop-\im g_0 \aop (\bdop+\bop)+\int_{-\omega_L}^\infty d \omega \gamma(\omega) \aoop(\omega,t),\\
&\dot \bop= -\im \omega_t b-\im g_0 \adop \aop,\\
&\dot \aop_0(\omega,t)= -\im \omega \aoop(\omega,t)-\gamma(\omega) \aop.
\end{split}
\ee
The tool that will be used in this section is a time-dependent driving field at the laser frequency $\omega=0$ (in the rotating frame). This can be incorporated by applying the following displacement to the output modes
\be
\aoop(\omega,t) \rightarrow \aoop(\omega,t) - \sqrt{\frac{\pi}{\kappa}}\Omega_L(t) \delta(\omega).
\ee
By formally integrating the equation of $\dot \aop_0(\omega,t)$, and using the Markov approximation $\gamma(\omega) \approx \sqrt{\kappa/\pi}$, the system $\eqref{eq:perfect1}$ reads
\be
\begin{split} \label{eq:perfect2}
&\dot \aop=-(\im \Delta_0+\kappa) \aop-\im g_0 \aop (\bdop+\bop)+\Omega_L(t) + \sqrt{2 \kappa} \ain(t),\\
&\dot \bop= -\im \omega_t b-\im g_0 \adop \aop,\\
&\dot \aop_0(\omega,t)= -\im \omega \aoop(\omega,t)-\gamma(\omega) \aop + \sqrt{\frac{\pi}{\kappa}}\dot \Omega_L(t) \delta(\omega),
\end{split}
\ee
where we have defined the so called input operator as $\ain(t)\equiv(2 \pi)^{-1/2} \int d \omega \aoop(\omega,0)e^{-\im \omega t}$. Next, we perform the following time dependent displacement
\be
\begin{split}
\aop(t) &\rightarrow\aop(t) + \alpha(t), \\
\bop(t)& \rightarrow\bop(t) + \beta(t),
\end{split}
\ee
and choose $\alpha(t)$ and $\beta(t)$, such that the constant terms in the equations for $\dot \aop(t)$ and $\dot \bop(t)$ vanish, that is
\be
\begin{split} \label{eq:tdisplacements}
\dot \alpha&= - (\im \Delta_0 + \kappa) \alpha - \im g_0 \alpha(\beta+\beta^\star) + \Omega_L, \\
\dot \beta &= -\im \omega_t \beta - \im g_0 |\alpha|^2.
\end{split}
\ee

 Then, we perform the following changes of variables
\be
\begin{split}
\aop(t) & \rightarrow \aop(t) e^{-\im \Delta_0 t}, \\
\bop(t)& \rightarrow \bop(t) e^{-\im \omega_t t}, \\
\alpha(t)&= \frac{g(t)}{g_0} e^{\im \xi},
\end{split}
\ee
($g(t)$ is real) and perform the RWA considering the red-detuned case $\Delta_0 = \omega_t$. Putting all these things together, Eqs.~\eqref{eq:perfect2} read
\be
\begin{split} \label{eq:perfect3}
&\dot \aop=-\kappa a - \im g(t) e^{\im \xi}b  + \sqrt{2 \kappa} \ain(t) e^{\im \Delta_0 t },\\
&\dot \bop=-\im g(t) e^{-\im \xi}\aop, \\
&\dot \aop_0(\omega,t)= -\im \omega \aoop(\omega,t)-\gamma(\omega)\left[ \aop e^{-\im \Delta_0 t } + \alpha(t) \right] \\ & \hspace{4 em} + \sqrt{\frac{\pi}{\kappa}}\dot \Omega_L(t) \delta(\omega).
\end{split}
\ee
We have neglected the small terms (not proportional to $\alpha$) $-\im g_0 a (\bop+\bdop)$ and $-\im g_0 \adop \aop$ in the equation of motion for $\dot b$. We have also neglected the term $-\im g_0 a (\beta+\beta^\star)$ in the equation for $\aop$. This term, which is smaller than $-\im g_0 a \Delta_0$, makes the equation decribing the shape of $g(t)$ (to be derived below) much more difficult to solve and is therefore also neglect since it does not change the physics of the problem. 

Finally, note that equations \eqref{eq:tdisplacements} give the solution of the time-dependent laser amplitude $\Omega_L(t)$ such that the time-dependent coupling $g(t)$ is implemented. In the next sections, we derive the pulse $g(t)$ for which any light state is absorbed into the cavity and therefore perfectly mapped into the mechanical system.

\subsubsection{Condition for perfect absorption}

The formal condition for perfect absorption can be derived as follows. After the transformations are made, the evolution equation for $\aoop(\omega,t)$ reads
\be
\begin{split}
\dot \aop_0(\omega,t) =& -\im \omega \aoop(\omega,t)-\gamma(\omega) \aop e^{- \im \Delta_0t }\\& -\gamma(\omega) \alpha(t)+ \sqrt{\frac{\pi}{\kappa}}\dot \Omega_L(t) \delta(\omega).
\end{split}
\ee
By formally integrating this equation for the initial condition $t=0$, as well as for the final condition $t=t_1$, and subtracting these two solutions after integrating over $\omega$, one obtains (using the approximation $\gamma(\omega)\approx \sqrt{\kappa/\pi}$ and that $\Omega_L(0)=\Omega_L(t_1)=0$)
\be
\begin{split} \label{eq:inout}
0=&\ain(t)- \frac{1}{\sqrt{2\pi}} \int_{-\omega_L}^{\infty} e^{-\im \omega (t-t_1)} \aoop(\omega,t_1) \\&- \sqrt{2 \kappa} \left( a(t)e^{-\im \Delta_0 t }+\alpha(t)\right).
\end{split}
\ee
This is the so called input-output relation~\cite{gardinerbook}, which relates the output field (the second term containing the $\aoop(\omega,t_1)$ modes) with the input field $\ain(t)$, the quantum field from the cavity $\aop(t)$, and its coherent part $\alpha(t)$. The condition for perfect absorption is that the output field only contains the coherent part from the cavity, that is
\be
 \frac{1}{\sqrt{2\pi}} \int_{-\omega_L}^{\infty} e^{-\im \omega (t-t_1)} \avg{\aoop(\omega,t_1)} =- \sqrt{2 \kappa} \alpha(t).
\ee
With this condition, Eq.~\eqref{eq:inout} reads
\be \label{eq:inoutperfect}
\avg{\ain(t)}= \sqrt{2 \kappa} \avg{a(t)} e^{-\im \Delta_0 t}.
\ee
One can now plug this condition into the Eqs.~\eqref{eq:perfect3} and obtains
\be
\begin{split} \label{eq:perfect4}
&\avg{\dot \aop(t)}= \kappa \avg{a(t)} - \im g(t) \avg{b(t)} e^{\im \xi},\\
& \avg{\dot \bop(t)}=-\im g(t) \avg{\aop(t)} e^{-\im \xi},
\end{split}
\ee
which can be further simplified to
\be \label{eq:gequation}
\eta(t)\dot g(t)-\dot \eta(t)g(t)+g^3(t) \avg{a(t)}=0,
\ee
where $\eta(t)=\kappa \avg{\aop(t)} - \avg{\dot \aop(t)}$.

\subsubsection{State-independent pulse}

The solution of the equation Eq.~\eqref{eq:gequation} yields the optomechanical pulse $g(t)$ necessary to perfectly transmit a light state into the mechanical system.  In order to obtain a state-independent solution,  we will assume that a coherent state with phase $\alpha_s$ is sent to the cavity and will show that the solution does not depend on $\alpha_s$. Therefore any linear combination of coherent states (and therefore any state since they form a complete basis) will be perfectly transmitted to the cavity with the pulse $g(t)$.  

\begin{figure}[t]
\centering
\includegraphics[scale=.7]{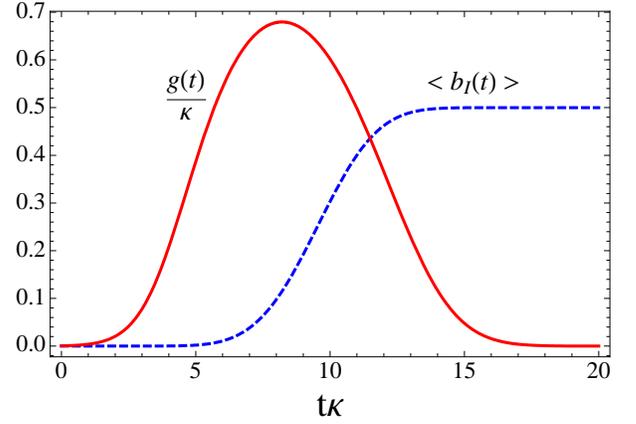}
\renewcommand{\figurename}{Fig.}
\caption{Perfect state transfer of a $\ket{0} + \ket{1}$ photonic state by sending a Gaussian light pulse of width $\sigma=2 \kappa/3$ from a distance $x_\text{in}=10 \kappa$ ($c=1$). We plot the time modulation of $g(t)/\kappa$ (solid red line) and $\avg{\bop(t)}$ (dashed blue line). After the modulation, when $g(t)=0$, one obtains that $\avg{\bop}=1/2$. This shows that the superposition state has been mapped to the mechanical system without requiring  a measurement.}\label{Figure21B1}
\end{figure}

The initial state is assumed to be
\be
\ket{\psi(0)}=\exp \left[ \alpha_s \int_{-\omega_L}^\infty  \phi^\star_{\text{in}}(\omega) \aodop(\omega)d \omega - \hc \right] \ket{00 \Omega},
\ee
where $\phi^\star_{\text{in}}(\omega)$ is the shape of the pulse. One can then obtain that $\avg{\ain(t)}= \alpha_s \tilde \phi^\star_\text{in}(t)$, and using Eq.~\eqref{eq:inoutperfect}, $\avg{\aop(t)}=\alpha_s \tilde \phi^\star_\text{in}(t) e^{\im \Delta_0 t}/\sqrt{2 \kappa}$. Then, Eq.~\eqref{eq:gequation} reads
\be \label{eq:eqg}
[\kappa \mu(t)-\dot \mu(t)] \dot g(t)- [\kappa \dot \mu(t)-\ddot \mu(t)] g(t) + \mu(t)  g^3(t)=0,
\ee
where $\mu(t) \equiv \tilde \phi^\star_\text{in}(t) e^{\im \Delta_0 t}$.  This is the main result of the section since its solution yields the time-dependent coupling $g(t)$ for perfect mapping of any light state into the mechanical system, since it does not dependent on the coherent phase $\alpha_s$. In Fig.~\ref{Figure21B1}, the solution $g(t)$ is plotted considering $\phi_\text{in}(\omega)$ to be the same Gaussian pulse used in the reflected one-photon protocol, see Eq.~\eqref{eq:gaussian}.

As an example,  let us assume that one wants to transfer a photon in a superposition state described by
\be
\ket{\psi} = \frac{1}{\sqrt{2}} \ket{00\Omega} + \frac{1}{\sqrt{2}} \int_{-\infty}^{\infty} \phi_\text{in}^\star(\omega) \aodop(\omega) \ket{00 \Omega}.
\ee
In Fig.~\ref{Figure21B1} the mean value of $\bop(t)$ is plotted using the $g(t)$ solution obtained for the Gaussian case. As expected,  $\avg{\bop(t)}$ attains the value $1/2$, showing that the superpostion state $(\ket{0}+\ket{1})/\sqrt{2}$ has been prepared. 

To sum up, this protocol enables one to perfectly map any state of light into the mechanical system without performing any measurement, merely by using a smooth modulation of the optomechanical coupling.

\subsection{Teleportation in the bad cavity limit} \label{subsec:teleportation}

\begin{figure}[t]
  \includegraphics[width=\linewidth]{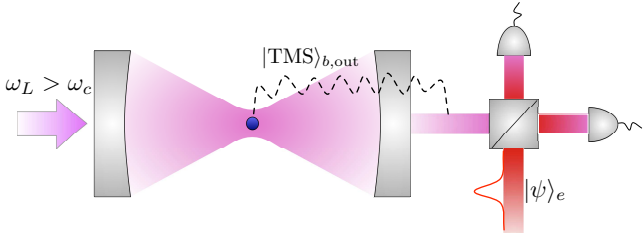}
\caption{Schematic representation of a light-mechanics interface of teleportation in the bad cavity limit. The cavity is driven by a blue-detuned laser which induces a two mode squeezing interaction between the cavity mode and the mechanical mode. Being in the bad cavity limit $\kappa > g$, the cavity photons, which are in a two mode squeezed state $\ket{\text{TMS}}$ with the mechanical phonons, rapidly leak out  of the cavity. The output field is combined in a beam-splitter together with the non-Gaussian state to be teleported $\ket{\psi}_e$. A measurement of the output quadratures would realize the Bell measurement required for teleportation \cite{Braunstein98}.}
\label{Fig:Teleportation}
\end{figure}

The two previous protocols required the strong coupling $g \sim \kappa$. In this section we provide a protocol,  announced in~\cite{romeroisart10}, to map non-Gaussian states which is also applicable in the bad cavity limit $\kappa > g$. The key ingredient of the protocol is to drive the cavity with a blue-detuned field in order to obtain a two mode squeezing interaction, see Eq.~\eqref{eq:tms}. The two mode squeezed state is then prepared by the optomechanical coupling between the mechanical mode and the cavity mode, which rapidly leaks out of the cavity. The output mode of the cavity, which is in a two mode squeezed state with the mechanical system, can be used as an entanglement channel to teleport~\cite{Braunstein98} a non-Gaussian state of light from outside the cavity into the mechanical system (see Fig.~\ref{Fig:Teleportation} for an illustration of the protocol). 
This protocol has first been introduced as an interface between quantum dots in optical cavities~\cite{Schwager10}. In reference~\cite{Schwager10}, a detailed discussion of the protocol is provided, which applies to our optomechanical setup in complete analogy. Thus, we will only summarize and remark the important aspects of the protocol here.

Using the Hamiltonian Eq.~\eqref{eq:tms}, one can obtain the equations of evolution for $\aop$ and $\bop$,
\be
\begin{split}
\dot a(t)&= - (\im \Delta_0 + \kappa) \aop(t)- \im g \bdop(t)  + \sqrt{2 \kappa}  \ain(t) \\
\dot b(t)&= -\im \omega_t \bop(t)- \im g \adop(t).
\end{split}
\ee
One can now move to the interaction picture ($a(t)=a_I(t) e^{-\im \Delta_0 t}$ and $b(t)=b_I(t) e^{-\im \omega_t t}$), and by considering the bad cavity limit ($\kappa \gg g$) one can adiabatically eliminate $\aop_I(t)$, by setting $\dot a_I(t)=0$. One obtains
\be
\dot b_I(t)= \frac{g^2}{\kappa} \bop_I(t) - \im g \sqrt{\frac{2}{\kappa}} \adin(t) e^{-\im \Delta_0 t}. 
\ee
Formally integrating this equation and using the initial conditions $\avg{\adin(t) \ain(t)}=\avg{\bdop_I(0) \bop_I(0)}=0$ (the mechanical initial state is assumed to be in the ground state) yields
\be
\avg{\bdop_I(t) \bop_I(t)} = e^{2 \frac{g^2}{\kappa} t}-1.
\ee
This can be used to obtain the squeezing parameter $r$ of the entangled state, which will provide the fidelity of the teleportation scheme. As proved in \cite{Schwager10}, the output mode of the cavity and the mechanical system are in the two mode squeezed state $\ket{\text{TMS}}_{b,\text{out}}$, defined by (in the displaced frame)
\be
\begin{split}
\ket{\text{TMS}}_{b,\text{out}}&= S\left(r e^{\im \phi} \right ) \ket{00} \\&= \frac{1}{\cosh r}\sum_{n=0}^\infty \left [ -e^{\im \phi} \tanh r\right]^n \ket{nn}_{b,\text{out}}\\
&\equiv \sum_{n=0}^{\infty}\Theta^n \ket{nn}_{b,\text{out}},
\end{split}
\ee
where, in our case, $\phi=\pi/2$ with the squeezing operator defined as $S\left(r e^{\im \phi} \right )= \exp[-r( e^{\im \phi}\adop \bdop -e^{-\im \phi} \aop \bop)]$. The squeezing parameter $r$ can be obtained using the relation
$ \avg{\bdop \bop} = \left( \cosh r -1\right)/2$, as
\be
r= \text{arcosh} \left( 2 e^{2 g^2 t/\kappa}-1\right).
\ee
The teleportation fidelity is given by $F=1/(1+e^{-2 r})$~\cite{Fiurasek02}.

Let us discuss the fact that the entangled state in the original frame is given by $D_b(\beta) D_\text{out}(\alpha_\text{out}) \ket{\text{TMS}}_{b,\text{out}}$. Here $D_\text{out}(\alpha_\text{out}) $ is the displacement operator of the output mode, which is displaced by $\alpha_\text{out}$ as a consequence of the displacement of the output operators $\aoop(\omega)$ by $\alpha_\omega$ (analogously to the discussion in App.~\ref{sec:photon}). 
First, let us generally define the teleportation scheme as the map $\Lambda$, such that it teleports a light state $\ket{\psi}_e$ as follows
\be
\Lambda \left[\ket{\text{TMS}}_{b,\text{out}} \otimes \ket{\psi}_e  \right]= \ket{\psi'}_{b}.
\ee
Here, the subindex $e$ labels the external system containing the state  that one wants to teleport. Let us remark that perfect teleportation $|\braket{\psi}{\psi'}|=1$ can only be achieved for the maximally entangled state $r\rightarrow \infty$. In order to determine the output state in the original frame, let us first transform the initial state
\be
\begin{split}
D_b  D_\text{out} \ket{\text{TMS}}_{b,\text{out}}  & =D_b\mathcal{O}_b \otimes D_\text{out} \sum_{n=0}^\infty\ket{nn}_{b,\text{out}} \\
&=[D_b\mathcal{O}_b D_\text{out}^\intercal ]_b \otimes \id \sum_{n=0}^\infty\ket{nn}_{b,\text{out}},
\end{split}
\ee
where $\mathcal{O}_b=\sum_{n=0}^\infty \Theta^n \ketbra{n}{n}$. The relation $A \otimes B\sum_n  \ket{nn}=AB^\intercal \otimes \id\sum_n  \ket{nn} $ has been used, where $B^\intercal$ denotes the transpose of $B$.
Using this relation, the output state of the teleportation scheme with the original state is given by
\be
\begin{split}
\Lambda \left[D_b  D_\text{out}  \ket{\text{TMS}}_{b,\text{out}}\otimes \ket{\psi}_e\right] = D_b\mathcal{O}_b D_\text{out}^\intercal\ket{\psi}_{b}.
\end{split}
\ee
This gives the final state of the teleportation protocol in the original frame. Note that $D_\text{out}^\intercal(\alpha_\text{out})= D_\text{out}^\dagger(\alpha^\star_\text{out})$. Therefore, one can get rid of this displacement by teleporting the state $D(\alpha^\star_\text{out})\ket{\psi}_e$,  such that the state teleported in the mechanical system is given by $D_b(\beta)\mathcal{O}\ket{\psi}_b$ (the displacement $D_b(\beta)$ can also be reduced by varying the center of the trap when switching off the cavity lasers). Besides, note that one can in principle also choose the appropriate initial state $\ket{\psi}$ in order to prepare a desired mechanical system $\ket{\phi}$, such that $D_b\ket{\phi}_b=D_b\mathcal{O}\ket{\psi}_b$.

\section{Mechanical tomography by time of flight} \label{sec:tomography}

After showing how to prepare non-Gaussian states, in this section we discuss how to measure them. In general quantum optomechanical systems, the proposal is to transfer the mechanical state into a well controlled quantum system, such as a qubit, and probe the state in that system. This could be analogously done in our setup by mapping the mechanical state to the cavity mode by using the enhanced beam-splitter interaction and perform full tomography of the output field. However, this technique would suffer from the drawback that the output field would contain a quantum state displaced by the large driving field and therefore, the signal-to-noise ratio would be challenging for experimental detection with present day technology. 

In this section we propose an alternative method to {\em directly} perform full tomography of the mechanical system. In particular, we exploit the analogy of levitated nanodielectric objects to atomic physics, more specifically to cold gases, where time of flight measurements are used to experimentally probe different many-body states~\cite{bloch08}. 
\begin{figure} [t]
  \includegraphics[width=0.75\linewidth]{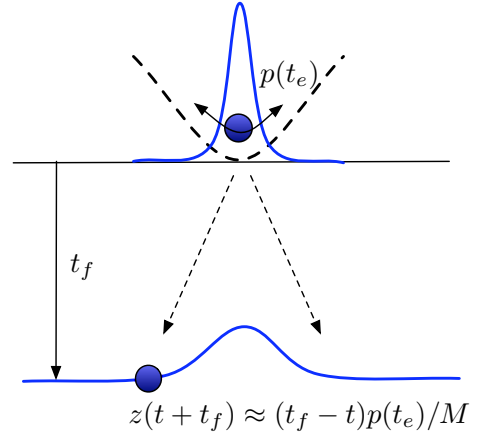}
\caption{Schematic illustration of the time of flight protocol to perform full tomography of the mechanical state. The momentum operator at times $t_e$, which corresponds to the rotated phase quadrature $\chi(\omega_t t_e + \pi/2)$, is determined by measuring the position of the dielectric after some time of flight. By repeating the experiment at different times $t_e$, one can perform full tomography of the mechanical state.}
\label{Fig:tof}
\end{figure} 
The time of flight protocol to perform direct full tomography of the mechanical state is the following (see Fig.~\ref{Fig:tof}):
\begin{enumerate}
\item We consider that at $t=0$ a particular state $\ket{\psi}$ in the mechanical system is prepared (for instance, a non-Gaussian state using the light-mechanics interface introduced in Sec.~\ref{sec:protocols}). Just after the preparation of the state, the cavity field is switched off and only the optical trapping remains on.  During these transient times the center of the trap has to be changed in order to account for the variation in the classical force created by the driving field, as discussed in the light-mechanics interface.
\item Then, during some given time $t_e$, the system is evolving within the harmonic potential, such that the mechanical momentum operator in the Heisenberg picture is given by
\be
p(t)= \im p_m (\bdop e^{\im \omega_t t} - \bop e^{-\im \omega_t t}), \;\; t \in [0,t_e],
\ee
where $p_m=(M \omega_t /2)^{1/2}$.
\item At $t=t_e$, the trap is switched off and the nanodielectric falls freely during the time of flight $t_f$, such that the distance from the center of the cavity along the cavity axis is given by
\be
\begin{split}
z(t_e+t_f)&=z(t_e) + (t_f-t_e)\frac{ p(t_e)}{M} \\& \sim (t_f-t_e) \frac{p(t_e)}{M},
\end{split}
\ee
where we assume that $t_f$ is sufficiently large such that $(t_f-t_e) p(t_e)/M \gg z(t_e)$.
\item At $t=t_e+t_f$, the position  $z(t_e+t_f)$ is measured (\eg by imaging the object and measuring the center of the light spot in the screen), which means that the in-trap momentum $p(t_e)$ is effectively measured.
\item The experiment is repeated in order to obtain statistics for any time $t_e \in [0,2 \pi /\omega_t]$.
\end{enumerate}

The key observation is that with the data obtained in this protocol, one has the statistical distribution of the rotated quadrature phase operator
\be \label{eq:rotatedquadrature}
\mathcal{X} (\theta) = e^{\im \theta} \bdop + e^{-\im \theta} \bop,
\ee
which permits the reconstruction of the Wigner function~\cite{Lvovsky01, lvovsky09}, and therefore contains all the information about the mechanical state $\ket{\Psi}$ (of course, it does not have to be a pure state). Indeed, there exists the following one-to-one relation between the momentum operator and the rotated quadrature phase operator,
\be
p(t_e)= \mathcal{X} (\omega_t t_e + \pi/2).
\ee

Let us now discuss some experimental considerations. First, we will estimate the order of magnitude of $t_f$ (and therefore the time of flight distance $d_f=g t_f^2 /2$, where $g$ is the gravitational acceleration). In particular, let us assume that after the time of flight the position can be measured with a resolution given by $\delta z$. This implies that the object has to spread over a distance much larger than $\delta z$, which means that $t_f  \gg M \delta z/p_m$ is required. Using the parameters given in App.~\ref{app:experimental}, one obtains that $t_f$ is of the order of tens of ms, which would require a time of flight distance of the order of one cm. Although this position resolution is very feasible, the requirement could be even relaxed with the same duration of time of flight. The idea is to amplify the oscillation via driving the field with a blue-detuned laser prior to letting the object fall. More specifically, let us assume that just after the preparation of the mechanical state, one impinges the cavity with a laser detuned to the blue sideband of the cavity. This corresponds to including an additional step (point $1.b$) between steps $1$ and $2$ in the previous protocol. The blue-detuned driving is performed during a certain time $\tau < 1/\Gamma$ (where $\Gamma$ is the decoherence rate when the cavity field is switched on). After this amplification, the momentum operator is transformed into
\be
p(\tau)= \im p_m p_+(\tau) (\bdop e^{\im \omega_t \tau} - \bop e^{-\im \omega_t \tau}) + p_\text{cav}(\tau) ,
\ee
where $p_+(\tau)$ is the amplifying parameter given by
\be
p_{+}(\tau) = e^{-\kappa t/2} \left[ \cosh(\chi\tau) + \frac{\kappa}{2 \chi} \sinh(\chi \tau)\right]
\ee
with $\chi=\sqrt{g^2 + \kappa^2/4}$. The term $p_\text{cav}(\tau)$ results from the entanglement of the mechanical system to the cavity field due to the two mode squeezing interaction. It reads $p_\text{cav}(\tau)= (q^\star(\tau) e^{\im \omega \tau}a_I(0) + \hc)$, where $q(t)=- \im g e^{-\kappa t/2} \sinh (\chi t)/\chi$, and fulfills $\avg{p_\text{cav}(\tau)}=0$ (the cavity field is empty at $t=0$) and $\avg{p^2_\text{cav}(\tau)}=|q(\tau)|^2$. After this amplification, step $2$ of the protocol follows. If one assumes $g=\kappa= 2 \pi \times 100 $ kHz, and $\tau=0.02$ ms, one obtains that $p_+(\tau)\sim10^3$ and hence with the same time of flight $t_f$ the required resolution is only $\delta z   \ll t_f p_m p_+(\tau) /M \sim 100 \mu$m; three orders of magnitude lower. Note that the amplification is restricted by keeping the nanodielectric object in the region, where it still sees the slope of the standing wave, i.e., the condition $x_0 p_+(\tau) < 1$ nm has to be fulfilled, where $x_0 \sim 10^{-12}$ m is the ground state size. 

Let us remark that the rotated quadrature $\cal \chi(\theta)$ could, in principle, also be measured by a quantum non-demolition measurement. This could be done by using the back-action evasion scheme proposed by Braginsky in the 80's \cite{braginsky80}, and recently revised from a quantum noise perspective \cite{clerk08}. This protocol would also benefit from the prominent property of levitating objects; being free of any thermal contact.  The key idea of this method is to impinge the cavity at the two motional sidebands, a scheme that has already been realized with trapped ions~\cite{cirac93, Vahala09}. 

The time of flight protocol presented in this section exploits the unique property of using levitating objects in quantum optomechanical systems; the mechanical resonator is unattached to other objects and therefore can fall.

\section{Conclusions \& overlook} \label{sec:conclusions}

We conclude by summarizing and giving an overlook of the contents presented in this article. First, we have developed a quantum theory to describe the coupling of light to the motion of dielectric objects inside a high finesse optical cavity. The main result is the derivation of a master equation  describing the joint state of the center of mass motion and the cavity field. In parallel, we have derived a quantum elasticity theory to show that the center of mass decouples from the internal vibrational modes for sufficiently small objects. This theory has been applied to describe the experimental proposal of using an optically levitating nanodielectric as a cavity optomechanical system~\cite{romeroisart10, chang10}. The master equation allows us to describe the coherent dynamics as well as the dissipative processes. More specifically, we have obtained the decoherence rate for the mechanical mode, the enhanced cavity decay rate due to light scattering, and a renormalization of the light-matter interaction Hamiltonian due to virtual photon exchange processes. This theory supports the statement that the center of mass motion of a levitating nanodielectric inside an optical cavity behaves as a mechanical resonator of very high quality. 

In the second part of the article, we have developed a light-mechanics interface with the aim of bringing levitating objects into the quantum regime. This can be used to prepare non-Gaussian states such as superpositions of Fock states. We have provided three protocols with different properties; first, the reflected one-photon, which requires strong coupling and a measurement of the output field. Second, the perfect mapping, which does not require the measurement by time-modulating the optomechanical coupling. Third, the teleportation in the bad cavity limit, which can be used in the weak coupling regime. These light-mechanics interfaces apply to other optomechanical systems and provide an effective way to obtain non-linearities. Besides, these input-output protocols required a formalism to be described in the Schr\"odinger picture in order to obtain the final state. Finally, we have proposed a method to perform direct full tomography of the mechanical state. This method exploits the levitation of the mechanical resonator, since it consists in measuring the center of mass position after letting the object fall. The position after time of flight carries information about the momentum in the trap, and by repeating the measurement at different times of evolution in the harmonic trap, one can perform full tomography. 

The theory and the protocols introduced in this article apply to a large variety of setups and dielectric objects, even to microorganisms~ \cite{romeroisart10}. This work opens many further directions that we are currently investigating. Some of them are the following:  (i) the possibility to apply time of flight experiments to prepare and measure macroscopic superpositions of the levitating object, that is, states in which the object is in two macroscopically distant (larger than its radius) positions. (ii) A thorough study of higher order light scattering processes in dielectric objects. (iii) The possibility to circumvent the decoherence processes due to light scattering by using magnetic levitation of micron objects. (iv) To reduce light scattering by using dielectrics of other shapes. (v) To address the internal modes of the object for high frequency resonator optomechanical purposes. (vi) To use objects with internal degrees of freedom, such as nanocristals with NV centers in order to couple the internal degree of freedom to the center of mass mode.  

As mentioned in the introduction, the project of cavity optomechanics with levitating objects aims at applying the quantum techniques developed to control and manipulate atoms back to the nanodielectrics that were first used by Ashkin. It is our hope that this article stimulates further theoretical and experimental research in this direction. The ultimate goal of these investigations is to explore the boundaries of quantum mechanics, which may reveal unexpected, and fascinating new insights.

We acknowledge funding by the Alexander von Humboldt foundation (O.R.I. and N. K.), the Elite Network of Bavaria (ENB) project QCCC (A.C.P.),
the DFG --FOR635 and the EU project AQUTE,  Spanish Ministry of Sciences through Grants TEC2007-60186/MIC and CSD2007-046-NanoLight.es, Fundaci{\'o} CELLEX Barcelona, Caixa Manresa, Austrian Science Fund (projects START, SFB FOQUS), European Research Council (ERC StG QOM), European Commission (MINOS), and Foundational Questions Institute FQXi. 

\appendix

\section{Light scattering equation} \label{app:scattering}

In this Appendix, we show how from the light-matter interaction Hamiltonian Eq.~\eqref{eq:Hamint}, one can derive the scattering equations which can be used to compute the total electric field inside the dielectric object.
We decompose the total electric field into
\be
 E(\bold{x})=\frac{\im}{(2\pi)^{3/2}}\int d^3k\sqrt{\frac{\omega_{\bold{k}}}{2\epsilon_0}}(e^{-i \bold{k} \bold{ x}}\aop(\bold{k})-\hc),
 \ee
  where $\omega_{\bold{k}}$ is the frequency of the different light modes and $\epsilon_0$ the vacuum permittivity.  Starting from the Hamiltonian \eqref{eq:Hamint} (including the free term $\Hffree$, Eq.~\eqref{eq:Hfreeterms}, of the modes $\aop(\bold{k})$), one can connect the electric field $E(\bold{x})$ to the electric field without the presence of the dielectric object, $E_{0}(\bold{x})$. To achieve this, we determine the equation of motion for the annihilation operator $a(\bold{k})$
\be \label{eq:adot}
\begin{split}
\dot a(\bold{k},t)=-\im\omega_{\bold{k}}a(\bold{k},t)\!-\!\frac{\im\alpha_p}{(2\pi)^{3/2}}\sqrt{\frac{\omega_{\bold{k}}}{2\epsilon_0}}\int d^3x  E(\bold{x},t) e^{i\bold{k}\bold{x}}.
\end{split}
\ee
In order to obtain $E(\bold{x})$, one needs to formally integrate  Eq.~\eqref{eq:adot}  over time, multiply both sides of the equation by $\im \sqrt{\omega_k/2\epsilon_0}/(2\pi)^{3/2}e^{-\im \bold{k}\bold{x}}$, and integrate it over $\kk$. Given that the electric field is varying with the laser frequency, one can move  to the rotating frame, where $\tilde{E}(\bold{x},t)$ is slowly varying, $\tilde{E}^{\pm}(\bold{x},t)=e^{\pm \im \omega_Lt}E^{\pm}(\bold{x})$. This justifies the assumption $\tilde{E}^{\pm}(\bold{x},t)\approx\tilde{E}^{\pm}(\bold{x},t')$ which permits to simplify the integration over time, leading to
\be
\begin{split}
E^{+}(\bold{x},t)=&E^{+}_{0}(\bold{x},t) \\
&+\int d^3x' \!\int \!d^3k \frac{\alpha_p\omega_{\bold{k}}}{2(2\pi)^{3}} e^{\im\bold{k}(\bold{x}'-\bold{x})} e^{-\im \omega_k t}\\
& \times \Big[\tilde{E}^{+}(\bold{x}',t)\int_0^t d\tau e^{-\im(\omega_L-\omega_k) \tau}\\
& \; \; \; \; \;\;+\tilde{E}^{-}(\bold{x}',t)\int_0^td\tau e^{\im(\omega_L+ \omega_k)\tau}\Big]. 
\end{split}
\label{effE}
\ee
The terms containing $\tilde{E}^{-}(\bold{x},t)$ are rotating fast compared to any other time scale and will hence be neglected within a rotating wave approximation (RWA).
Carrying out the integration $d^3 k$ in Eq.~\eqref{effE} gives us a function that decays very quickly in $\tau$. This permits us to extend the upper integration boundary $t$ to $\infty$ and hence to obtain
\be
\begin{split}
E(\bold{x},t)&=E_{0}(\bold{x},t)+\alpha_p \int d^3x' \mathcal{G}(\bold{x}', \bold{x})E(\bold{x}',t),\label{elfield}
\end{split}
\ee
where $\mathcal{G}(\bold{x}, \bold{x}')= k_{\rm L}^2 \sin(k_{\rm L}|\bold{x}-\bold{x}'|)/|\bold{x}-\bold{x}'|$. This equation has the same structure as a scattering equation where $\mathcal{G}(\bold{x}, \bold{x}')$ takes the role of the propagator. 

Let us remark here that in order to determine $E(\bold{x},t)$ for any size and shape of the object, the scattering wave equation~\eqref{elfield} has to be solved. However, this equation can only be solved approximately (see~\cite{abajo07, purcell73} for a solution). Merely in the special case of a spherical object, the electric field can be determined exactly by expansion in spherical waves, the Mie solution~\cite{strattonbook, bohrenbook}. Indeed, this solution coincides with the discrete-dipole aproximation for perfectly spherical objects~\cite{purcell73}. In the limit of very large spheres, $R\gg \lambda$, a ray optics approach has to be used to determine the forces on the sphere~\cite{ashkin92}.

\section{Optomechanical parameters} \label{app:trappingandcoupling}

In this Appendix we show how from the light-matter interaction term of the total Hamiltonian Eq.~\eqref{eq:Hamint}, one can easily obtain the optomechanical Hamiltonian discussed in Sec.~\ref{sec:hamiltonian}.

\subsection{External trapping with optical tweezers}\label{exttrap}
The trapping of the sphere can be achieved either by using optical tweezers~\cite{Ashkin86} or two optical cavity modes \cite{romeroisart10, chang10}. 
For the optical tweezers, we assume a Gaussian beam 
\be
[\Et]=E_0 \frac{W_t}{W(y)}\exp\left(-\frac{x^2+z^2}{W(y)^2}\right),
\ee
where $E_0=[P_t/(\epsilon_0 c \pi W_t^2)]^{1/2}$, $P_t$ is the laser power, $W_t$ is the laser beam waist, $W(y)=W_t[1+(y\lambda/(\pi W_t^2))^2]^{1/2}$ and assume the beam is aligned as sketched in Fig.~\ref{Fig:Cavitytweezers}.
The interaction between the sphere and the light field is described by Eq.~\eqref{eq:Hamint}. If the object is smaller than the laser waist and is placed close to the beam center, one obtains after integrating over a sphere of radius $R$, mass $M$, density $\rho$ and relative dielectric constant $\epsilon_r$, that the Hamiltonian is of the form of a harmonic oscillator with frequency $\omega_t$ given by
\be \label{eq:trappingsphere}
\omega_t^2=\frac{4  \epsilon_{\rm c}}{\rho c} \frac{I}{W_t^2}\approx \frac{ \epsilon_{\rm c} }{\rho c} Ik^2 \mathcal{N}^2,
\ee
corresponding to the $x-$ and $z-$ direction in  our setup. Here, $I$ is the field intensity, the laser waist can be approximated by $W_t\approx\lambda/(\pi \mathcal{N})$, $\mathcal{N}$ is the numerical aperture, and $k=2\pi/\lambda$ the wave vector. In the direction of light propagation, the $y-$direction in our configuration, the trapping frequency is
\be
\omega_{\parallel}^2\approx \frac{2 \epsilon_{\rm c}}{\rho c}  \frac{I \mathcal{N}^2}{W_t^2},
\ee
which is reduced by a factor of $\mathcal{N}^2/2$ compared to the trapping in the $x-$ and $z-$ direction. This lower trapping frequency can be enhanced by the use of a second optical tweezers perpendicular to the first one. Besides, the scattering force will change the equilibrium position  of the object in the direction of light propagation. To circumvent this, a second tweezers of the same intensity and waist, but with a different polarization, can be used.

\subsection{The optomechanical coupling}\label{Sec:optocoupl}
The optomechanical coupling arises from plugging the cavity mode into Eq.~\eqref{eq:Hamint}. 
 For small spheres, we choose a TEM $00$ mode as the cavity mode in the presence of the sphere.  Given that the sphere has a radius smaller than the laser waist and is placed close to the center of the cavity, one can approximate the square of the electromagnetic field close to the center of the beam by
\be
[\Ec]^2\approx \frac{\omega_c}{\epsilon_0 V_c} \left (1-\frac{2(x^2+y^2)}{W_c^2}\right) \cos^2\left(k_c z-\varphi\right) a^{\dagger}  a.\label{eq:efield}
\ee
  Here, $V_c=\pi W_c^2 L/4$ is the cavity volume, $W_c=[\lambda L/(2 \pi)]^{1/2}$ is the laser's waist at the center of  a confocal cavity, $L$ the cavity length, $\omega_{\rm c}$ the cavity's resonance frequency, $\lambda$ the laser wavelength and $\aop (\adop)$ the annihilation (creation) operator of cavity photons.
Furthermore, we presume that the laser is aligned such that the wave vector of the cavity mode $\bold{k}_{\rm c}$ points in $z$-direction.
The integration over the volume $V$ around the center of mass position $ \bold r=(x,y,z)$ leads to
\be
-\frac{\epsilon_c}{2}\int_{V({\bold{r}})}d\xx[\Ec]^2= \omega_c \adop \aop f(\bold{r}),
\ee
 with
 \be \label{eq:optocoupl}
 f(\bold{r})=\frac{ V\epsilon_{\rm c}  \left[ W_c^2 - 2 (x^2+y^2)\right] \cos^2 \left(\frac{\omega_c z}{c}-\varphi \right)}{V_c W_c^2},
 \ee
 where $V$ is the volume of the sphere, $z$ the center of mass position in $z$-direction and $\varphi$ a phase shift.
If the sphere is trapped at the maximum slope of the standing wave, that is, $x_0=y_0=0$ and $z_0=0, \varphi=\pi/4$, it is justified to expand Eq.~\eqref{eq:optocoupl} to first order in the $z$-coordinate. The zeroth order contribution leads to a constant shift of the trapping frequency given by 
\be
\tilde{\omega}_c=\omega_c\left(1-\frac{ \epsilon_c  V}{c  V_c}\right),
\ee
 which we will always use in the article without explicitly denoting the tilde. After quantization of the z-coordinate, the first order contribution yields the optomechnical coupling $g_0 \adop \aop (\bdop + \bop)$, where the optomechanical coupling reads
\be \label{optocoupl}
g_0=-\frac{1}{\sqrt{2 M \omega_t}} \frac{ \epsilon_{\rm c}\omega_{\rm c}^2 V}{c  V_c}.
\ee
\subsection{Shift due to interference of the tweezers and the cavity field} \label{app:shift}

In this section, we show that the term of the shift Hamiltonian $\HS$ discussed in Sec.~\ref{sec:scattering} yields a shift in the trapping frequency as well as in the equilibrium position of the dielectric object. We first recall the term leading to this shift,
\be
\HS =- \epsilon_c \epsilon_0 \int_{V(\rr)} d^3 x \Ecc \Et,
\ee
where
\be
\Ecc =2 \sqrt{\frac{\omega_c}{ \epsilon_0 V_c}} \cos (k_c z -\varphi) |\alpha|.
\ee
Then, by expanding up to second order around the equilibrium position of the center of mass coordinate, one obtains
\be
\HS= \omega_\text{sh} \bdop \bop + \xi_\text{sh} (\bdop+\bop),
\ee
where the parameters are given by
\be
\begin{split}
\omega_\text{sh}^2 &= \frac{\epsilon_c |\alpha|}{\rho} \sqrt{\frac{\omega_c P_t}{V_c c \pi}}  \frac{k_c W_t^2 + 2}{W_t^3}, \\
\xi_\text{sh} &= - \epsilon_c V |\alpha|k_c z_0 \sqrt{\frac{\omega_c P_t}{V_c c \pi W_t^2}}.
\end{split}
\ee
Therefore, the total trapping frequency of the object is given by $\omega_t'=\omega_t+\omega_\text{sh}$. The change in the equilibrium position is considered when displacing the mechanical mode in Sec.~\ref{sec:transformed}.
\section{Ground state cooling} \label{app:cooling}
Using the theory of ground state sideband cooling in optomechanical systems, \cite{Mancini98, Marquardt07, WilsonRae07, WilsonRae08, Genes08}, the minimal number of phonons attainable is given by
\be
 n_{\rm M}^0=\left(\frac{\kappa + \kappa_\text{sc}}{4\omega_{\rm t}}\right)^2.
\ee
Taking into account heating mechanisms, such as the recoil heating due to light scattering $\Gamma_\text{sc}$, and others $\Gamma_\text{others}$ (\eg Brownian motion heating due to the surrounding gas, laser noise, blackbody radiation, etc.~\cite{romeroisart10,chang10}), which can be shown to fulfill $\Gamma_\text{others} \ll \Gamma_\text{sc}$, the final phonon occupation number is given by
 \be
 n_{\rm M}=n_{\rm M}^0+\frac{\Gamma_\text{sc}+\Gamma_\text{others}}{\Gamma_-}.
 \ee
For $g<\kappa$, the maximal achievable cooling rate is given by
\be
 \Gamma_{-}=4\frac{(g_0|\alpha|)^2}{\kappa}\frac{\Delta}{\omega_{\rm t}},
 \ee 
 where the detuning is $\Delta \sim \omega_t$. By using the experimental parameters described in App.~\ref{app:experimental}, $n_{\rm M}$ can be made much smaller than one.

\section{Displacement of the output modes} \label{sec:outputmodes}

In this section we show how the expression of the displacement of the output modes, $\alpha_\omega$, appears naturally by computing the steady state obtained when the driving field is switched on. Then we discuss how to measure a photon created on top of the coherent cavity field in the output field. To simplify the problem we assume a cavity of resonance frequency $\omega_c$, driven by a laser at $\omega_L$. In the rotating frame at the laser frequency, and by defining $\Delta=\omega_c -\omega_L$, the Hamiltonian reads
\be
\begin{split}
H =& \Delta \adop \aop +\int_{-\omega_L}^\infty \omega \aodop(\omega) \aoop(\omega)d\omega \\&+ \im \int_{-\omega_L}^\infty\gamma(\omega) (\adop  \aoop(\omega) -\hc)d\omega.
\end{split}
\ee

\subsection{Steady-state with a driving field}
The initial state of the system is assumed to be
\be
\ket{\text{in}}=\ket{\alpha} \otimes \int_{-\omega_L}^\infty \delta(\omega) D(\alpha_0) \ket{\Omega},
\ee
that is, the cavity state is in a coherent state with phase $\alpha$, and all the output modes are empty, only the laser mode is in a coherent state with phase $\alpha_0$ (which is related to the laser power). In the following, we aim at computing the final state $\ket{\text{st}}=\lim_{t\rightarrow \infty}\exp[-\im H t] \ket{\text{in}}$.

First, let us write the Heisenberg evolution equations for the cavity mode and the output modes:
\be
\begin{split}
\dot \aop(t) &= - \im \Delta \aop(t) +\int_{-\omega_L}^\infty\gamma(\omega)  \aoop(\omega,t),\\
\dot \aoop(\omega,t) &= - \im \omega \aoop(\omega,t) -\gamma(\omega) a(t).
\end{split}
\ee
Then, one can formally integrate the differential equation for $\aoop(\omega,t)$
\be
\aoop(\omega,t)= e^{-\im \omega t} \aoop(\omega,0) - \gamma(\omega) \int_0^t dÊ\tau a(\tau) e^{-\im \omega (t-\tau)}.
\ee
This solution can be introduced into the differential equation for $\aop(t)$. By using the approximation $\gamma(\omega) \approx \gamma(0) = \sqrt{\kappa/\pi}$, one gets
\be
\dot a(t)=- (\im \Delta + \kappa) \aop(t) +\int_{-\omega_L}^\infty \gamma(\omega)e^{-\im \omega t} \aoop(\omega,0),
\ee
which can be trivially integrated to
\be
\begin{split}
a(t)=&e^{-(\im \Delta + \kappa)t}a(0)\\& + \int_0^t d\tau \int_{-\omega_L}^\infty \gamma(\omega)e^{-\im \omega \tau} \aoop(\omega,0) e^{-(\im \Delta + \kappa)(t-\tau)}.
\end{split}
\ee
By taking the average value of this expression, using that $\avg{\aop(0)}=\alpha$ and $\avg{\aoop(\omega,0)}=\alpha_0$, one gets
\be
\avg{a(t)}= e^{-(\im \Delta + \kappa)t} \alpha +\gamma(0) \alpha_0  \frac{1- e^{-(\im \Delta + \kappa)t}}{\im \Delta + \kappa}.
\ee
In the steady state, one obtains
\be
\alpha \equiv \lim_{t \rightarrow \infty} \avg{a(t)}=  \frac{\gamma(0) \alpha_0 }{\im \Delta + \kappa} =   \frac{\Omega_L }{\im \Delta + \kappa}. 
\ee
Note that we have assumed that the initial coherent state of the cavity is equal to the steady state obtained when driving the cavity with the laser. We have also related $\alpha_0$ to the usual frequency $\Omega_L=\sqrt{2 P_c \kappa/\omega_L}$, $P_c$ being the laser power. Let us now compute the mean value of the output modes, which after some algebra is given by
\be
\begin{split}
\avg{\aoop(\omega,t)}&
= \alpha_0 \delta(\omega) - \gamma(\omega) \int_0^t dÊ\tau \avg{a(\tau)}e^{-\im \omega (t-\tau)} \\
&= \alpha_0 \delta(\omega) -\alpha \gamma(\omega)\int_0^t dÊ\tau e^{-\im \omega \tau}. 
\end{split}
\ee
Then, the steady state phase of the output modes can be expressed  by
\be
\begin{split}
\alpha_\omega&=\lim_{t\rightarrow \infty} \avg{\aoop(\omega,t)} \\
&=  (\alpha_0 - \pi \alpha \gamma(0) ) \delta(\omega) +\im\alpha \gamma(\omega)  \mathcal{P} \left(\frac{1}{\omega} \right),
\end{split}
\ee
which coincides with the expression used in Eq.~\eqref{eq:displacements}.

It is trivial to show that the Hamiltonian is invariant under the displacement operation $\mathcal{D}=D_a D_\text{out}$, such that $D_a^\dagger \aop D_a= \aop+\alpha$, and $D_\text{out}^\dagger \aoop(\omega) D_\text{out}=\aoop(\omega)+\alpha_\omega$. By using that $\mathcal{P} \int_{-\infty}^{\infty} \omega^{-1} d\omega=0$, one can check that
\be
\mathcal{D}^\dagger H \mathcal{D}=H.
\ee
This implies that the steady state
\be
\ket{\text{in}}=\mathcal{D} \ket{0\Omega}= \ket{\alpha} \otimes \int_{-\omega_L}^\infty d\omega D(\alpha_\omega) \ket{0 \Omega}
\ee
is indeed an eigenstate of the Hamiltonian:
\be
H \ket{\text{in}}= \mathcal{D} \mathcal{D}^\dagger H \mathcal{D} \ket{0 \Omega} =\mathcal{D} H \ket{0 \Omega}=0.
\ee
\subsection{Measurement of a photon} \label{sec:photon}
In this section we want to compute the displacement of the output mode of the cavity. To do so, we assume that at $t=0$ a photon is present inside the cavity in the displaced frame, that is,
\be
\ket{\psi(0)}=\ket{1\Omega}.
\ee
Therefore, using the Wigner-Weisskopf formalism, one can obtain the state at some later time
\be
\ket{\psi(t)}= c_a(t) \ket{1 \Omega} + \int_{-\omega_L}^\infty d \omega c(\omega,t) \aodop(\omega) \ket{0 \Omega},
\ee
where the coefficients are given by 
\be
\begin{split}
c_a(t) &= e^{-(\im \Delta_0 + \kappa)t}, \\
c(\omega,t)&= \frac{\gamma(\omega) \left( e^{-\im \omega t} - e^{-(\im \Delta_0 + \kappa)t}\right)}{\im (\omega-\Delta_0) + \kappa}.
\end{split}
\ee
For large $t$, the final state is given by
$\ket{\psi(t)}= A^\dagger_\text{out,t} \ket{0 \Omega}$,
where the collective output mode is defined as $A_\text{out,t}=\int\phi_\text{out}(\omega)e^{\im \omega t} \aoop(\omega)d\omega$, with the mode function
\be
\phi_\text{out}(\omega)=\frac{\gamma(\omega)}{\kappa-\im (\omega-\Delta_0) }.
\ee
Let us now compute how many photons will be encountered in this collective photon mode after transforming back to the non-displaced frame.  By using the expression of the displacement of the output modes $\alpha_\omega$, one can obtain after some careful manipulation that the displacement of the output mode $A_\text{out,t}$, defined as 
\be
\alpha_\text{out}=\int_{-\omega_L}^\infty \phi_\text{out}(\omega)e^{\im \omega t} \alpha_\omega d\omega
\ee
is just given by
$\alpha_\text{out}= \alpha$.

\subsection{Switching off the driving field} \label{app:off}

In this Appendix we want to discuss the final state of the one-photon protocol once the driving field has been switched off. The Hamiltonian in the frame rotating with the laser frequency $\omega_L$, is given by
\be
\begin{split}
\Htot'= &\omega_t \bdop \bop + \Delta \adop \aop +  \int_{-\omega_L}^\infty \omega \aodop(\omega) \aoop(\omega)d\omega \\
& +g_0 \adop \aop (\bdop + \bop)+  \im \int_{-\omega_L}^\infty\gamma(\omega) (\adop  \aoop(\omega) -\hc)d\omega \\
&+ \lambda(t) (\bdop +\bop),
\end{split}
\ee
where the term with $\lambda(t)$ accounts for the variation of the center of the harmonic trap. By writing the Langevin equations, and considering that there is no input fields since they have already been switched off, one obtains
\be
\begin{split}
\dot a(t)&=-\im \Delta a(t) - \kappa a(t) - \im g_0 \aop (\bdop + \bop), \\
\dot b(t) &= - \im \omega_t \bop - \im g_0 \adop \aop - \im \lambda(t).
\end{split}
\ee
By displacing the operators by $\aop'=\aop + \alpha(t)$, and choosing the restriction
\be
\begin{split}
 \dot \alpha&=-\im \Delta \alpha - \kappa \alpha, \\
0 &= - \im g_0 |\alpha|^2 - \im \lambda(t),
\end{split}
\ee
one obtains the following equations:
\be
\begin{split}
\dot a(t)&=-\im \Delta a(t) - \kappa a(t) - \im g_0  |\alpha|(\bdop + \bop), \\
\dot b(t) &= - \im \omega_t \bop - \im g_0 |\alpha |(\adop +  \aop). 
\end{split}
\ee
In the interaction picture one can perform the RWA in order to get
\be
\begin{split}
\dot a(t)&= - \kappa a(t) - \im g(t)\bop, \\
\dot b(t) &= - \im g(t)   \aop, 
\end{split}
\ee
where
\be
g(t)= g_0|\alpha(0)| e^{-\kappa t}.
\ee
The equation for $\bop(t)$ is then given by
\be
\ddot b(t) - \dot b(t) \left( \frac{\dot g(t)}{g(t)}-\kappa\right) + \bop(t) g^2(t)=0.
\ee
This can be solved in order to predict the variation of mechanical state by switching off the driving field.

\section{Experimental parameters}\label{app:experimental}

\begin{figure} 
\includegraphics[width=.9\linewidth]{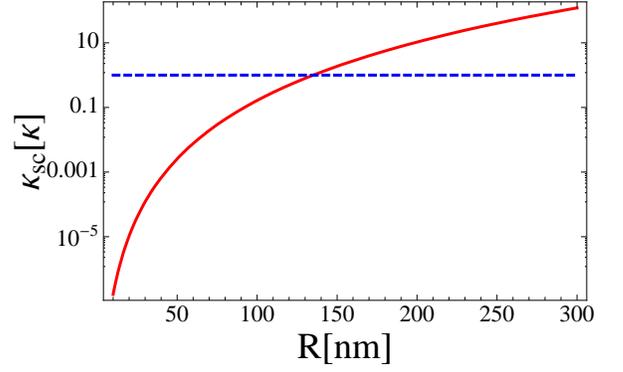}
\caption{The ration between the decay rate of the cavity due to light scattering $\kappa_\text{sc}$ over the standard decay rate of a cavity of finesse $\mathcal{F}=5 \times 10^5$ is plotted as a function of the radius of the sphere. As can be observed, photon losses due to scattering are dominating for objects larger than $150$ nm.}
\label{Fig:decay}
\end{figure}

\begin{figure} 
\includegraphics[width=.9\linewidth]{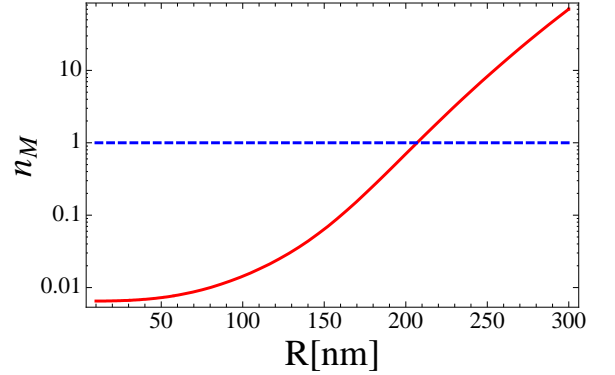}
\caption{Final phonon occupation number using spheres of radius $R$ and for the parameters given in the text. Ground state cooling is possible for spheres smaller than $R \sim 200$ nm.}
\label{Fig:cooling}
\end{figure}

In order to illustrate the experimental feasibility of the proposal, we will choose a set of experimental parameters.
\begin{enumerate}

\item {\bf Dielectric object:} We assume nanospheres fabricated of fused silica with a radius $R=100\rm nm$, density $\rho=2201\rm kg/m^3$  and a dielectric constant $\Re[\epsilon_r]=2.1$ and $\Im[\epsilon_r] \sim 2.5 \times 10^{-10}$. Their Young modulus is $Y=73~\rm GPa$ and their Poisson constant $\sigma=0.17$, giving internal vibrational modes with frequencies of the order of $\sim 10^{11}$ Hz. 

\item {\bf Cavity:} We assume a confocal high finesse cavity of length $L=4~\rm mm$ and finesse $\mathcal{F}=5 \times 10^5$ leading to a cavity decay rate $\kappa=c\pi/2\mathcal{F}L=2\pi \times 44\rm kHz$.
\item {\bf Lasers:}
The optical tweezers is constructed with a laser of power $P_t=15\rm mW$ at a wavelength $\lambda=1064~\rm nm$ and a lense of high numerical aperture $\mathcal{N}=0.8$. The cavity is impinged by a laser of power $P_c=0.1~\rm mW$, wavelength $\lambda=1064~\rm nm$, which gives a waist of $W_c=\sqrt{\lambda d/2\pi}\approx 26~\mu \rm m$.
\item {\bf Dissipation due light scattering:} The decay rate of the cavity due to the light scattering is $\kappa_\text{sc}=2 \pi \times 6$ kHz (see  Fig.~\ref{Fig:decay} for the dependence with the radius of the sphere). The mechanical motion decoherence rate is given by $\Gamma_\text{sc}=2 \pi \times 9$ kHz. 
\item {\bf Optomechanical parameters:} The tweezers supplies a harmonic trap for the object of frequency $\omega_{\rm t}=2\pi \times136~\rm kHz$ in the transversal direction and $\omega_{\parallel}=2\pi \times 77~\rm kHz$ in the direction of light propagation. The steady state photon number is $|\alpha|^2\approx 3.7 \times 10^8$.  The optomechanical coupling to the CM degree of freedom of the sphere is given by $g_0\approx -2\pi \times 7~\rm Hz$, which is enhanced by a factor of $|\alpha|$ to $g=-2\pi \times 131~\rm kHz$. The frequency of the cavity photons is given by $\omega_c=2 \pi \times 2.8 \times 10^{14}$ Hz. Note that $\omega_t/(\kappa+\kappa_\text{sc}) \approx 3$ (condition for the RWA), $g/(\kappa+\kappa_\text{sc}) \approx 1$ (the strong coupling regime used in the light-mechanics interface), $g/\Gamma_\text{sc} \approx 15$, and  $\omega_t/\Gamma_\text{sc} \approx 17$ (number of coherent oscillations).
\item {\bf Ground state cooling:} The cooling rate is given by $\Gamma_-=2 \pi \times 1.6 $ MHz. The final occupation number is given by $n_M=0.014$, see Fig.~\ref{Fig:cooling} for the dependence with the radius of the sphere.

\end{enumerate}

\begin{thebibliography}{78}
\expandafter\ifx\csname natexlab\endcsname\relax\def\natexlab#1{#1}\fi
\expandafter\ifx\csname bibnamefont\endcsname\relax
  \def\bibnamefont#1{#1}\fi
\expandafter\ifx\csname bibfnamefont\endcsname\relax
  \def\bibfnamefont#1{#1}\fi
\expandafter\ifx\csname citenamefont\endcsname\relax
  \def\citenamefont#1{#1}\fi
\expandafter\ifx\csname url\endcsname\relax
  \def\url#1{\texttt{#1}}\fi
\expandafter\ifx\csname urlprefix\endcsname\relax\def\urlprefix{URL }\fi
\providecommand{\bibinfo}[2]{#2}
\providecommand{\eprint}[2][]{\url{#2}}

\bibitem[{\citenamefont{Romero-Isart et~al.}(2010)\citenamefont{Romero-Isart,
  Juan, Quidant, and Cirac}}]{romeroisart10}
\bibinfo{author}{\bibfnamefont{O.}~\bibnamefont{Romero-Isart}},
  \bibinfo{author}{\bibfnamefont{M.~L.} \bibnamefont{Juan}},
  \bibinfo{author}{\bibfnamefont{R.}~\bibnamefont{Quidant}}, \bibnamefont{and}
  \bibinfo{author}{\bibfnamefont{J.~I.} \bibnamefont{Cirac}},
  \bibinfo{journal}{New J. Phys.} \textbf{\bibinfo{volume}{12}},
  \bibinfo{pages}{033015} (\bibinfo{year}{2010}).

\bibitem[{\citenamefont{Chang et~al.}(2010)\citenamefont{Chang, Regal, Papp,
  Wilson, Painter, Kimble, and Zoller}}]{chang10}
\bibinfo{author}{\bibfnamefont{D.~E.} \bibnamefont{Chang}},
  \bibinfo{author}{\bibfnamefont{C.~A.} \bibnamefont{Regal}},
  \bibinfo{author}{\bibfnamefont{S.~B.} \bibnamefont{Papp}},
  \bibinfo{author}{\bibfnamefont{D.~J.} \bibnamefont{Wilson}},
  \bibinfo{author}{\bibfnamefont{O.}~\bibnamefont{Painter}},
  \bibinfo{author}{\bibfnamefont{H.~J.} \bibnamefont{Kimble}},
  \bibnamefont{and} \bibinfo{author}{\bibfnamefont{P.}~\bibnamefont{Zoller}},
  \bibinfo{journal}{Proc. Natl. Acad. Sci. U.S.A.}
  \textbf{\bibinfo{volume}{107}}, \bibinfo{pages}{1005} (\bibinfo{year}{2010}).

\bibitem[{\citenamefont{Ashkin}(1970)}]{Ashkin70}
\bibinfo{author}{\bibfnamefont{A.}~\bibnamefont{Ashkin}},
  \bibinfo{journal}{Phys. Rev. Lett.} \textbf{\bibinfo{volume}{24}},
  \bibinfo{pages}{156} (\bibinfo{year}{1970}).

\bibitem[{\citenamefont{Ashkin}(2006)}]{ashkinbook}
\bibinfo{author}{\bibfnamefont{A.}~\bibnamefont{Ashkin}},
  \emph{\bibinfo{title}{Optical trapping and manipulation of neutral particles
  using lasers}} (\bibinfo{publisher}{World Scientific},
  \bibinfo{address}{Singapore}, \bibinfo{year}{2006}).

\bibitem[{\citenamefont{Cronin et~al.}(2009)\citenamefont{Cronin, Schmiedmayer,
  and Pritchard}}]{Cronin09}
\bibinfo{author}{\bibfnamefont{A.~D.} \bibnamefont{Cronin}},
  \bibinfo{author}{\bibfnamefont{J.}~\bibnamefont{Schmiedmayer}},
  \bibnamefont{and} \bibinfo{author}{\bibfnamefont{D.~E.}
  \bibnamefont{Pritchard}}, \bibinfo{journal}{Rev. Mod. Phys.}
  \textbf{\bibinfo{volume}{81}}, \bibinfo{pages}{1051} (\bibinfo{year}{2009}).

\bibitem[{\citenamefont{{Bloch} et~al.}(2008)\citenamefont{{Bloch}, {Dalibard},
  and {Zwerger}}}]{bloch08}
\bibinfo{author}{\bibfnamefont{I.}~\bibnamefont{{Bloch}}},
  \bibinfo{author}{\bibfnamefont{J.}~\bibnamefont{{Dalibard}}},
  \bibnamefont{and}
  \bibinfo{author}{\bibfnamefont{W.}~\bibnamefont{{Zwerger}}},
  \bibinfo{journal}{Rev. Mod. Phys.} \textbf{\bibinfo{volume}{80}},
  \bibinfo{pages}{885} (\bibinfo{year}{2008}).

\bibitem[{\citenamefont{Zoller et~al.}(2005)\citenamefont{Zoller, Beth, Binosi,
  Blatt, Briegel, Bruss, Calarco, Cirac, Deutsch, Eisert et~al.}}]{Zoller05}
\bibinfo{author}{\bibfnamefont{P.}~\bibnamefont{Zoller}},
  \bibinfo{author}{\bibfnamefont{T.}~\bibnamefont{Beth}},
  \bibinfo{author}{\bibfnamefont{D.}~\bibnamefont{Binosi}},
  \bibinfo{author}{\bibfnamefont{R.}~\bibnamefont{Blatt}},
  \bibinfo{author}{\bibfnamefont{H.}~\bibnamefont{Briegel}},
  \bibinfo{author}{\bibfnamefont{D.}~\bibnamefont{Bruss}},
  \bibinfo{author}{\bibfnamefont{T.}~\bibnamefont{Calarco}},
  \bibinfo{author}{\bibfnamefont{J.~I.} \bibnamefont{Cirac}},
  \bibinfo{author}{\bibfnamefont{D.}~\bibnamefont{Deutsch}},
  \bibinfo{author}{\bibfnamefont{J.}~\bibnamefont{Eisert}},
  \bibnamefont{et~al.}, \bibinfo{journal}{Eur. Phys. J. D}
  \textbf{\bibinfo{volume}{36}}, \bibinfo{pages}{203} (\bibinfo{year}{2005}).

\bibitem[{\citenamefont{Kippenberg and Vahala}(2008)}]{Kippenberg08}
\bibinfo{author}{\bibfnamefont{T.}~\bibnamefont{Kippenberg}} \bibnamefont{and}
  \bibinfo{author}{\bibfnamefont{K.}~\bibnamefont{Vahala}},
  \bibinfo{journal}{Science} \textbf{\bibinfo{volume}{321}},
  \bibinfo{pages}{1172} (\bibinfo{year}{2008}).

\bibitem[{\citenamefont{Marquardt and Girvin}(2009)}]{Marquardt09}
\bibinfo{author}{\bibfnamefont{F.}~\bibnamefont{Marquardt}} \bibnamefont{and}
  \bibinfo{author}{\bibfnamefont{S.}~\bibnamefont{Girvin}},
  \bibinfo{journal}{Physics} \textbf{\bibinfo{volume}{2}}, \bibinfo{pages}{40}
  (\bibinfo{year}{2009}).

\bibitem[{\citenamefont{Favero and Karrai}(2009)}]{Karrai09}
\bibinfo{author}{\bibfnamefont{I.}~\bibnamefont{Favero}} \bibnamefont{and}
  \bibinfo{author}{\bibfnamefont{K.}~\bibnamefont{Karrai}},
  \bibinfo{journal}{Nature Photonics} \textbf{\bibinfo{volume}{3}},
  \bibinfo{pages}{201} (\bibinfo{year}{2009}).

\bibitem[{\citenamefont{{Genes} et~al.}(2009)\citenamefont{{Genes}, {Mari},
  {Vitali}, and {Tombesi}}}]{genes09}
\bibinfo{author}{\bibfnamefont{C.}~\bibnamefont{{Genes}}},
  \bibinfo{author}{\bibfnamefont{A.}~\bibnamefont{{Mari}}},
  \bibinfo{author}{\bibfnamefont{D.}~\bibnamefont{{Vitali}}}, \bibnamefont{and}
  \bibinfo{author}{\bibfnamefont{P.}~\bibnamefont{{Tombesi}}},
  \bibinfo{journal}{Adv. At. Mol. Opt. Phys.} \textbf{\bibinfo{volume}{57}},
  \bibinfo{pages}{33} (\bibinfo{year}{2009}).

\bibitem[{\citenamefont{Aspelmeyer et~al.}(2010)\citenamefont{Aspelmeyer,
  Gr\"{o}blacher, Hammerer, and Kiesel}}]{Aspelmeyer10}
\bibinfo{author}{\bibfnamefont{M.}~\bibnamefont{Aspelmeyer}},
  \bibinfo{author}{\bibfnamefont{S.}~\bibnamefont{Gr\"{o}blacher}},
  \bibinfo{author}{\bibfnamefont{K.}~\bibnamefont{Hammerer}}, \bibnamefont{and}
  \bibinfo{author}{\bibfnamefont{N.}~\bibnamefont{Kiesel}},
  \bibinfo{journal}{J. Opt. Soc. Am. B} \textbf{\bibinfo{volume}{27}},
  \bibinfo{pages}{A189} (\bibinfo{year}{2010}).

\bibitem[{\citenamefont{Aspelmeyer and Schwab}(2008)}]{Asp08}
\bibinfo{author}{\bibfnamefont{M.}~\bibnamefont{Aspelmeyer}} \bibnamefont{and}
  \bibinfo{author}{\bibfnamefont{K.}~\bibnamefont{Schwab}},
  \bibinfo{journal}{New J. Phys.} \textbf{\bibinfo{volume}{10}},
  \bibinfo{pages}{095001} (\bibinfo{year}{2008}),

\bibitem[{\citenamefont{{van Thourhout} and {Roels}}(2010)}]{Thourhout10}
\bibinfo{author}{\bibfnamefont{D.}~\bibnamefont{{van Thourhout}}}
  \bibnamefont{and} \bibinfo{author}{\bibfnamefont{J.}~\bibnamefont{{Roels}}},
  \bibinfo{journal}{Nature Photon.} \textbf{\bibinfo{volume}{4}},
  \bibinfo{pages}{211} (\bibinfo{year}{2010}).

\bibitem[{\citenamefont{Giscard}(2009)\citenamefont{Giscard, Bhattacharaya,
  Meystre}}]{Giscard09}
\bibinfo{author}{\bibfnamefont{P.-L.}~\bibnamefont{Giscard}},
  \bibinfo{author}{\bibfnamefont{M.}~\bibnamefont{Bhattacharaya}},
  \bibinfo{author}{\bibfnamefont{P.} \bibnamefont{Meystre}},
  \eprint{arXiv:0905.1081}.

\bibitem[{\citenamefont{Geraci et~al.}(2010)\citenamefont{Geraci, Papp, and
  Kitching}}]{Geraci10}
\bibinfo{author}{\bibfnamefont{A.~A.} \bibnamefont{Geraci}},
  \bibinfo{author}{\bibfnamefont{S.~B.} \bibnamefont{Papp}}, \bibnamefont{and}
  \bibinfo{author}{\bibfnamefont{J.}~\bibnamefont{Kitching}},
  \bibinfo{journal}{Phys. Rev. Lett.} \textbf{\bibinfo{volume}{105}},
  \bibinfo{pages}{101101} (\bibinfo{year}{2010}).

\bibitem[{\citenamefont{Stannigel et~al.}(2010)\citenamefont{Stannigel, Rabl,
  Sorensen, Zoller, and Lukin}}]{Stannigel2010}
\bibinfo{author}{\bibfnamefont{K.}~\bibnamefont{Stannigel}},
  \bibinfo{author}{\bibfnamefont{P.}~\bibnamefont{Rabl}},
  \bibinfo{author}{\bibfnamefont{A.~S.} \bibnamefont{Sorensen}},
  \bibinfo{author}{\bibfnamefont{P.}~\bibnamefont{Zoller}}, \bibnamefont{and}
  \bibinfo{author}{\bibfnamefont{M.~D.} \bibnamefont{Lukin}},
  \eprint{arXiv:1006.4361}.

\bibitem[{\citenamefont{{Rabl} et~al.}(2010)\citenamefont{{Rabl}, {Kolkowitz},
  {Koppens}, {Harris}, {Zoller}, and {Lukin}}}]{Rabl10}
\bibinfo{author}{\bibfnamefont{P.}~\bibnamefont{{Rabl}}},
  \bibinfo{author}{\bibfnamefont{S.~J.} \bibnamefont{{Kolkowitz}}},
  \bibinfo{author}{\bibfnamefont{F.~H.~L.} \bibnamefont{{Koppens}}},
  \bibinfo{author}{\bibfnamefont{J.~G.~E.} \bibnamefont{{Harris}}},
  \bibinfo{author}{\bibfnamefont{P.}~\bibnamefont{{Zoller}}}, \bibnamefont{and}
  \bibinfo{author}{\bibfnamefont{M.~D.} \bibnamefont{{Lukin}}},
  \bibinfo{journal}{Nature Phys.} \textbf{\bibinfo{volume}{6}},
  \bibinfo{pages}{602} (\bibinfo{year}{2010}).

\bibitem[{\citenamefont{{Hammerer} et~al.}(2009)\citenamefont{{Hammerer},
  {Wallquist}, {Genes}, {Ludwig}, {Marquardt}, {Treutlein}, {Zoller}, {Ye}, and
  {Kimble}}}]{Hammerer09}
\bibinfo{author}{\bibfnamefont{K.}~\bibnamefont{{Hammerer}}},
  \bibinfo{author}{\bibfnamefont{M.}~\bibnamefont{{Wallquist}}},
  \bibinfo{author}{\bibfnamefont{C.}~\bibnamefont{{Genes}}},
  \bibinfo{author}{\bibfnamefont{M.}~\bibnamefont{{Ludwig}}},
  \bibinfo{author}{\bibfnamefont{F.}~\bibnamefont{{Marquardt}}},
  \bibinfo{author}{\bibfnamefont{P.}~\bibnamefont{{Treutlein}}},
  \bibinfo{author}{\bibfnamefont{P.}~\bibnamefont{{Zoller}}},
  \bibinfo{author}{\bibfnamefont{J.}~\bibnamefont{{Ye}}}, \bibnamefont{and}
  \bibinfo{author}{\bibfnamefont{H.~J.} \bibnamefont{{Kimble}}},
  \bibinfo{journal}{Phys. Rev. Lett.} \textbf{\bibinfo{volume}{103}},
  \bibinfo{pages}{063005} (\bibinfo{year}{2009}).

\bibitem[{\citenamefont{Cleland and Geller}(2004)}]{Cleland04}
\bibinfo{author}{\bibfnamefont{A.~N.} \bibnamefont{Cleland}} \bibnamefont{and}
  \bibinfo{author}{\bibfnamefont{M.~R.} \bibnamefont{Geller}},
  \bibinfo{journal}{Phys. Rev. Lett.} \textbf{\bibinfo{volume}{93}},
  \bibinfo{pages}{070501} (\bibinfo{year}{2004}).

\bibitem[{\citenamefont{Marshall et~al.}(2003)\citenamefont{Marshall, Simon,
  Penrose, and Bouwmeester}}]{Bouwmeester03}
\bibinfo{author}{\bibfnamefont{W.}~\bibnamefont{Marshall}},
  \bibinfo{author}{\bibfnamefont{C.}~\bibnamefont{Simon}},
  \bibinfo{author}{\bibfnamefont{R.}~\bibnamefont{Penrose}}, \bibnamefont{and}
  \bibinfo{author}{\bibfnamefont{D.}~\bibnamefont{Bouwmeester}},
  \bibinfo{journal}{Phys. Rev. Lett.} \textbf{\bibinfo{volume}{91}},
  \bibinfo{pages}{130401} (\bibinfo{year}{2003}).

\bibitem[{\citenamefont{Kleckner et~al.}(2008)\citenamefont{Kleckner, Pikovski,
  Jeffrey, Ament, Eliel, van~den Brink, and Bouwmeester}}]{kleckner08}
\bibinfo{author}{\bibfnamefont{D.}~\bibnamefont{Kleckner}},
  \bibinfo{author}{\bibfnamefont{I.}~\bibnamefont{Pikovski}},
  \bibinfo{author}{\bibfnamefont{E.}~\bibnamefont{Jeffrey}},
  \bibinfo{author}{\bibfnamefont{L.}~\bibnamefont{Ament}},
  \bibinfo{author}{\bibfnamefont{E.}~\bibnamefont{Eliel}},
  \bibinfo{author}{\bibfnamefont{J.}~\bibnamefont{van~den Brink}},
  \bibnamefont{and}
  \bibinfo{author}{\bibfnamefont{D.}~\bibnamefont{Bouwmeester}},
  \bibinfo{journal}{New J. Phys.} \textbf{\bibinfo{volume}{10}},
  \bibinfo{pages}{095020} (\bibinfo{year}{2008}).

\bibitem[{\citenamefont{Barker and Shneider}(2010)}]{Barker10}
\bibinfo{author}{\bibfnamefont{P.~F.} \bibnamefont{Barker}} \bibnamefont{and}
  \bibinfo{author}{\bibfnamefont{M.~N.} \bibnamefont{Shneider}},
  \bibinfo{journal}{Phys. Rev. A} \textbf{\bibinfo{volume}{81}},
  \bibinfo{pages}{023826} (\bibinfo{year}{2010}).

\bibitem[{\citenamefont{{Singh} et~al.}(2010)\citenamefont{{Singh}, {Phelps},
  {Goldbaum}, {Wright}, and {Meystre}}}]{Singh10}
\bibinfo{author}{\bibfnamefont{S.}~\bibnamefont{{Singh}}},
  \bibinfo{author}{\bibfnamefont{G.~A.} \bibnamefont{{Phelps}}},
  \bibinfo{author}{\bibfnamefont{D.~S.} \bibnamefont{{Goldbaum}}},
  \bibinfo{author}{\bibfnamefont{E.~M.} \bibnamefont{{Wright}}},
  \bibnamefont{and}
  \bibinfo{author}{\bibfnamefont{P.}~\bibnamefont{{Meystre}}},
  \eprint{arXiv:1005.3568}.

\bibitem[{\citenamefont{{Mancini} et~al.}(1998)\citenamefont{{Mancini},
  {Vitali}, and {Tombesi}}}]{Mancini98}
\bibinfo{author}{\bibfnamefont{S.}~\bibnamefont{{Mancini}}},
  \bibinfo{author}{\bibfnamefont{D.}~\bibnamefont{{Vitali}}}, \bibnamefont{and}
  \bibinfo{author}{\bibfnamefont{P.}~\bibnamefont{{Tombesi}}},
  \bibinfo{journal}{Phys. Rev. Lett.} \textbf{\bibinfo{volume}{80}},
  \bibinfo{pages}{688} (\bibinfo{year}{1998}).

\bibitem[{\citenamefont{Marquardt et~al.}(2007)\citenamefont{Marquardt, Chen,
  Clerk, and Girvin}}]{Marquardt07}
\bibinfo{author}{\bibfnamefont{F.}~\bibnamefont{Marquardt}},
  \bibinfo{author}{\bibfnamefont{J.~P.} \bibnamefont{Chen}},
  \bibinfo{author}{\bibfnamefont{A.~A.} \bibnamefont{Clerk}}, \bibnamefont{and}
  \bibinfo{author}{\bibfnamefont{S.~M.} \bibnamefont{Girvin}},
  \bibinfo{journal}{Phys. Rev. Lett.} \textbf{\bibinfo{volume}{99}},
  \bibinfo{pages}{093902} (\bibinfo{year}{2007}).

\bibitem[{\citenamefont{Wilson-Rae et~al.}(2007)\citenamefont{Wilson-Rae,
  Nooshi, Zwerger, and Kippenberg}}]{WilsonRae07}
\bibinfo{author}{\bibfnamefont{I.}~\bibnamefont{Wilson-Rae}},
  \bibinfo{author}{\bibfnamefont{N.}~\bibnamefont{Nooshi}},
  \bibinfo{author}{\bibfnamefont{W.}~\bibnamefont{Zwerger}}, \bibnamefont{and}
  \bibinfo{author}{\bibfnamefont{T.~J.} \bibnamefont{Kippenberg}},
  \bibinfo{journal}{Phys. Rev. Lett.} \textbf{\bibinfo{volume}{99}},
  \bibinfo{pages}{093901} (\bibinfo{year}{2007}).

\bibitem[{\citenamefont{Wilson-Rae et~al.}(2008)\citenamefont{Wilson-Rae,
  Nooshi, Dobrindt, Kippenberg, and Zwerger}}]{WilsonRae08}
\bibinfo{author}{\bibfnamefont{I.}~\bibnamefont{Wilson-Rae}},
  \bibinfo{author}{\bibfnamefont{N.}~\bibnamefont{Nooshi}},
  \bibinfo{author}{\bibfnamefont{J.}~\bibnamefont{Dobrindt}},
  \bibinfo{author}{\bibfnamefont{T.~J.} \bibnamefont{Kippenberg}},
  \bibnamefont{and} \bibinfo{author}{\bibfnamefont{W.}~\bibnamefont{Zwerger}},
  \bibinfo{journal}{New J. Phys.} \textbf{\bibinfo{volume}{10}},
  \bibinfo{pages}{095007} (\bibinfo{year}{2008}).

\bibitem[{\citenamefont{Genes et~al.}(2008)\citenamefont{Genes, Vitali,
  Tombesi, Gigan, and Aspelmeyer}}]{Genes08}
\bibinfo{author}{\bibfnamefont{C.}~\bibnamefont{Genes}},
  \bibinfo{author}{\bibfnamefont{D.}~\bibnamefont{Vitali}},
  \bibinfo{author}{\bibfnamefont{P.}~\bibnamefont{Tombesi}},
  \bibinfo{author}{\bibfnamefont{S.}~\bibnamefont{Gigan}}, \bibnamefont{and}
  \bibinfo{author}{\bibfnamefont{M.}~\bibnamefont{Aspelmeyer}},
  \bibinfo{journal}{Phys. Rev. A} \textbf{\bibinfo{volume}{77}},
  \bibinfo{pages}{033804} (\bibinfo{year}{2008}).

\bibitem[{\citenamefont{Li et~al.}(2010)\citenamefont{Li, Kheifets, Medellin,
  and Raizen}}]{tongcang10}
\bibinfo{author}{\bibfnamefont{T.}~\bibnamefont{Li}},
  \bibinfo{author}{\bibfnamefont{S.}~\bibnamefont{Kheifets}},
  \bibinfo{author}{\bibfnamefont{D.}~\bibnamefont{Medellin}}, \bibnamefont{and}
  \bibinfo{author}{\bibfnamefont{M.~G.} \bibnamefont{Raizen}},
  \bibinfo{journal}{Science} \textbf{\bibinfo{volume}{328}},
  \bibinfo{pages}{1673} (\bibinfo{year}{2010}).

\bibitem[{\citenamefont{{Barker}}(2010)}]{Barker10B}
\bibinfo{author}{\bibfnamefont{P.~F.} \bibnamefont{{Barker}}},
  \bibinfo{journal}{Phys. Rev. Lett.} \textbf{\bibinfo{volume}{105}},
  \bibinfo{pages}{073002} (\bibinfo{year}{2010}).

\bibitem[{\citenamefont{{Yin} et~al.}(2010)\citenamefont{{Yin}, {Li}, and
  {Feng}}}]{Yin10}
\bibinfo{author}{\bibfnamefont{Z.}~\bibnamefont{{Yin}}},
  \bibinfo{author}{\bibfnamefont{T.}~\bibnamefont{{Li}}}, \bibnamefont{and}
  \bibinfo{author}{\bibfnamefont{M.}~\bibnamefont{{Feng}}},
  \eprint{arXiv:1007.0827}.

\bibitem[{\citenamefont{Schulze et~al.}(2010)\citenamefont{Schulze, Genes, and
  Ritsch}}]{Schulze10}
\bibinfo{author}{\bibfnamefont{R.~J.} \bibnamefont{Schulze}},
  \bibinfo{author}{\bibfnamefont{C.}~\bibnamefont{Genes}}, \bibnamefont{and}
  \bibinfo{author}{\bibfnamefont{H.}~\bibnamefont{Ritsch}},
  \bibinfo{journal}{Phys. Rev. A} \textbf{\bibinfo{volume}{81}},
  \bibinfo{pages}{063820} (\bibinfo{year}{2010}).

\bibitem[{\citenamefont{Miller et~al.}(2005)\citenamefont{Miller, Northup,
  Birnbaum, Boca, Boozer, and Kimble}}]{miller05}
\bibinfo{author}{\bibfnamefont{R.}~\bibnamefont{Miller}},
  \bibinfo{author}{\bibfnamefont{T.~E.} \bibnamefont{Northup}},
  \bibinfo{author}{\bibfnamefont{K.~M.} \bibnamefont{Birnbaum}},
  \bibinfo{author}{\bibfnamefont{A.}~\bibnamefont{Boca}},
  \bibinfo{author}{\bibfnamefont{A.~D.} \bibnamefont{Boozer}},
  \bibnamefont{and} \bibinfo{author}{\bibfnamefont{H.~J.}
  \bibnamefont{Kimble}}, \bibinfo{journal}{J. Phys. B: At. Mol. Opt. Phys.}
  \textbf{\bibinfo{volume}{38}}, \bibinfo{pages}{S551} (\bibinfo{year}{2005}).

\bibitem[{\citenamefont{Gangl and Ritsch}(1999)}]{Gan99}
\bibinfo{author}{\bibfnamefont{M.}~\bibnamefont{Gangl}} \bibnamefont{and}
  \bibinfo{author}{\bibfnamefont{H.}~\bibnamefont{Ritsch}},
  \bibinfo{journal}{Phys. Rev. A} \textbf{\bibinfo{volume}{61}},
  \bibinfo{pages}{011402} (\bibinfo{year}{1999}).

\bibitem[{\citenamefont{Vuleti\'{c} and Chu}(2000)}]{Vul00}
\bibinfo{author}{\bibfnamefont{V.}~\bibnamefont{Vuleti\'{c}}} \bibnamefont{and}
  \bibinfo{author}{\bibfnamefont{S.}~\bibnamefont{Chu}},
  \bibinfo{journal}{Phys. Rev. Lett.} \textbf{\bibinfo{volume}{84}},
  \bibinfo{pages}{3787} (\bibinfo{year}{2000}).

\bibitem[{\citenamefont{Vuleti\'{c} et~al.}(2001)\citenamefont{Vuleti\'{c},
  Chan, and Black}}]{Vul01}
\bibinfo{author}{\bibfnamefont{V.}~\bibnamefont{Vuleti\'{c}}},
  \bibinfo{author}{\bibfnamefont{H.~W.} \bibnamefont{Chan}}, \bibnamefont{and}
  \bibinfo{author}{\bibfnamefont{A.~T.} \bibnamefont{Black}},
  \bibinfo{journal}{Phys. Rev. A} \textbf{\bibinfo{volume}{64}},
  \bibinfo{pages}{033405} (\bibinfo{year}{2001}).

\bibitem[{\citenamefont{Ashkin and Dziedzic}(1971)}]{Ashkin71}
\bibinfo{author}{\bibfnamefont{A.}~\bibnamefont{Ashkin}} \bibnamefont{and}
  \bibinfo{author}{\bibfnamefont{J.~M.} \bibnamefont{Dziedzic}},
  \bibinfo{journal}{Appl. Phys. Lett.} \textbf{\bibinfo{volume}{19}},
  \bibinfo{pages}{283} (\bibinfo{year}{1971}).

\bibitem[{\citenamefont{Ashkin and Dziedzic}(1974)}]{Ashkin74}
\bibinfo{author}{\bibfnamefont{A.}~\bibnamefont{Ashkin}} \bibnamefont{and}
  \bibinfo{author}{\bibfnamefont{J.~M.} \bibnamefont{Dziedzic}},
  \bibinfo{journal}{Appl. Phys. Lett.} \textbf{\bibinfo{volume}{24}},
  \bibinfo{pages}{586} (\bibinfo{year}{1974}).

\bibitem[{\citenamefont{Ashkin and Dziedzic}(1976)}]{Ashkin76}
\bibinfo{author}{\bibfnamefont{A.}~\bibnamefont{Ashkin}} \bibnamefont{and}
  \bibinfo{author}{\bibfnamefont{J.~M.} \bibnamefont{Dziedzic}},
  \bibinfo{journal}{Appl. Phys. Lett.} \textbf{\bibinfo{volume}{28}},
  \bibinfo{pages}{333} (\bibinfo{year}{1976}).

\bibitem[{\citenamefont{{Sankey} et~al.}(2010)\citenamefont{{Sankey}, {Yang},
  {Zwickl}, {Jayich}, and {Harris}}}]{sankey10}
\bibinfo{author}{\bibfnamefont{J.~C.} \bibnamefont{{Sankey}}},
  \bibinfo{author}{\bibfnamefont{C.}~\bibnamefont{{Yang}}},
  \bibinfo{author}{\bibfnamefont{B.~M.} \bibnamefont{{Zwickl}}},
  \bibinfo{author}{\bibfnamefont{A.~M.} \bibnamefont{{Jayich}}},
  \bibnamefont{and} \bibinfo{author}{\bibfnamefont{J.~G.~E.}
  \bibnamefont{{Harris}}},
 \eprint{arXiv:1002.4158}.

\bibitem[{\citenamefont{Gardiner and Zoller}(2004)}]{gardinerbook}
\bibinfo{author}{\bibfnamefont{C.~W.} \bibnamefont{Gardiner}} \bibnamefont{and}
  \bibinfo{author}{\bibfnamefont{P.}~\bibnamefont{Zoller}},
  \emph{\bibinfo{title}{Quantum Noise}} (\bibinfo{publisher}{Springer},
  \bibinfo{address}{Berlin}, \bibinfo{year}{2004}).

\bibitem[{\citenamefont{Pflanzer}()}]{pflanzer11}
\bibinfo{author}{\bibfnamefont{A.~C.} \bibnamefont{Pflanzer}},
  \bibinfo{journal}{in preparation}.

\bibitem[{\citenamefont{{Dobrindt} et~al.}(2008)\citenamefont{{Dobrindt},
  {Wilson-Rae}, and {Kippenberg}}}]{dobrindt08}
\bibinfo{author}{\bibfnamefont{J.~M.} \bibnamefont{{Dobrindt}}},
  \bibinfo{author}{\bibfnamefont{I.}~\bibnamefont{{Wilson-Rae}}},
  \bibnamefont{and} \bibinfo{author}{\bibfnamefont{T.~J.}
  \bibnamefont{{Kippenberg}}}, \bibinfo{journal}{Phys. Rev. Lett.}
  \textbf{\bibinfo{volume}{101}}, \bibinfo{pages}{263602}
  (\bibinfo{year}{2008}).

\bibitem[{\citenamefont{Lvovsky et~al.}(2001)\citenamefont{Lvovsky, Hansen,
  Aichele, Benson, Mlynek, and Schiller}}]{Lvovsky01}
\bibinfo{author}{\bibfnamefont{A.~I.} \bibnamefont{Lvovsky}},
  \bibinfo{author}{\bibfnamefont{H.}~\bibnamefont{Hansen}},
  \bibinfo{author}{\bibfnamefont{T.}~\bibnamefont{Aichele}},
  \bibinfo{author}{\bibfnamefont{O.}~\bibnamefont{Benson}},
  \bibinfo{author}{\bibfnamefont{J.}~\bibnamefont{Mlynek}}, \bibnamefont{and}
  \bibinfo{author}{\bibfnamefont{S.}~\bibnamefont{Schiller}},
  \bibinfo{journal}{Phys. Rev. Lett.} \textbf{\bibinfo{volume}{87}},
  \bibinfo{pages}{050402} (\bibinfo{year}{2001}).

\bibitem[{\citenamefont{Dutra}(2005)}]{dutrabook}
\bibinfo{author}{\bibfnamefont{S.~M.} \bibnamefont{Dutra}},
  \emph{\bibinfo{title}{Cavity Quantum Electrodynamics}}
  (\bibinfo{publisher}{Wiley}, \bibinfo{address}{New York},
  \bibinfo{year}{2005}).

\bibitem[{\citenamefont{{Born} and {Wolf}}()}]{born_wolf}
\bibinfo{author}{\bibfnamefont{M.}~\bibnamefont{{Born}}} \bibnamefont{and}
  \bibinfo{author}{\bibfnamefont{E.}~\bibnamefont{{Wolf}}},
  \emph{\bibinfo{title}{Principles of Optics}} (\bibinfo{publisher}{Pergamon
  Press}, 1980), \bibinfo{edition}{6th} ed.

\bibitem[{\citenamefont{Jackson}(1998)}]{jackson}
\bibinfo{author}{\bibfnamefont{J.~D.} \bibnamefont{Jackson}},
  \emph{\bibinfo{title}{Classical Electrodynamics}}
  (\bibinfo{publisher}{Wiley}, \bibinfo{address}{New York},
  \bibinfo{year}{1998}), \bibinfo{edition}{3rd} ed.

\bibitem[{\citenamefont{Ashkin et~al.}(1986)\citenamefont{Ashkin, Dziedzic,
  Bjorkholm, and Chu}}]{Ashkin86}
\bibinfo{author}{\bibfnamefont{A.}~\bibnamefont{Ashkin}},
  \bibinfo{author}{\bibfnamefont{J.~M.} \bibnamefont{Dziedzic}},
  \bibinfo{author}{\bibfnamefont{J.~E.} \bibnamefont{Bjorkholm}},
  \bibnamefont{and} \bibinfo{author}{\bibfnamefont{S.}~\bibnamefont{Chu}},
  \bibinfo{journal}{Opt. Lett.} \textbf{\bibinfo{volume}{11}},
  \bibinfo{pages}{288} (\bibinfo{year}{1986}).

\bibitem[{\citenamefont{{Joos} and {Zeh}}(1985)}]{joos85}
\bibinfo{author}{\bibfnamefont{E.}~\bibnamefont{{Joos}}} \bibnamefont{and}
  \bibinfo{author}{\bibfnamefont{H.~D.} \bibnamefont{{Zeh}}},
  \bibinfo{journal}{Z. Phys. B} \textbf{\bibinfo{volume}{59}},
  \bibinfo{pages}{223} (\bibinfo{year}{1985}).

\bibitem[{\citenamefont{{Joos} et~al.}(2003)\citenamefont{{Joos}, {Zeh},
  {Kiefer}, {Giulini}, {Kupsch}, and {Stamatescu}}}]{joos_decoherence}
\bibinfo{author}{\bibfnamefont{E.}~\bibnamefont{{Joos}}},
  \bibinfo{author}{\bibfnamefont{H.}~\bibnamefont{{Zeh}}},
  \bibinfo{author}{\bibfnamefont{C.}~\bibnamefont{{Kiefer}}},
  \bibinfo{author}{\bibfnamefont{D.}~\bibnamefont{{Giulini}}},
  \bibinfo{author}{\bibfnamefont{J.}~\bibnamefont{{Kupsch}}}, \bibnamefont{and}
  \bibinfo{author}{\bibfnamefont{I.-O.} \bibnamefont{{Stamatescu}}},
  \emph{\bibinfo{title}{Decoherence and the Appearance of a Classical World in
  Quantum Theory}} (\bibinfo{publisher}{Springer}, \bibinfo{address}{New York},
  \bibinfo{year}{2003}).

\bibitem[{\citenamefont{{Schlosshauer}}(2008)}]{schlosshauer_decoherence}
\bibinfo{author}{\bibfnamefont{M.}~\bibnamefont{{Schlosshauer}}},
  \emph{\bibinfo{title}{Decoherence and the Quantum-to-Classical Transition}}
  (\bibinfo{publisher}{Springer}, \bibinfo{address}{Heidelberg},
  \bibinfo{year}{2008}).

\bibitem[{\citenamefont{Landau and Lifschitz}(1970)}]{landaubook7}
\bibinfo{author}{\bibfnamefont{L.}~\bibnamefont{Landau}} \bibnamefont{and}
  \bibinfo{author}{\bibfnamefont{E.}~\bibnamefont{Lifschitz}},
  \emph{\bibinfo{title}{Theory of Elasticity, 2nd Edition, Vol.7 of Course of
  Theoretical Physics}} (\bibinfo{publisher}{Pergamonn Press},
  \bibinfo{address}{Oxford}, \bibinfo{year}{1970}).

\bibitem[{\citenamefont{Goldstein et~al.}(2002)\citenamefont{Goldstein, Poole,
  and Safko}}]{goldsteinbook}
\bibinfo{author}{\bibfnamefont{H.}~\bibnamefont{Goldstein}},
  \bibinfo{author}{\bibfnamefont{C.}~\bibnamefont{Poole}}, \bibnamefont{and}
  \bibinfo{author}{\bibfnamefont{J.}~\bibnamefont{Safko}},
  \emph{\bibinfo{title}{Classical Mechanics, 3rd Edition}}
  (\bibinfo{publisher}{Addison Wesley}, \bibinfo{address}{San Francisco},
  \bibinfo{year}{2002}).

\bibitem[{\citenamefont{Bose et~al.}(1997)\citenamefont{Bose, Jacobs, and
  Knight}}]{bose97}
\bibinfo{author}{\bibfnamefont{S.}~\bibnamefont{Bose}},
  \bibinfo{author}{\bibfnamefont{K.}~\bibnamefont{Jacobs}}, \bibnamefont{and}
  \bibinfo{author}{\bibfnamefont{P.}~\bibnamefont{Knight}},
  \bibinfo{journal}{Phys Rev A} \textbf{\bibinfo{volume}{56}},
  \bibinfo{pages}{4175} (\bibinfo{year}{1997}).

\bibitem[{\citenamefont{Mancini et~al.}(1997)\citenamefont{Mancini, Manko, and
  Tombesi}}]{mancini97}
\bibinfo{author}{\bibfnamefont{S.}~\bibnamefont{Mancini}},
  \bibinfo{author}{\bibfnamefont{V.}~\bibnamefont{Manko}}, \bibnamefont{and}
  \bibinfo{author}{\bibfnamefont{P.}~\bibnamefont{Tombesi}},
  \bibinfo{journal}{Phys. Rev. A} \textbf{\bibinfo{volume}{55}},
  \bibinfo{pages}{3042} (\bibinfo{year}{1997}).

\bibitem[{\citenamefont{Bose et~al.}(1999)\citenamefont{Bose, Jacobs, and
  Knight}}]{bose99}
\bibinfo{author}{\bibfnamefont{S.}~\bibnamefont{Bose}},
  \bibinfo{author}{\bibfnamefont{K.}~\bibnamefont{Jacobs}}, \bibnamefont{and}
  \bibinfo{author}{\bibfnamefont{P.}~\bibnamefont{Knight}},
  \bibinfo{journal}{Phys. Rev. A} \textbf{\bibinfo{volume}{59}},
  \bibinfo{pages}{3204} (\bibinfo{year}{1999}).

\bibitem[{\citenamefont{Armour et~al.}(2002)\citenamefont{Armour, Blencowe, and
  Schwab}}]{armour02}
\bibinfo{author}{\bibfnamefont{A.}~\bibnamefont{Armour}},
  \bibinfo{author}{\bibfnamefont{M.}~\bibnamefont{Blencowe}}, \bibnamefont{and}
  \bibinfo{author}{\bibfnamefont{K.}~\bibnamefont{Schwab}},
  \bibinfo{journal}{Phys. Rev. Lett} \textbf{\bibinfo{volume}{88}},
  \bibinfo{pages}{148301} (\bibinfo{year}{2002}).

\bibitem[{\citenamefont{O'Connell et~al.}(2010)\citenamefont{O'Connell,
  Hofheinz, Ansmann, Bialczak, Lenander, Lucero, Neeley, Sank, Wang, Weides
  et~al.}}]{O'Connell2010a}
\bibinfo{author}{\bibfnamefont{A.~D.} \bibnamefont{O'Connell}},
  \bibinfo{author}{\bibfnamefont{M.}~\bibnamefont{Hofheinz}},
  \bibinfo{author}{\bibfnamefont{M.}~\bibnamefont{Ansmann}},
  \bibinfo{author}{\bibfnamefont{R.~C.} \bibnamefont{Bialczak}},
  \bibinfo{author}{\bibfnamefont{M.}~\bibnamefont{Lenander}},
  \bibinfo{author}{\bibfnamefont{E.}~\bibnamefont{Lucero}},
  \bibinfo{author}{\bibfnamefont{M.}~\bibnamefont{Neeley}},
  \bibinfo{author}{\bibfnamefont{D.}~\bibnamefont{Sank}},
  \bibinfo{author}{\bibfnamefont{H.}~\bibnamefont{Wang}},
  \bibinfo{author}{\bibfnamefont{M.}~\bibnamefont{Weides}},
  \bibnamefont{et~al.}, \bibinfo{journal}{Nature}
  \textbf{\bibinfo{volume}{464}}, \bibinfo{pages}{697} (\bibinfo{year}{2010}).

\bibitem[{\citenamefont{Groeblacher et~al.}(2009)\citenamefont{Groeblacher,
  Hammerer, Vanner, and Aspelmeyer}}]{Groeblacher09b}
\bibinfo{author}{\bibfnamefont{S.}~\bibnamefont{Groeblacher}},
  \bibinfo{author}{\bibfnamefont{K.}~\bibnamefont{Hammerer}},
  \bibinfo{author}{\bibfnamefont{M.~R.} \bibnamefont{Vanner}},
  \bibnamefont{and}
  \bibinfo{author}{\bibfnamefont{M.}~\bibnamefont{Aspelmeyer}},
  \bibinfo{journal}{Nature} \textbf{\bibinfo{volume}{460}},
  \bibinfo{pages}{724} (\bibinfo{year}{2009}).

\bibitem[{\citenamefont{{Akram} et~al.}(2010)\citenamefont{{Akram}, {Kiesel},
  {Aspelmeyer}, and {Milburn}}}]{akram10}
\bibinfo{author}{\bibfnamefont{U.}~\bibnamefont{{Akram}}},
  \bibinfo{author}{\bibfnamefont{N.}~\bibnamefont{{Kiesel}}},
  \bibinfo{author}{\bibfnamefont{M.}~\bibnamefont{{Aspelmeyer}}},
  \bibnamefont{and} \bibinfo{author}{\bibfnamefont{G.~J.}
  \bibnamefont{{Milburn}}}, \bibinfo{journal}{New J. Phys.}
  \textbf{\bibinfo{volume}{12}}, \bibinfo{pages}{083030}
  (\bibinfo{year}{2010}).

\bibitem[{\citenamefont{{Khalili} et~al.}(2010)\citenamefont{{Khalili},
  {Danilishin}, {Miao}, {M{\"u}ller-Ebhardt}, {Yang}, and {Chen}}}]{Khalili10}
\bibinfo{author}{\bibfnamefont{F.}~\bibnamefont{{Khalili}}},
  \bibinfo{author}{\bibfnamefont{S.}~\bibnamefont{{Danilishin}}},
  \bibinfo{author}{\bibfnamefont{H.}~\bibnamefont{{Miao}}},
  \bibinfo{author}{\bibfnamefont{H.}~\bibnamefont{{M{\"u}ller-Ebhardt}}},
  \bibinfo{author}{\bibfnamefont{H.}~\bibnamefont{{Yang}}}, \bibnamefont{and}
  \bibinfo{author}{\bibfnamefont{Y.}~\bibnamefont{{Chen}}},
  \bibinfo{journal}{Phys. Rev. Lett.} \textbf{\bibinfo{volume}{105}},
  \bibinfo{pages}{070403} (\bibinfo{year}{2010}).

\bibitem[{\citenamefont{{Lloyd} and {Braunstein}}(1999)}]{lloyd99}
\bibinfo{author}{\bibfnamefont{S.}~\bibnamefont{{Lloyd}}} \bibnamefont{and}
  \bibinfo{author}{\bibfnamefont{S.~L.} \bibnamefont{{Braunstein}}},
  \bibinfo{journal}{Phys. Rev. Lett.} \textbf{\bibinfo{volume}{82}},
  \bibinfo{pages}{1784} (\bibinfo{year}{1999}).

\bibitem[{\citenamefont{{Knill} et~al.}(2001)\citenamefont{{Knill}, {Laflamme},
  and {Milburn}}}]{knill01}
\bibinfo{author}{\bibfnamefont{E.}~\bibnamefont{{Knill}}},
  \bibinfo{author}{\bibfnamefont{R.}~\bibnamefont{{Laflamme}}},
  \bibnamefont{and} \bibinfo{author}{\bibfnamefont{G.~J.}
  \bibnamefont{{Milburn}}}, \bibinfo{journal}{Nature}
  \textbf{\bibinfo{volume}{409}}, \bibinfo{pages}{46} (\bibinfo{year}{2001}).

\bibitem[{\citenamefont{{Duan} and {Kimble}}(2004)}]{Duan04}
\bibinfo{author}{\bibfnamefont{L.}~\bibnamefont{{Duan}}} \bibnamefont{and}
  \bibinfo{author}{\bibfnamefont{H.~J.} \bibnamefont{{Kimble}}},
  \bibinfo{journal}{Phys. Rev. Lett.} \textbf{\bibinfo{volume}{92}},
  \bibinfo{pages}{127902} (\bibinfo{year}{2004}).

\bibitem[{\citenamefont{Cirac et~al.}(1997)\citenamefont{Cirac, Zoller, Kimble,
  and Mabuchi}}]{Cirac97}
\bibinfo{author}{\bibfnamefont{J.~I.} \bibnamefont{Cirac}},
  \bibinfo{author}{\bibfnamefont{P.}~\bibnamefont{Zoller}},
  \bibinfo{author}{\bibfnamefont{H.~J.} \bibnamefont{Kimble}},
  \bibnamefont{and} \bibinfo{author}{\bibfnamefont{H.}~\bibnamefont{Mabuchi}},
  \bibinfo{journal}{Phys. Rev. Lett.} \textbf{\bibinfo{volume}{78}},
  \bibinfo{pages}{3221} (\bibinfo{year}{1997}).

\bibitem[{\citenamefont{Braunstein and Kimble}(1998)}]{Braunstein98}
\bibinfo{author}{\bibfnamefont{S.~L.} \bibnamefont{Braunstein}}
  \bibnamefont{and} \bibinfo{author}{\bibfnamefont{H.~J.}
  \bibnamefont{Kimble}}, \bibinfo{journal}{Phys. Rev. Lett.}
  \textbf{\bibinfo{volume}{80}}, \bibinfo{pages}{869} (\bibinfo{year}{1998}).

\bibitem[{\citenamefont{{Schwager} et~al.}(2010)\citenamefont{{Schwager},
  {Cirac}, and {Giedke}}}]{Schwager10}
\bibinfo{author}{\bibfnamefont{H.}~\bibnamefont{{Schwager}}},
  \bibinfo{author}{\bibfnamefont{J.~I.} \bibnamefont{{Cirac}}},
  \bibnamefont{and} \bibinfo{author}{\bibfnamefont{G.}~\bibnamefont{{Giedke}}},
  \bibinfo{journal}{New J. Phys.} \textbf{\bibinfo{volume}{12}},
  \bibinfo{pages}{043026} (\bibinfo{year}{2010}).

\bibitem[{\citenamefont{Fiur\'a\ifmmode~\check{s}\else
  \v{s}\fi{}ek}(2002)}]{Fiurasek02}
\bibinfo{author}{\bibfnamefont{J.}~\bibnamefont{Fiur\'a\ifmmode~\check{s}\else
  \v{s}\fi{}ek}}, \bibinfo{journal}{Phys. Rev. A}
  \textbf{\bibinfo{volume}{66}}, \bibinfo{pages}{012304}
  (\bibinfo{year}{2002}).

\bibitem[{\citenamefont{{Lvovsky} and {Raymer}}(2009)}]{lvovsky09}
\bibinfo{author}{\bibfnamefont{A.~I.} \bibnamefont{{Lvovsky}}}
  \bibnamefont{and} \bibinfo{author}{\bibfnamefont{M.~G.}
  \bibnamefont{{Raymer}}}, \bibinfo{journal}{Rev. Mod. Phys.}
  \textbf{\bibinfo{volume}{81}}, \bibinfo{pages}{299} (\bibinfo{year}{2009}).

\bibitem[{\citenamefont{Braginsky et~al.}(1980)\citenamefont{Braginsky,
  Vorontsov, and Thorne}}]{braginsky80}
\bibinfo{author}{\bibfnamefont{V.~B.} \bibnamefont{Braginsky}},
  \bibinfo{author}{\bibfnamefont{Y.~I.} \bibnamefont{Vorontsov}},
  \bibnamefont{and} \bibinfo{author}{\bibfnamefont{K.~S.}
  \bibnamefont{Thorne}}, \bibinfo{journal}{Science}
  \textbf{\bibinfo{volume}{209}}, \bibinfo{pages}{547} (\bibinfo{year}{1980}).

\bibitem[{\citenamefont{Clerk et~al.}(2008)\citenamefont{Clerk, Marquardt, and
  Jacobs}}]{clerk08}
\bibinfo{author}{\bibfnamefont{A.~A.} \bibnamefont{Clerk}},
  \bibinfo{author}{\bibfnamefont{F.}~\bibnamefont{Marquardt}},
  \bibnamefont{and} \bibinfo{author}{\bibfnamefont{K.}~\bibnamefont{Jacobs}},
  \bibinfo{journal}{New J. Phys.} \textbf{\bibinfo{volume}{10}},
  \bibinfo{pages}{095010} (\bibinfo{year}{2008}).

\bibitem[{\citenamefont{{Vahala} et~al.}(2009)\citenamefont{{Vahala},
  {Herrmann}, {Kn{\"u}nz}, {Batteiger}, {Saathoff}, {H{\"a}nsch}, and
  {Udem}}}]{Vahala09}
\bibinfo{author}{\bibfnamefont{K.}~\bibnamefont{{Vahala}}},
  \bibinfo{author}{\bibfnamefont{M.}~\bibnamefont{{Herrmann}}},
  \bibinfo{author}{\bibfnamefont{S.}~\bibnamefont{{Kn{\"u}nz}}},
  \bibinfo{author}{\bibfnamefont{V.}~\bibnamefont{{Batteiger}}},
  \bibinfo{author}{\bibfnamefont{G.}~\bibnamefont{{Saathoff}}},
  \bibinfo{author}{\bibfnamefont{T.~W.} \bibnamefont{{H{\"a}nsch}}},
  \bibnamefont{and} \bibinfo{author}{\bibfnamefont{T.}~\bibnamefont{{Udem}}},
  \bibinfo{journal}{Nature Phys.} \textbf{\bibinfo{volume}{5}},
  \bibinfo{pages}{682} (\bibinfo{year}{2009}).

\bibitem[{\citenamefont{Cirac et~al.}(1993)\citenamefont{Cirac, Parkins, Blatt,
  and Zoller}}]{cirac93}
\bibinfo{author}{\bibfnamefont{J.~I.} \bibnamefont{Cirac}},
  \bibinfo{author}{\bibfnamefont{A.~S.} \bibnamefont{Parkins}},
  \bibinfo{author}{\bibfnamefont{R.}~\bibnamefont{Blatt}}, \bibnamefont{and}
  \bibinfo{author}{\bibfnamefont{P.}~\bibnamefont{Zoller}},
  \bibinfo{journal}{Phys. Rev. Lett.} \textbf{\bibinfo{volume}{70}},
  \bibinfo{pages}{556} (\bibinfo{year}{1993}).

\bibitem[{\citenamefont{Garc\'\i{}a~de Abajo}(2007)}]{abajo07}
\bibinfo{author}{\bibfnamefont{F.~J.} \bibnamefont{Garc\'\i{}a~de Abajo}},
  \bibinfo{journal}{Rev. Mod. Phys.} \textbf{\bibinfo{volume}{79}},
  \bibinfo{pages}{1267} (\bibinfo{year}{2007}).

\bibitem[{\citenamefont{{Purcell} and {Pennyparker}}(1973)}]{purcell73}
\bibinfo{author}{\bibfnamefont{E.~M.} \bibnamefont{{Purcell}}}
  \bibnamefont{and} \bibinfo{author}{\bibfnamefont{C.~R.}
  \bibnamefont{{Pennyparker}}}, \bibinfo{journal}{The Astrophysical Journal}
  \textbf{\bibinfo{volume}{186}}, \bibinfo{pages}{705} (\bibinfo{year}{1973}).

\bibitem[{\citenamefont{{Stratton}}(1941)}]{strattonbook}
\bibinfo{author}{\bibfnamefont{J.~A.} \bibnamefont{{Stratton}}},
  \emph{\bibinfo{title}{Electromagnetic Theory}}
  (\bibinfo{publisher}{McGraw-Hill}, \bibinfo{address}{New York},
  \bibinfo{year}{1941}).

\bibitem[{\citenamefont{{Bohren} and {Huffman}}(1983)}]{bohrenbook}
\bibinfo{author}{\bibfnamefont{C.~F.} \bibnamefont{{Bohren}}} \bibnamefont{and}
  \bibinfo{author}{\bibfnamefont{D.~R.} \bibnamefont{{Huffman}}},
  \emph{\bibinfo{title}{Absorption and Scattering of Light by Small Particles}}
  (\bibinfo{publisher}{Wiley}, \bibinfo{address}{New York},
  \bibinfo{year}{1983}).

\bibitem[{\citenamefont{{Ashkin}}(1992)}]{ashkin92}
\bibinfo{author}{\bibfnamefont{A.}~\bibnamefont{{Ashkin}}},
  \bibinfo{journal}{Biophys. J.} \textbf{\bibinfo{volume}{61}},
  \bibinfo{pages}{569} (\bibinfo{year}{1992}).

\end{thebibliography}
\end{document}